\documentclass[aps,prd,showpacs,twocolumn,superscriptaddress,nofootinbib,preprintnumbers]{revtex4-1} 

\makeatletter
\def\p@subsection{}
\makeatother

\usepackage{color,graphicx,amsmath,amssymb,lipsum,bm}
\usepackage[colorlinks=true,citecolor=blue,linkcolor=blue,urlcolor=blue, backref=false,pdfborder={0 0 0}]{hyperref}

\usepackage[normalem]{ulem}
\usepackage[utf8]{inputenc} 
\usepackage{mathtools}

\usepackage{float}
\usepackage{aas_macros}
\usepackage{multirow}

\usepackage{ulem}
\usepackage{diagbox}

\usepackage[dvipsnames]{xcolor}
\usepackage{mathrsfs}

\definecolor{xlinkcolor}{rgb}{0.7752941176470588, 0.22078431372549023, 0.2262745098039215}

\newcommand{\be}{\begin{equation}}
\newcommand{\ee}{\end{equation}}
\newcommand{\beqa}{\begin{eqnarray}}
\newcommand{\eeqa}{\end{eqnarray}}

\newcommand\p{{\bm p}}
\renewcommand\k{{\bm k}}
\newcommand\q{\bm{q}}

\newcommand\GG{\Gamma_3}
\newcommand\G{\mathcal{G}_2}

\newcommand{\bseq}{\begin{subequations}}
\newcommand{\eseq}{\end{subequations}}

\def\vpsi{{\boldsymbol{\psi}}}

\def\gsim{\raise0.3ex\hbox{$\;>$\kern-0.75em\raise-1.1ex\hbox{$\sim\;$}}}
\def\lsim{\raise0.3ex\hbox{$\;<$\kern-0.75em\raise-1.1ex\hbox{$\sim\;$}}}

\def\beqn#1{\begin{equation}\label{#1}}
\def\eeqn{\end{equation}}

\def\beqa#1{\begin{eqnarray}\label{#1}}
\def\eeqa{\end{eqnarray}}

\def\kmax{{k_\text{max}}}
\def\hMpc{h{\text{Mpc}}^{-1}}

\def\Z2{$\mathcal{Z_2}$}

\newcommand {\ignore}[1]{}

\begin{document}

\preprint{MIT-CTP/5867}

\title{High-redshift Millennium and Astrid galaxies 
in effective field theory at the field level
}

\author{James M. Sullivan}
\email{jms3@mit.edu}\thanks{Brinson Prize Fellow} 
\affiliation{Center for Theoretical Physics, Massachusetts Institute of Technology, 
Cambridge, MA 02139, USA}

\author{Carolina Cuesta-Lazaro}
\email{cuestalz@mit.edu}
 \affiliation{The NSF AI Institute for Artificial Intelligence and Fundamental Interactions, Cambridge, MA 02139, USA}
\affiliation{Department of Physics, Massachusetts Institute of Technology, Cambridge, MA 02139, USA}
\affiliation{Center for Astrophysics $\vert$ Harvard \& Smithsonian, 60 Garden St, Cambridge, MA 02138, USA}

\author{Mikhail M. Ivanov }
\email{ivanov99@mit.edu}
\affiliation{Center for Theoretical Physics, Massachusetts Institute of Technology, 
Cambridge, MA 02139, USA} 
 \affiliation{The NSF AI Institute for Artificial Intelligence and Fundamental Interactions, Cambridge, MA 02139, USA}

\author{Yueying Ni}
\affiliation{Center for Astrophysics $\vert$ Harvard \& Smithsonian, 60 Garden St, Cambridge, MA 02138, USA}

 \author{Sownak Bose}
\affiliation{Institute for Computational Cosmology, Department of Physics, Durham University, South Road, Durham DH1 3LE, UK} 

\author{Boryana Hadzhiyska}
\affiliation{Lawrence Berkeley National Laboratory, 1 Cyclotron Road, Berkeley, CA 94720, USA}
\affiliation{Berkeley Center for Cosmological Physics, Department of Physics,
University of California, Berkeley, CA 94720, USA}
\affiliation{Miller Institute for Basic Research in Science, University of California, Berkeley, CA, 94720, USA}

\author{C\'esar Hern\'andez-Aguayo}
\affiliation{Max-Planck-Institut f\"ur Astrophysik, Karl-Schwarzschild-Str. 1, D-85748, Garching, Germany}
\affiliation{
Excellence Cluster ORIGINS, Boltzmannstrasse 2, D-85748 Garching, Germany}

\author{Lars Hernquist}
\affiliation{Center for Astrophysics $\vert$ Harvard \& Smithsonian, 60 Garden St, Cambridge, MA 02138, USA}

\author{Rahul Kannan}
\affiliation{Department of Physics and Astronomy, York University, 4700 Keele Street, Toronto, ON M3J 1P3, Canada}

\begin{abstract} 
Effective Field Theory (EFT) modeling is expected to be a useful tool in the era of future higher-redshift galaxy surveys 
such as DESI-II and Spec-S5 due to its robust description of various large-scale structure tracers.
However, large values of EFT bias parameters of higher-redshift galaxies could jeopardize the convergence of the perturbative expansion.
In this paper we measure the bias parameters and other EFT coefficients from samples of two types of star-forming galaxies in the state-of-the-art MilleniumTNG and Astrid hydrodynamical simulations. Our measurements
are based on the field-level EFT forward model that allows for precision EFT parameter measurements by virtue of cosmic variance cancellation. 
Specifically, we consider approximately representative samples of Lyman-break galaxies (LBGs) and Lyman-alpha emitters (LAEs) that are consistent with the observed (angular) clustering and number density of these galaxies at $z=3$.
Reproducing the linear biases and number densities observed from existing LAE and LBG data, we find quadratic bias parameters that are roughly consistent with those predicted from the halo model coupled with a simple halo occupation distribution model.
We also find non-perturbative velocity contributions (Fingers of God) of a similar size for LBGs to the familiar case of Luminous Red Galaxies.
However, these contributions are quite small for LAEs despite their large satellite fraction values of up to $\sim 30\%$.
Our results indicate that the effective momentum reach $\kmax$ at $z=3$ for LAEs (LBGs) will be in the range $0.3-0.6 ~h\rm{Mpc}^{-1}$ 
($0.2-0.8~h\rm{Mpc}^{-1}$), suggesting that EFT will perform well for high redshift galaxy clustering.
This work provides the first step toward obtaining realistic simulation-based priors on EFT parameters for LAEs and LBGs.
\end{abstract}

\maketitle

\section{Introduction \label{sec:intro}}
Upcoming ground-based spectroscopic galaxy surveys including DESI-II \cite{Schlegel:2022_spec_roadmap} as well as the proposed Spec-S5 mission \cite{2025:spec_s5} will map the high-redshift universe with unprecedented coverage.
At higher redshifts, the dark matter density and velocity fields are significantly less non-linear at a given length scale, as structure has not had as much time to grow.
This would suggest that for these surveys, perturbation theory-based approaches to modeling Large-scale Structure (LSS) tracers, such as Effective Field Theory (EFT) methods \cite{Baumann:2010tm,Carrasco:2012cv,Ivanov:2022mrd} with the large-scale biasing framework \cite{Desjacques:2016bnm} would lead to 
more accessible cosmological information than for existing lower redshift surveys. 
However, simple estimates of halo bias, e.g. from the peak-background split argument \cite{Bardeen:1985tr,Kaiser:1984}, suggest that the non-linear bias contributions to the galaxy density field grow rapidly with redshift.
If the bias of galaxies observed by high redshift surveys retains the same relationship to halo bias as in their low redshift counterparts, then uncertainties in galaxy formation may limit the potential of these upcoming surveys on all but the largest scales.\footnote{In the peak-background split model the redshift growth of the bias parameters is exactly compensated by the suppression of the matter density field at high redshifts. An additional enhancement of bias parameters with redshift, however, takes place if the halo mass selection is redshift-dependent, which can be the case for a realistic galaxy sample.}

This picture is complicated by the fact that galaxy populations at high redshift are quite different from those that have traditionally been used for LSS analysis at low redshift.
In particular, the galaxies that will be selected for these cosmological samples are, unlike the familiar case of Luminous Red Galaxies (LRGs) \cite{2001:eisenstein_lrgs}, undergoing significant active star formation.
These star-forming galaxies (SFGs) are also different from the Emission Line Galaxies (ELGs) of current ground-based surveys 
such as DESI \cite{Aghamousa:2016zmz}, which are selected largely based on their [OII] emission \cite{Raichoor:2023_elg}.
Instead, spectral features related to Lyman
series 
emission and absorption are used to identify the higher-redshift galaxy populations of interest, as these rest-frame UV features are observed in optical photometric bands.
These high-redshift ($z\approx2.5-5.0$) galaxies can be defined according to their selection properties into two populations. 

Lyman-break galaxies (LBGs) are star-forming galaxies that exhibit a relative flux excess at wavelengths greater than the Lyman-$\alpha$ line relative to their flux blueward of the same line\footnote{Of course, at all redshifts, flux in the rest-frame Lyman continuum is suppressed, while at lower redshifts, there is higher emitted flux between the Lyman limit and the Lyman-$\alpha$ line, though in both cases, the break at Lyman-$\alpha$ is significant.}.
Roughly, LBGs can be selected when a galaxy ``drops out'' of a particular bluer photometric band (e.g. the SDSS $u$ band) while retaining flux in other redder bands
\cite{Steidel1992:lyman_break_technique,Shapley:2003_lbg,Moscardini:1998_early_lbg_acf,Wilson_White:2019_lbg, Ebina_White:2024_forecast}.
The limit to the number of observed galaxies at a given redshift 
depends on the magnitude limit of the imaging, which can be associated with a minimum galaxy stellar (halo) mass threshold. 
As such, LBGs will be the most massive and luminous SFGs \cite{2025:spec_s5,Park:2016_lbg_hod_sam_small}, and we expect them to have similar halo occupation properties as very bright low-redshift galaxies (such as Luminous Red Galaxies, or LRGs).
We will see below that this is approximately the case in our simulated samples.

Lyman-$\alpha$ emitters (LAEs) are also star-forming galaxies\footnote{There is significant overlap between the LBG and LAE populations - galaxies selected as LBGs very often have strong Ly$\alpha$ emission features (see e.g. Fig.~8 of \cite{Ruhlmann-Kleider:2024_desi_lbg}), though differences in star formation in each type of galaxy have recently been observed in the ODIN survey \cite{Firestone:2025_ODIN_SFH} at a range of stellar masses.} exhibiting a strong Lyman-$\alpha$ emission feature largely due to recombination of ionizing photons from recently-formed O/B stars that are incident on the ISM \cite{Partridge:1967_peebles,Furlanetto:2005_lae_igm}. 
LAEs are thought to occupy lower mass halos than LBGs, though the detailed physical link between Ly$\alpha$ emitting galaxies, dust, the intergalactic medium, the details of radiative transfer and the observed spectra defies a simple explanation \cite{ouchi:2020_AnnRev}.
However, a typical LAE is characterized by a large Lyman-$\alpha$ line flux excess over UV continuum flux, or high Lyman-$\alpha$ equivalent width (EW).
Practically, LAEs can easily be selected by equivalent width using narrow-band photometry\footnote{This has been the strategy of LAE samples such as those associated with ODIN \cite{Lee:2024_ODIN_initial} or SILVERRUSH \cite{Kikuta:2023_SILVERRUSH_catalog}, though it is possible to us an alternative selection strategy such as that employed by HETDEX \cite{Davis:2023_hetdex_classif,Landriau:2025_desi_hetdex_line}} - in that case, the narrow-band center will align with Lyman-$\alpha$ emission for LAEs at a particular redshift\footnote{For a more realistic description of this selection process, including rest-frame equivalent width [REW] cuts, see, e.g., Ref.~\cite{Firestone:2024_odin_selection}}.
The EW of a particular LAE depends on many factors, including their metallicity, dust distribution and geometry, and whether the galaxy is currently emitting (Lyman-$\alpha$ duty cycle).
Due to this selection process, the minimal effective physical picture of these galaxies is therefore significantly more complicated than that of LBGs, which affects the LAE halo occupation, and, therefore, their clustering properties.
As we will show, such complexity leads to halo occupation properties for our simulated samples that may require modifying halo occupation modeling for LAEs beyond the most popularly used forms for low-redshift galaxies.

The goal of this work is to use the MTNG and Astrid state-of-the-art large-volume hydrodynamical galaxy formation simulations to inform the plausible range of EFT model parameters that correspond to these high-redshift galaxies.
Specifically, we extend the precision field-level determination of bias parameters in these simulations applied in Ref.~\cite{Ivanov:2024_mtngeft} to high redshift star-forming galaxies at $z=3$ using a selection procedure that replicates existing angular clustering data and number densities.
Both the values of these parameters and the spatial and angular structure of the resulting EFT model then inform the clustering properties of the simulated sample selections.
Crucially, we are then able to suggest the expected range of validity of EFT modeling (in terms of $k_{\rm{max}}$ in real and redshift space) for the high-redshift galaxies that will drive the next decade of large-scale structure cosmology.

This paper is organized as follows.
In Section~\ref{sec:selection} we first describe our simulated galaxy selection process for star-forming galaxies similar to LBG and LAE populations relevant for upcoming LSS surveys, then detail the halo occupation properties of LAEs and LBGs, including  satellite fractions, and compare to LAE and LBG angular clustering measurements.
In Section~\ref{sec:eft_params}, we present non-linear bias parameter measurements obtained from field level fits to our samples in redshift space as well as matter cross-correlation coefficients and discuss the EFT ranges of validity they imply.
We conclude in Section~\ref{sec:conclusions}.

\section {Galaxy Selection \label{sec:selection} }

\begin{figure*}
\centering
\includegraphics[width=0.45\textwidth]{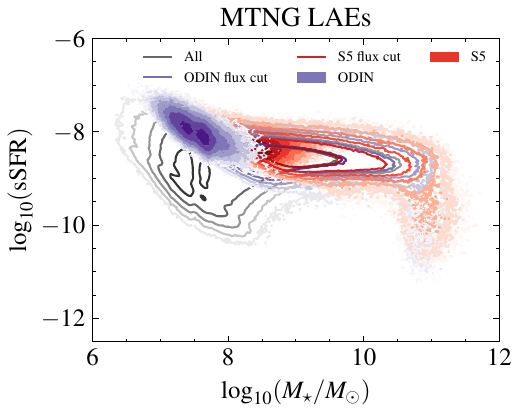}
\includegraphics[width=0.45\textwidth]{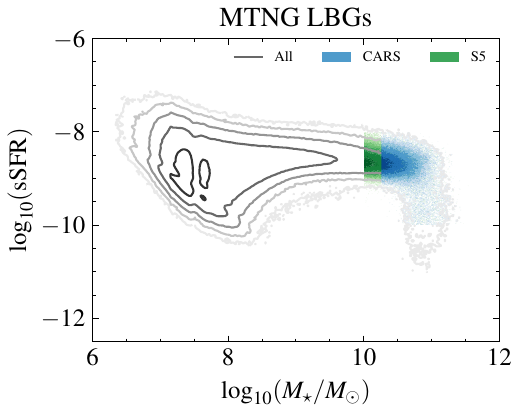}
\includegraphics[width=0.45\textwidth]{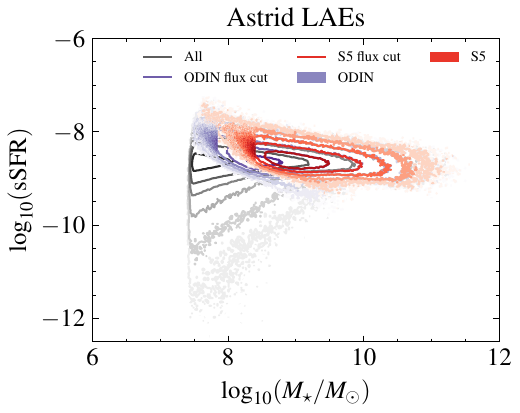}
\includegraphics[width=0.45\textwidth]{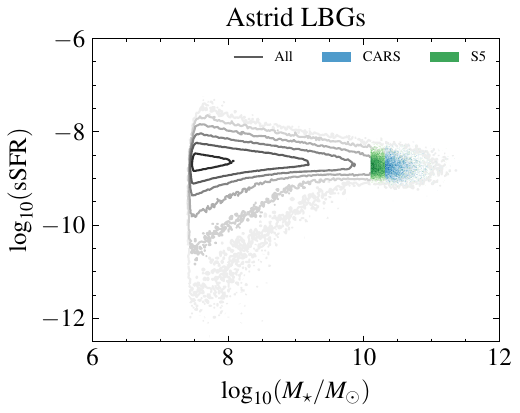}
   \caption{
   Stellar mass-specific star formation rate plane for the different samples used in this work. 
   The upper row of panels shows MTNG galaxies while the lower row of panels shows Astrid galaxies.
   All simulated galaxies (with non-zero SFR) are shown in gray unfilled contours in both panels.
   \textit{Left:} LAE sample selections used in this work. 
   Purple and red unfilled contours show galaxies that satisfy the ODIN and S5 flux cuts, respectively. 
   Purple and red filled contours show galaxies that both satisfy these flux cuts and our additional (fiducial) maximum metallicity (MTNG) and stellar mass (Astrid) cuts.
   These samples approximately match the number density and linear bias of observed ODIN LAEs and that expected of LAEs from a Stage 5 spectroscopic survey, respectively.
   \textit{Right:} LBG sample selections for two different minimum stellar mass cuts that reproduce the number density of LBGs expected from a Stage 5 spectroscopic survey (S5, green filled contours) and that which is compatible with the CARS data (CARS, blue filled contours). 
    } \label{fig:selection}
\end{figure*}

We consider the two high-redshift galaxy tracer populations that will be most relevant for upcoming spectroscopic galaxy redshift surveys - LBGs and LAEs.
We select simulated galaxies from MilleniumTNG740 (MTNG) \cite{Pakmor:2022yyn} and Astrid \cite{Bird:2022ulj,Ni2022} that are meant to approximate observed LBGs/LAEs based on their properties as described in the literature (for an overview of properties of MTNG and Astrid see Ref.~\cite{Ivanov:2024_mtngeft,Pakmor:2022yyn,Hernandez-Aguayo:2022xcl,Ni2022,Ni2024:Astrid2}).
Both simulations provide galaxy populations at both high and low redshifts with a continuous set of modeling choices.
However, we highlight two important differences between the two simulations for the purposes of this work.
The first is the difference in resolution between the two simulations.
Astrid uses a smaller simulation volume ($250 h^{-1}\rm{Mpc}$) and higher mass resolution ($2\times5500^3$ particles), while MTNG uses a larger volume ($500 h^{-1} \rm{Mpc}$) and lower mass resolution ($2\times4230^3$ particles)
The second is the set of choices that go into each galaxy formation model.
A significant fraction of existing studies are concerned with LBGs/LAEs at very high redshifts ($z\gtrsim 5-6$) relevant for studying reionization,  \cite[e.g.,][]{mcquinn_lae_07,Ouchi2010:highzlae_7,Konno2014:highz_lae,Ouchi2018:silverrush_highz,Inoue2018:silverrush_sims_highz,Ning2022:magellan_highz_lae,Umeda2025highzlae,Runnholm2025:jwst_highz_lae,2015Sobacchi:highz_lae} - we instead here focus only on $z=3$, which we take to be roughly representative of the star-forming galaxies targeted by upcoming surveys.

\subsection{Simulated star-forming galaxies}

\subsubsection{Physical properties of LBGs \& LAEs}
Since both populations are star-forming galaxies, LBGs and LAEs share several features.
Both LBGs and LAEs are compact, disk-like, and both likely have a similar distribution of star formation events \cite{Firestone:2025_ODIN_SFH}, and both LAEs and LBGs are situated on or close to a region of fixed specific star formation rate (sSFR) over a range of stellar masses (the star-forming ``main sequence'').
The key difference between the two for upcoming LSS surveys is that LBGs are observed with the Lyman-break technique \cite{Steidel1992:lyman_break_technique}, which requires deep
(blue) imaging, practically limiting the sample to only the brightest galaxies.
As a result, LBGs that will be used in cosmological surveys will be the most luminous ones at a particular redshift, and will tend to occupy high-mass halos with corresponding large values of linear bias.
For this reason, we would naively expect LBGs to behave like the more familiar lower redshift LRGs, which similarly occupy the most massive halos.

In contrast, LAEs are selected based on their Lyman-$\alpha$ emission detection, roughly corresponding to a minimum Ly$\alpha$ flux and a threshold in (rest-frame) equivalent Ly$\alpha$ width (REW, or, here, EW).
This selection 
reveals galaxies that are not simply the most luminous or massive, but rather those with high Ly$\alpha$ excess relative to continuum UV flux.
High-$z$ SFGs selected in this way appear to occupy lower halo masses than LBGs (the range of stellar masses for LBGs is, very roughly, $M_\star\sim 10^{10}-10^{11} M_\odot$, and for LAEs is $M_\star\sim 10^{8}-10^{9} M_\odot$ \cite{ouchi:2020_AnnRev}), and it is less clear that these galaxies will match naive expectations about the bias parameters of high-redshift galaxies.

Compared to the observational studies of LBGs and LAEs in which galaxy properties are inferred from stellar population synthesis models fit to (stacked) spectra/SEDs,  there are relatively few numerical simulations modeling such galaxies. 
Ref.~\cite{Nagamine:2010_sim_sfgal} studied galaxies at $z=3$ in a suite of hydrodynamical simulations \cite{Nagamine:2004_LBG,Night:2006_LBG}, extracting lessons for LAEs and LBGs by comparing physical quantities of their simulated SFGs to those constrained from the literature (e.g using SPS models)\footnote{Ref.~\cite{Im:2024_sim_lbg_lae} also recently explored modeling LAEs/LBGs the Horizon run 5 simulations \cite{Lee:2021_horizon_5}, however the details of the prescriptions they used to semi-analytically assign LAE equivalent widths and UV/Ly-$\alpha$ luminosities as well as LBG colors are not yet available.
}.
Recently, Ref.~\cite{Ravi:2024} used MTNG to build on this earlier work, selecting simulated LAEs from the simulated MTNG SFG population in a manner inspired by actual observational selections of equivalent width and Ly-$\alpha$ flux.
These selections effectively amount to a minimum star formation rate cut 
motivated by observable Ly-$\alpha$ flux limit as well as the same selection with an additional maximum stellar mass cut
\footnote{Ref.~\cite{Ravi:2024} also considers a ``default+REW'' cut based on rest-frame equivalent width of the galaxies. 
However, the REW-related quantity assigned to the galaxies there is not connected with any physical properties of the galaxy, and instead subsamples the ``default'' sample, as the authors note. We comment on subsampling in this way in Appendix~\ref{app:downsampling}, effectively changing the Ly-$\alpha$ effective escape fraction at fixed number density.}.
Their stellar mass maximum sample is motivated by the observations that do not support LAEs living in very massive halos.
Indeed, Ref.~\cite{White:2024_odin} found that the ODIN survey LAEs \cite{Lee:2024_ODIN_initial,Firestone:2024_odin_selection} occupied lower mass halos than the default sample identified in Ref.~\cite{Ravi:2024} (both when fitting the standard Zheng07 Halo Occupation Distribution (HOD) form \cite{Zheng:2007zg}).

The reason for the high stellar mass cutoff in the observed LAE population remains unclear.
However, it may be due to a variety of factors, including a limited Ly$\alpha$ escape fraction at higher masses (due, e.g., to changes in effective escape fraction of Ly$\alpha$ emission, the ISM ionization state, amount of dust, dust and ISM clumpiness, dust geometry, or attenuation by the IGM - and it is likely a combination of all of these effects with different strengths \cite{ouchi:2020_AnnRev}).
None of these effects have been directly modeled for LBGs/LAEs in cosmological hydrodynamical simulations, to the best of our knowledge (though see Ref.~\cite{Smith:2021hqg} for an example of Lyman-$\alpha$ post-processing applied to IllustrisTNG \cite{Nelson_TNG}).
Our ability to realistically model the selection of LAEs is fundamentally limited given our use of hydrodynamical simulations without radiative transfer effects, which are computationally infeasible to include in cosmological volume simulations.

In the absence of hydrodynamical simulations in cosmological volumes with radiative transfer effects, we use galaxy formation model quantities (stellar mass $M_\star$, specific star formation rate sSFR, and gas-phase metallicity $Z$\footnote{While stellar metallicity $Z_{\star}$ is more tightly correlated with stellar mass than gas metallicity $Z$ in MTNG, the latter is the relevant quantity for recombination in the ISM, though the distributions of both quantities are mostly qualitatively similar. Unless otherwise specified, ``metallicity'' will refer to gas metallicity.}) to construct samples approximating LAEs for which some constraints from stellar population models exist in the literature.
Since Lyman-$\alpha$ EW is highly-dependent on metallicity \cite{Kerutt2022:EW_metallicity_LAE}, qualitatively, restricting the low metallicity galaxies should crudely model the higher Lyman-$\alpha$ EW of such galaxies.

\subsubsection{Modeling LBGs \& LAEs in hydrodynamical simulations }
Following Refs.~\cite{Ravi:2024, Dijkstra:2010}, we adapt a modified version of the procedure used there to assign Lyman-$\alpha$ luminosities (and therefore fluxes) based on physical star formation rate:
\begin{equation} \label{eqn:SFR_Dijkstra}
    L_{\mathrm{Ly}\alpha}^{\mathrm{obs}} = \frac{1.1 \times 10^{42} ~\mathrm{erg} ~\mathrm{s}^{-1}}{\mathcal{M}} \left(\frac{\mathrm{SFR}}{M_\odot~\mathrm{yr}^{-1}}\right),
\end{equation}
where $\mathcal{M}$ has a log-normal distribution (with parameters provided in Ref.~\cite{Dijkstra:2010}) and is defined in Ref.~\cite{Dijkstra:2010} as the ratio of the UV-derived SFR to the Ly-$\alpha$-derived SFR.
The procedure of Ref.~\cite{Dijkstra:2010} for assigning $L_{\mathrm{Ly}\alpha}^{\mathrm{int}}$, however is only meant to be a rough approximation as the authors note. 
This procedure relies on fits to values for SFR(UV) and SFR(Ly$\alpha$) drawn from the literature, and computes the distribution of their ratio $\mathcal{M}$\footnote{This is done by combining estimates from $z=3$ and $z=5.7$, so using the fitted distribution $P(\mathcal{M})$ may introduce a systematic error in our flux assignment, as the distribution may shift if fitted only to LAEs at $z=3$.}. 
These star formation rate estimates that are fitted to observed $L_{\mathrm{Ly}\alpha}^{\rm{obs}}, ~L_{\mathrm{UV}}^{\rm{obs}}$ are computed without regard to the physical processes affecting photon propagation to the observer (including dust, IGM absorption) or additional sources of Lyman-$\alpha$ photons beyond the ISM.
Neglecting the latter, the former can be compressed in the effective escape fraction \cite{Nagamine:2010_sim_sfgal}
\begin{equation} \label{eqn:eff_escape}
f_{\mathrm{esc}}^{\mathrm{eff},\mathrm{Ly}\alpha} = \frac{L_{\mathrm{Ly}\alpha}^{\mathrm{obs}}}{L_{\mathrm{Ly}\alpha}^{\mathrm{int}}}.
\end{equation}

The conversion from physical SFR to \textit{intrinsic} Lyman alpha luminosity $L_{\mathrm{Ly}\alpha}^{\mathrm{int}}$ (before applying the empirically-calibrated $\mathcal{M}$) is itself very uncertain. 
However, the usual procedure is to assume the relationship appropriate for Case B recombination for a Salpeter Initial Mass Function (IMF) assuming constant star formation over a period of $100~ \mathrm{Myr}$\footnote{For $z=3$ ODIN LAEs, although star formation appears to be ongoing, the star formation a the time of observation is either somewhere in the peak or at the end of a burst \cite{Firestone:2025_ODIN_SFH}.}
\cite{Furlanetto:2005_lae_igm,Kennnicut:1998, Salpeter:1955}, which gives a prescription for the \textit{intrinsic} Lyman-$\alpha$ luminosity
\begin{equation} \label{eqn:SFR}
    L_{\mathrm{Ly}\alpha}^{\mathrm{int}} = 1.1 \times 10^{42} ~\mathrm{erg} ~\mathrm{s}^{-1} \left(\frac{\mathrm{SFR}}{M_\odot~\mathrm{yr}^{-1}}\right)
\end{equation}
It is well-known that this conversion is significantly metallicity dependent \cite{Schaerer:2003}, which will significantly impact LAEs.
Further, even assuming no metallicity dependence, a different choice of low-mass limit for the \textit{same} IMF can change the conversion factor by a factor of a few \cite{Dijkstra:2017_notes}. 
To roughly model the metallicity dependence of the intrinsic Lyman-$\alpha$ luminosity, we replace the numerical coefficient eqn.~\ref{eqn:SFR} with an interpolation of Table~4 of Ref.~\cite{Schaerer:2003} (a rapidly declining function of $Z$) to assign a metallicity-dependent SFR-to-Ly$\alpha$ flux coefficient based on \textit{each} galaxy gas particle metallicity
\begin{equation} \label{eqn:SFR_za}
    L_{\mathrm{Ly}\alpha}^{\mathrm{obs}} = \frac{f_{\mathrm{Ly}\alpha}(Z)}{\mathcal{M}} \left(\frac{\mathrm{SFR}}{M_\odot~\mathrm{yr}^{-1}}\right).
\end{equation}
We do not claim that this is a precision operation, as the conversion factor is uncertain within factors of a few, and note that the value of Ref.~\cite{Schaerer:2003} at solar metallicity is higher than that of eqn.~\ref{eqn:SFR}, but this procedure reflects the physical effect of increasing Lyman-$\alpha$ flux with decreasing metallicity.

For the redshift $z=3$ snapshots that we use, defining LAE samples in both MTNG and Astrid provides complementary advantages.
Though in a smaller box than MTNG, Astrid is a higher resolution simulation and contains significantly more star particles per galaxy, which is relevant especially at the low-stellar-mass extremes populated by some of the LAE samples.
On the other hand, MTNG provides metallicity information for each galaxy at $z=3$, which allows us to use eqn.~\ref{eqn:SFR_za} to reflect the known physical anticorrelation between EW (and so selection probability) and gas-phase metallicity.
For Astrid, gas metallicity information is not available for each galaxy and so we instead make cuts based on maximum stellar mass (see Appendix~\ref{app:mtng_res} for further discussion)\footnote{This is done with a SFR-Ly$\alpha$ flux conversion rate of $f_{\mathrm{Ly}\alpha}(Z=0.2~Z_\odot)$ (roughly following \cite{Finkelstein2011:low_ZZsun_LyaE}), and implicitly using the Astrid minimum star particle cut of 100 star particles per galaxy (see also Appendix~\ref{app:mtng_res})}.

\subsection{Sample construction}
Our fiducial samples are constructed to be representative of future LSS survey observations of LBGs and LAEs and are informed by smaller existing datasets.
We consider four samples, two for LAEs and two for LBGs.
For the LAEs, we consider a sample that approximately matches the reported linear bias $b_1$ and number density $\bar{n}$ of ODIN LAEs in Ref.~\cite{White:2024_odin} (LAE ``ODIN'' sample).
For LBGs, we similarly consider a sample that approximately matches the existing observed angular clustering (and linear bias $b_1$) of the CARS survey of $u-$drop LBGs in Ref.~\cite{Hildebrandt:2009_CARS_LBG} (LBG ``CARS'' sample).
For each type of galaxy population, we also include a sample that roughly reproduces the projected possible number density and linear bias of an upcoming Stage-5 spectroscopic survey (LAE/LBG ``S5'' samples).
The number densities and biases in this case are drawn from Ref.~\cite{Ebina_White:2024_forecast}\footnote{We note that these values are roughly a factor of 2 different from those quoted at $z=3$ in Ref.~\cite{Schlegel:2022_spec_roadmap},
possibly due to the factor of few uncertainty in the luminosity functions considered.}, though we note the LAE sample is still relevant for the recently-completed HETDEX, which are projected to have roughly similar LAE number densities \cite{Schlegel:2022_spec_roadmap,2008:Hill_hetdex}, and both samples are still relevant for DESI-II \cite{Schlegel:2022_spec_roadmap,Ruhlmann-Kleider:2024_desi_lbg}.
The selections for these fiducial samples are shown in the specific star formation rate (sSFR)-stellar mass ($M_\star$) plane in Fig.~\ref{fig:selection}.

We provide values of the number densities, mean parent halo masses, approximate linear bias values of angular clustering (from the simulations), 1D velocity dispersion (for MTNG), and satellite fractions in Tables~\ref{tab:samples_mtng},\ref{tab:samples_astrid}.

\subsubsection{LAEs}

\textbf{ODIN sample:}
We construct the ODIN sample to reproduce the observed clustering and number density of ODIN LAEs \cite{White:2024_odin}.
We build the samples in both simulations by assigning Lyman-$\alpha$ flux to every galaxy using eqn.~\ref{eqn:SFR_za} and then applying the reported ODIN flux cut of $f_{\mathrm{Ly}\alpha} > 1.8 \times 10^{-17} ~ \mathrm{erg}~\mathrm{s}^{-1} \mathrm{cm}^{-2}$.

\textit{MTNG:}
In MTNG, we use the metallicity-aware version of the procedure of Ref.~\cite{Dijkstra:2010} of eqn.~\ref{eqn:SFR_za}, which does not assume that all SFGs are LAEs.
We then apply an aggressive maximum metallicity cut (at the 5th percentile of \texttt{SubhaloGasMetallicity} for galaxies satisfying the flux cut, $Z \geq 0.04 ~Z_\odot$) to the galaxies satisfying this flux cut to ensure that we obtain a sample that approximately matches the number density and angular clustering of the ODIN observations.
The values of the resulting properties of this sample listed in the first row of Table~\ref{tab:samples_mtng} then approximately reproduce the measured ODIN values of $\bar{n} = 1.0 \times 10^{-3}~[h^{-1}~\mathrm{Mpc}]^{-3}$ and $b_1 = 2.0 \pm 2.0$.
The values of mean host halo mass are also in agreement with those reported for LAEs in Ref.~\cite{ouchi:2020_AnnRev}.

The metallicity maximum cut, at $Z = 0.04 ~Z_\odot$, produces galaxies with metallicities significantly lower than those estimated for LAEs using metal features in spectra of very small samples of galaxies \cite{ouchi:2020_AnnRev}.
However, recent work studying a small number of central galaxies using zoom-ins of the THESAN simulations \cite{2025:Kannan_THESANZOOM,Smith:2021hqg} indicates that LAEs of approximately the stellar masses we consider have metallicities consistent with $\frac{Z}{Z_\odot} \sim 0.03-0.2$ (when converting from O/H to $Z$) for all LAEs above $z=3$ (with little redshift dependence, as estimated from $\approx600$ uncontaminated high-resolution galaxies that constitute the $z=3$ curve of Fig. 13 of Ref.~\cite{2025:Kannan_THESANZOOM}).
For the stellar mass range of the ODIN sample, this corresponds to metallicities of roughly$\frac{Z}{Z_\odot} \sim 0.03-0.12$.
We note that the uncertainty in the SFR to $L_{\mathrm{Ly}\alpha}^{(\mathrm{int})}$ conversion coefficient, $f_{\mathrm{Ly}\alpha}(Z)$ combined with the steep decay of the halo mass function means that a lower choice of $f_{\mathrm{Ly}\alpha}$  would lead to fewer SFGs making it past the flux cut, and due to the SFR-$M_\star$ correlation, would lead to galaxies with higher stellar mass, and overall higher halo mass. 
This would suggest that the linear halo bias of these galaxies would then \textit{increase} - leading to the need for an even more aggressive cut (e.g. in metallicity) to match the relatively low linear bias of ODIN galaxies. 
In this sense the low value of the ODIN bias, under the assumptions made here, seems to favor a \textit{more efficient} conversion of SFR to $L_{\mathrm{Ly}\alpha}$.

\textit{Astrid:}
For the Astrid sample, we use the same flux limit as for the MTNG galaxies described above. 
However, for defining the Astrid LAE samples there are two significant differences: 1) with respect to the SFR-Ly$\alpha$ flux conversion, and 2) with respect to the metallicity cut.
In Astrid data catalog, we do not have (gas or stellar) metallicity information available for each galaxy.
For assigning flux, we then choose $f(Z=0.2~Z_\odot)$ for use in eqn.~\ref{eqn:SFR}.
Without the metallicities we also cannot define a metallicity cut, which was inspired by the anti-correlation of metallicity and equivalent width.
Instead, we select galaxies using a maximum stellar mass cut (e.g., as briefly mentioned in Ref.~\cite{Ravi:2024}) to match the target number densities.
Specifically, we use a stellar mass maximum for the ODIN sample of  $M_{\star} \leq 7 \times 10^{7} M_{\odot}$.
While stellar mass anti-correlates with metallicity, there is significant spread in stellar mass at fixed metallicity, meaning that it is possible our use of a stellar mass cut instead of a metallicity cut restricts the range of stellar mass in a less realistic way than a metallicity cut does.
However, using a stellar mass maximum cut still provides a reasonable proxy for Lyman-$\alpha$ escape fraction due to the expected higher amount of dust in high stellar mass galaxies \cite{Oyarzun:2016stellarmasslae}. 
This also has implications for the galaxy population satellite fraction\footnote{A comparison of a stellar mass and metallicity cut for the S5 number density LAEs in MTNG (where both metallicity and stellar mass information is available) indicates that substituting a $Z$ cut for a $M_\star$ cut changes the satellite fraction by 28\% without significantly changing the mean parent halo mass.
We expect the effect would be even more dramatic for the ODIN sample.}.

\textbf{S5 sample:} 
For the S5 LAE sample, we proceed similarly to the ODIN sample, but instead use a flux limit of $f_{\mathrm{Ly}\alpha} >  5.0 \times 10^{-17} ~ \mathrm{erg}~\mathrm{s}^{-1} \mathrm{cm}^{-2}$ and a maximum metallicity cut at the 10th percentile ($Z<0.19~Z_\odot$). 
Compared with the THESAN-ZOOM simulations the stellar mass range of the S5 LAEs corresponds to metallicities of roughly$\frac{Z}{Z_\odot} \sim 0.07-0.32$.
This sample achieves a mean number density of $\bar{n} = 1.3 \times 10^{-3}~[h^{-1}~\rm{Mpc}]^{-3}$
For the Astrid sample, we use a S5 stellar mass maximum of $M_{\star} \leq 2.6 \times 10^{8} M_{\odot}$.

\subsubsection{LBGs}
For both LBG samples, we select galaxies with specific star formation rate between $10^{-10} \leq \mathrm{sSFR} < 10^{-8} ~ h~\mathrm{yr}^{-1}$.

\begin{table}[] 
\begin{tabular}{|ll|ccccc|} 
\hline 
&         & $\bar{n} ~[h^{-1}\mathrm{Mpc}]^{-3}$ & $ \log_{10} \left( \frac{\langle M_{h}\rangle}{h^{-1}M_{\odot}}\right) $ & $b_{1}^{w(\theta)}$  & $f_{\mathrm{sat}}$ & $\langle \frac{\sigma}{\mathrm{km}/\mathrm{s}} \rangle$\\
\hline
\multirow{2}{*}{\textbf{LAE}} & ODIN &  $1.2 \times 10^{-3}$     &   11.40   &   2.1  &    $30\%$  & 60   \\
& S5   &  $1.3 \times 10^{-3}$   &   11.51    & 2.4  &  $15\%$ & 88 \\
\hline
\multirow{2}{*}{\textbf{LBG}} & CARS &  $6.7 \times 10^{-4}$   &  12.37  & 3.9  &  $12\%$ & 200\\
& S5   &    $1.3 \times 10^{-3}$     &     12.26    &  3.6  &  $10\%$ & 179\\
\hline
\end{tabular}
\caption{\textbf{MTNG sample properties}: sample number density $\bar{n}$, mean parent halo mass $\langle M_{h} \rangle$, and satellite fraction $f_{\mathrm{sat}}$. Also, $b_{1}^{w(\theta)}$ is the rough estimate of the linear bias from the angular clustering, which is how $b_1$ has largely been studied in the literature for these galaxies. 
We also quote the 1D mean velocity dispersion $\sigma$ supplied by MTNG, which gives a qualitative indication of strength of Fingers of God for each sample. \label{tab:samples_mtng}}
\end{table}

\begin{table}[] 
\begin{tabular}{|ll|cccc|} 
\hline 
&         & $\bar{n} ~[h^{-1}\mathrm{Mpc}]^{-3}$ & $ \log_{10} \left( \frac{\langle M_{h}\rangle}{h^{-1}M_{\odot}}\right) $ & $b_{1}^{w(\theta)}$  & $f_{\mathrm{sat}}$ \\
\hline
\multirow{2}{*}{\textbf{LAE}} & ODIN &   $1.3 \times 10^{-3}$     &  11.98   &   2.0  &    $40\%$    \\
& S5   &  $1.3 \times 10^{-3}$   &  11.96    & 2.1  &  $33\%$  \\
\hline
\multirow{2}{*}{\textbf{LBG}} & CARS &  $6.7 \times 10^{-4}$   &  12.63  & 3.8 &  $12\%$ \\
& S5   &    $1.3 \times 10^{-3}$     &  12.52  &  3.8  &  $13\%$\\
\hline
\end{tabular}
\caption{\textbf{Astrid sample properties}: Similar to Table~\ref{tab:samples_mtng}, but for the simulated Astrid galaxies. See text for differences in selection from the MTNG samples. \label{tab:samples_astrid}}
\end{table}

\textbf{CARS sample:}
For the CARS sample, we attempt to match the angular clustering of the CARS data of Ref.~\cite{Hildebrandt:2009_CARS_LBG}.
Specifically, we consider their $u-$drop sample with a limiting magnitude of $23.0<r<24.5$ (in an effective are of $\approx 3~\mathrm{deg}^2$).
For MTNG, we use a minimum stellar mass cut of  $ M_\star > 1.8 \times 10^{10} ~M_{\odot}~h^{-1}$ for the CARS sample\footnote{We can also obtain a sample with similar agreement with the CARS angular clustering on large scales by making a stellar mass cut of $M_\star > 10^{8} ~ h^{-1}~M_{\odot}$ and a narrower specific star formation rate window of $10$, enforcing less scatter away from the SFR main sequence, which produces a similar large-scale bias at the level of the angular clustering.}.
Applying this selection produces a sample with mean number density $\bar{n} = 6.7 \times 10^{-4} ~[\hMpc]^3$.
For the Astrid CARS LBG sample we use the same procedure except that the minimum stellar mass cut is $ M_\star > 2.1 \times 10^{10} ~M_{\odot}~h^{-1}$.
Similar to as described in Ref.~\cite{Ivanov:2024_mtngeft} for the lower redshift LRG sample, to match the same number density between MTNG and Astrid for LBGs, we need to use two different stellar mass minimum cuts in the two simulations\footnote{This is also the case for the S5 sample below.}.
Interestingly, opposite to the case of LRGs at $z=0.5$ where MTNG required a higher stellar mass minimum, for the LBGs at $z=3$ we need a lower stellar mass minimum in MTNG.

\textbf{S5 sample:} 
For the MTNG S5 LBG sample, we use a stellar mass minimum cut $M_{\star} > 1.0 \times 10^{10} ~M_{\odot}~h^{-1}$ and for the Astrid S5 LBG sample we use $M_{\star} > 1.3 \times 10^{10} ~M_{\odot}~h^{-1}$.
Both samples achieve a number density of $\bar{n} = 1.3 \times 10^{-3} ~[\hMpc]^3$.

\subsection{Sample properties \label{subsec:sample_properties}}

\textit{Stellar mass and specific star formation rate:}
Fig.~\ref{fig:selection} shows the selected simulated galaxies for all the SFG samples in the sSFR-$M_\star$ plane. 
All galaxies at $z=3$ are shown in gray. 
For all galaxies, the two simulations show qualitative differences at the high and low stellar mass ends - at higher stellar mass MTNG has a significant population of low sSFR galaxies while the Astrid galaxies appear to show more spread at lower masses than MTNG (up to the Astrid star particle resolution limit).

The left panels show the populations relevant for LAEs in MTNG (top) and Astrid (bottom).
Simply applying a flux cut (for the ODIN flux limit in purple unfilled contours, and for S5 in red unfilled contours) removes the very lowest $M_\star$ and sSFR galaxies, with the higher S5 minimum flux cut occupying higher stellar masses and slightly higher sSFRs than the ODIN sample.
After applying the additional metallicity cuts, we obtain our fiducial ODIN and S5 samples.
These samples select galaxies at very low stellar mass, as stellar mass correlates with metallicity.
For the ODIN sample (purple filled contour), the selected galaxies have lower stellar masses and slightly higher, though similar, sSFRs to the S5 sample (red filled contour).
The LAE regions of the $M_\star-$sSFR plane in both MTNG and Astrid are roughly similar, up to qualitative differences in the overall galaxy populations.
However, the MTNG LAE samples span a wider range of stellar masses, while the Astrid LAEs occupy a narrower rage of stellar masses.
This difference is due to our use of a stellar mass maximum cut in Astrid while we used the metallicity cut in MTNG.
The LAE samples in both simulations attain similar values of linear bias (especially the ODIN samples), illustrating the uncertainty in the physical properties of the galaxies that produce the observed clustering.
The simpler stellar mass maximum cut excludes high-mass, low-metallicity galaxies that might very well be good proxies for real LAEs with high EW.

In the right panels, we see that the LBG samples are again qualitatively similar between Astrid and MTNG.
The only noticeable difference is the reduced spread in sSFR of all Astrid galaxies that is inherited by the LBG samples.
The two target sample selections (CARS, S5) only differ by a simple selection in minimum stellar mass, which is clearly visible in the selection, with the high number density sample (S5) extending to lower stellar masses.

\textit{Halo summary quantities:}
The LAEs in both Astrid and MTNG show very high satellite fractions ($f_{\rm{sat}} \approx 15\%-40 \%$) especially for the ODIN samples.
For MTNG, these satellite galaxies live in halos that have centrals at similar (or somewhat lower) sSFR and significantly higher stellar masses\footnote{Astrid lacks the necessary catalog information needed to match satellite galaxies to their neighboring centrals in their parent halo.}.
Rough approximations to the linear bias (from angular clustering) are consistent with $b_1 \sim 2$.
We note that the linear bias estimates in Tables~\ref{tab:samples_mtng},\ref{tab:samples_astrid} are only approximate and we provide a more complete discussion of linear bias in Section~\ref{sec:eft_params}.

For LBGs, the S5 sample attains a slightly higher number density and lower linear bias and parent halo mass, than the CARS sample, but is qualitatively similar, with similar satellite fraction and mean host halo mass.
Both the LAE and LBG samples occupy relatively low mean halo masses, and these values are somewhat lower in the MTNG simulations than in the Astrid simulations. 
For LBGs, the satellite fraction is approximately $f_{\rm{sat}} \approx 10\%-13 \%$, closer to what is expected at low redshift for LRGs.

The satellite fraction of both LBGs and LAE samples for upcoming spectroscopic surveys remains very uncertain, but if these high values for LAEs are at even roughly indicative of the galaxies that will be measured with upcoming spectroscopic surveys, a reevaluation of halo occupation modeling may be necessary.
We note that the satellite fraction of $f_{\mathrm{sat}} \sim 30\%$ of LAEs differs significantly from the low satellite fractions (~6\%) of the HOD mocks used in Ref.~\cite{White:2024_odin}, though this may be to the use of the Zheng07 HOD model in Ref.~\cite{White:2024_odin}.
The presence of a high satellite fraction would, for high mass halos, suggest a large Fingers-of-God (FoG) effect, the size of which we discuss in Section~\ref{sec:eft_params}.

\begin{figure*}
\centering
\includegraphics[width=0.49\textwidth]{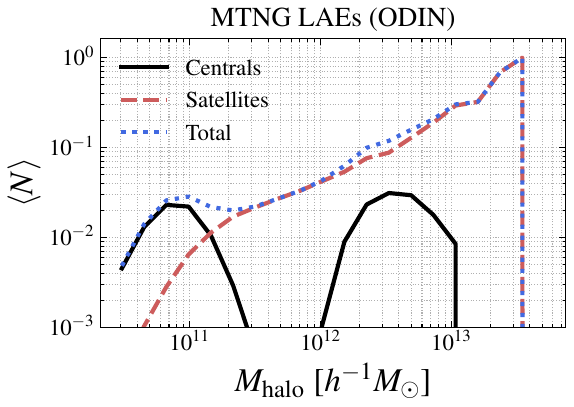}
\includegraphics[width=0.49\textwidth]{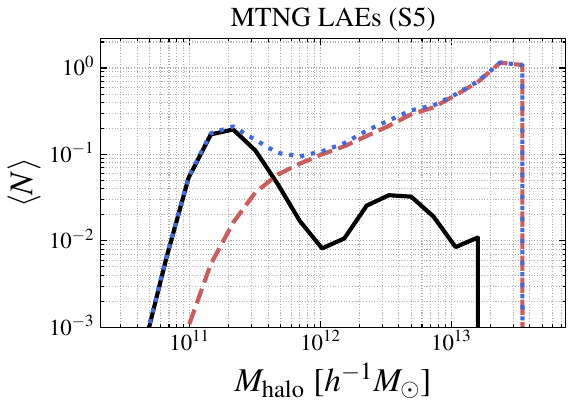}
\includegraphics[width=0.49\textwidth]{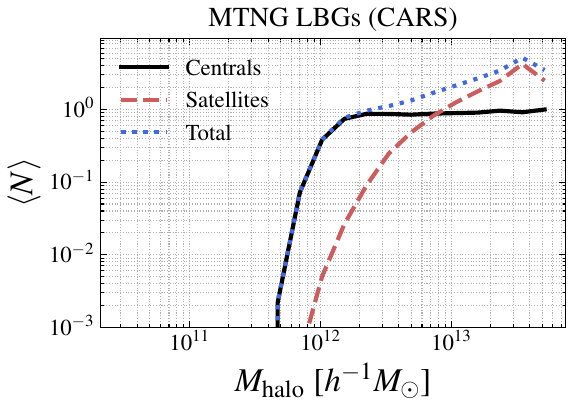}
\includegraphics[width=0.49\textwidth]{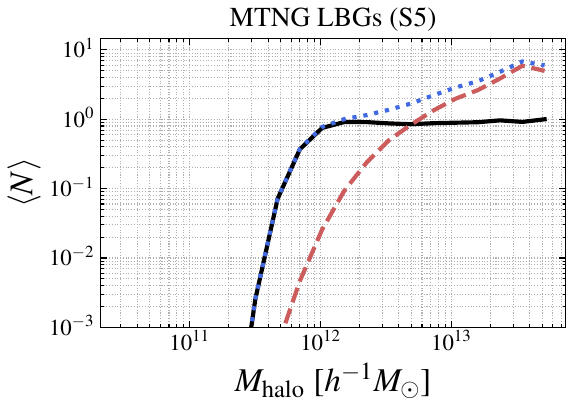}
   \caption{
   Mean halo occupation as a function of halo mass for the MTNG samples defined in Table~\ref{tab:samples_mtng}.
   \textit{Top Left:} LAEs defined with the ODIN flux cut of $f_{\mathrm{Ly}\alpha} \geq 1.8 \times 10^{-17}~\mathrm{erg}~\mathrm{s}^{-1}~\mathrm{cm}^{-2}$ and the fiducial maximum metallicity cut.
   \textit{Top Right:} LAEs defined with the S5 flux cut of $f_{\mathrm{Ly}\alpha} \geq 5 \times 10^{-17}~\mathrm{erg}~\mathrm{s}^{-1}~\mathrm{cm}^{-2}$ and 20th percentile maximum metallicity cut.
   \textit{Bottom Left:} LBGs defined with the CARS stellar mass cut of $M_\star \geq 1.8 \times 10^{10}~h^{-1}~M_{\odot}$ on the sSFR main sequence ($10^{-10} h~\mathrm{yr}^{-1} < \mathrm{sSFR} < 10^{-8} h~\mathrm{yr}^{-1}$).
   \textit{Bottom Right:} LBGs defined with the S5 stellar mass cut of $M_\star \geq 1 \times 10^{10}~h^{-1}~M_{\odot}$ on the sSFR main sequence ($10^{-10} h~\mathrm{yr}^{-1} < \mathrm{sSFR} < 10^{-8} h~\mathrm{yr}^{-1}$).
    } \label{fig:mtng_hod}
\end{figure*}

\begin{figure*}
\centering
\includegraphics[width=0.49\textwidth]{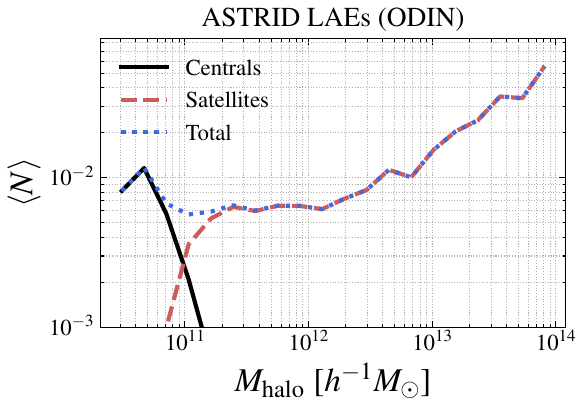}
\includegraphics[width=0.49\textwidth]{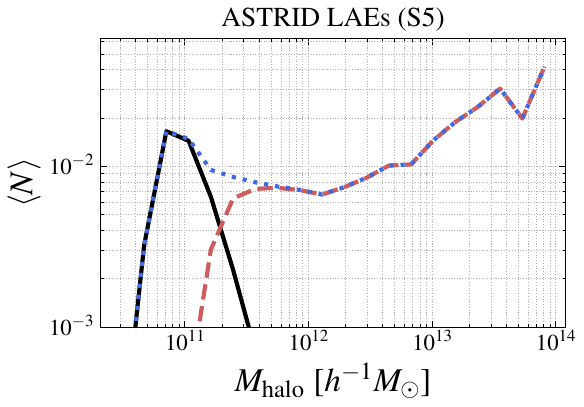}
\includegraphics[width=0.49\textwidth]{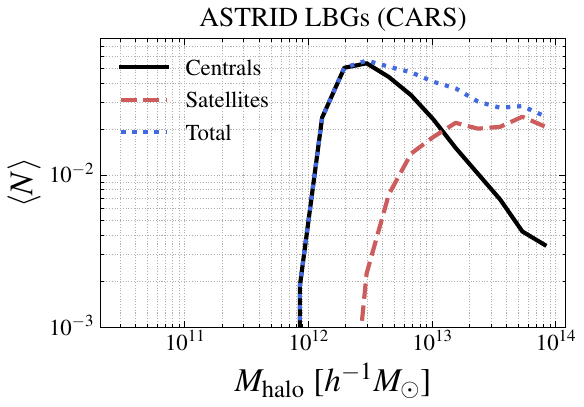}
\includegraphics[width=0.49\textwidth]{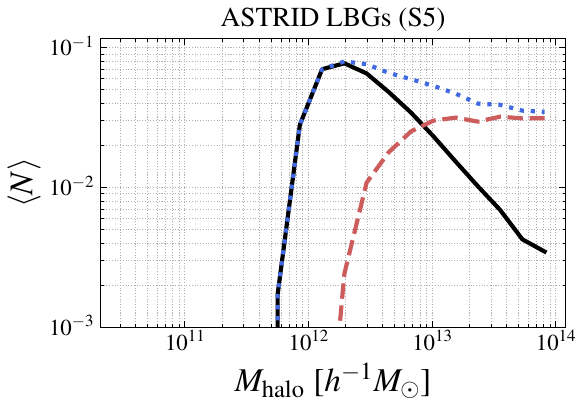}
   \caption{
   Mean halo occupation as a function of halo mass for the Astrid samples defined in Table~\ref{tab:samples_astrid}.
   \textit{Top Left:} LAEs defined with the ODIN flux cut of $f_{\mathrm{Ly}\alpha} \geq 1.8 \times 10^{-17}~\mathrm{erg}~\mathrm{s}^{-1}~\mathrm{cm}^{-2}$ and $M_\star \leq 7 \times 10^7 M_\odot$.
   \textit{Top Right:} LAEs defined with the S5 flux cut of $f_{\mathrm{Ly}\alpha} \geq 5 \times 10^{-17}~\mathrm{erg}~\mathrm{s}^{-1}~\mathrm{cm}^{-2}$ and $M_\star \leq 2.6 \times 10^8 M_\odot$.
   \textit{Bottom Left:} LBGs defined with the CARS stellar mass cut of $M_\star \geq 1.8 \times 10^{10}~h^{-1}~M_{\odot}$ on the sSFR main sequence ($10^{-10} h~\mathrm{yr}^{-1} < \mathrm{sSFR} < 10^{-8} h~\mathrm{yr}^{-1}$).
   \textit{Bottom Right:} LBGs defined with the S5 stellar mass cut of $M_\star \geq 1 \times 10^{10}~h^{-1}~M_{\odot}$ on the sSFR main sequence ($10^{-10} h~\mathrm{yr}^{-1} < \mathrm{sSFR} < 10^{-8} h~\mathrm{yr}^{-1}$).
    } \label{fig:astrid_hod}
\end{figure*}

\textit{HODs:}
The top two panels of Figs.~\ref{fig:mtng_hod},\ref{fig:astrid_hod} show the HODs for the LAE ODIN and S5 samples (for MTNG and Astrid, respectively).
The functional form of these HODs are quite unlike the commonly-used Zheng07 HOD used for LRGs.
The form is, however, not too dissimilar from the HMQ HOD model used for DESI ELGs \cite{Yuan:2022rsc,2021:Hadzhiyska_tng_elg}, albeit at lower masses since we are working at higher redshift.
As for ELGs, since the halo mass function is rapidly falling off with halo mass, the majority of halos hosting galaxies are toward the low mass end.
While we similarly find for the S5 sample that the mean occupation number is less than 1 (at about 0.1), for the ODIN sample, it is even below $10^{-2}$ at some halo masses.
In both cases, if we expect that LAEs have a relatively low effective escape fraction, as suggested in Ref.~\cite{Nagamine:2010_sim_sfgal}, such a low occupation fraction is to be expected.

The presence of the bump at higher halo masses for both LAE samples in MTNG is also distinct from the case of the HMQ model.
The origin of this bump can be traced to a bimodality in the distribution of stellar mass conditioned on fixed metallicity at low metallicities.
Using a stellar metallicity cut instead of a gas metallicity cut produces an HOD without a high-mass sub-dominant subpopulation (the second high-mass bump in Fig.~\ref{fig:mtng_hod}).
While this does not appreciably change the mean parent halo mass of the sample or the satellite fraction for the ODIN flux cut, and so is unlikely to have a significant influence on the bias parameters, the satellite fraction does change (from 14\% to 17\%) for the S5 flux cut.
See Appendix~\ref{app:mtng_res} for a discussion of the MTNG stellar mass resolution on our results for the ODIN sample.
For the Astrid samples (Fig.~\ref{fig:astrid_hod}), we do not see such behavior, and both LAE samples are even more satellite dominated than those of MTNG.
We note that there are differences in the halo finders used by MTNG (CompaSO \cite{Hadzhiyska:2021zbd}) and Astrid (FoF+SUBFIND \cite{1985ApJ...292..371D,springel_subfind}) that could be influencing our results, though as discussed in Section 3.3 of Ref.~\cite{Ivanov:2024_mtngeft}, these differences are likely subdominant to the differences in galaxy formation modeling choices between the two simulations.

For LBGs, both MTNG simulated LBG samples seem to be well-described by the standard Zheng07 HOD used for LRGs at low redshift.
Under the simplistic assumption here that a stellar mass cut is able to approximately reproduce the observed selection of LAEs, this is not surprising.
However, the Astrid simulated LBG samples appear to deviate somewhat from the Zheng07 at high mass (centrals occupation turns over and satellites are flat) while the maximum mean occupation is lower than that of MTNG.
Whether or not real LBGs are well-described (at the level of 3D clustering statistics) by a simple stellar mass cut remains to be seen, though for the Astrid and MTNG galaxies we study here, a minimum stellar mass cut is roughly equivalent to a minimum SFR cut, which may be more appropriate for the observational selection of LBGs (e.g. as very blue galaxies).
A more careful modeling of LBG selection with photometric bands would be particularly interesting, similar to the investigation of simulated ELGs performed in Ref.~\cite{2021:Hadzhiyska_tng_elg}.

\section{EFT parameters \label{sec:eft_params}}

\subsection{Recap of the EFT forward model}
The basis for our EFT forward 
model~\cite{Schmittfull:2018yuk,Schmittfull:2020trd}
is the EFT bias 
expansion 
for the galaxy density field
$\delta_g$ relevant at the 
one-loop power spectrum level,
~\cite{Assassi:2014fva,Desjacques:2016bnm,Ivanov:2019pdj}, 
\be 
\label{eq:naive_eft}
\begin{split}
\delta^{\rm EFT}_g\Big|_{\rm 2-pf}(\k) = b_1\delta + \frac{b_2}{2}\delta^2 +b_{\mathcal{G}_2}\mathcal{G}_2 \\
+b_{\Gamma_3}\Gamma_3 - b'_{\nabla^2\delta} {\nabla^2}\delta
+\epsilon\,,
\end{split}
\ee 
where $\delta$ is the non-linear matter density field, and the non-linear integral kernels are defined as 
\be 
\label{eq:G2}
\begin{split}
\mathcal{G}_2(\k) & = \int_{\bm p} 
F_{\mathcal{G}_2}(\p,\k-\p)
\delta({\bm p})\delta(\k-{\bm p})\,,\\
 F_{\mathcal{G}_2}(\k_1,\k_2) & =\frac{({\bm k_1}\cdot \k_2)^2}{k_1^2 k_2^2}-1\,,
\end{split}
\ee 
$\int_{\k}\equiv \int \frac{d^3\k}{(2\pi)^3}$
and $\GG$ is the Galileon tidal operator, 
\be 
\begin{split}
F_{\GG}=\frac{4}{7}\left(1-\frac{(\k_1\cdot\k_2)^2}{k_1^2k_2^2}\right)
\left(\frac{((\k_1+\k_2)\cdot \k_3)^2}{(\k_1+\k_2)^2k_3^2}-1\right)\,.
\end{split}
\ee 
$b_1,b_2,b'_{\nabla^2\delta},b_{\Gamma_3}$ above
are free coefficients
(for a given redshift and galaxy sample) called EFT bias parameters.
The field $\delta$ is subject to a non-linear expansion over the linear density field $\delta_1$:
\be 
\label{eq:d1_spt}
\begin{split}
\delta = &\sum_{m=1}^3\left(
\prod_{n=1}^m
\int_{\k_n}\delta_1(\k_n)
\right)(2\pi)^3\delta_D^{(3)}(\k-\k_{1...n})F_n\\
&-c_s k^2 \delta_1(\k)\,,
\end{split}
\ee 
where we limited ourselves to the one-loop order, $c_s$ is the dark matter sound speed~\cite{Carrasco:2012cv}, and $F_n$ are the so-called standard perturbation theory density kernels~\cite{Bernardeau:2001qr}.
The field $\epsilon$ in eq.~\eqref{eq:naive_eft} is the stochastic density component responsible for the shot noise. 
Plugging ~\eqref{eq:d1_spt} produces a nested perturbative expansion which unobservably shifts the observed values of some of the EFT coefficients.
In particular, the higher order counterterm is shifted as 
\be 
b_{\nabla^2\delta} = \frac{-b_1c_s+b'_{\nabla^2\delta}}{k^2_{\rm norm}}~\,,
\ee 
where $k_{\rm norm}=1~\hMpc$ is a normalization scale that makes $b_{\nabla^2\delta}$ dimensionless. 
For other EFT parameters, however, we will use their original definitions ~\eqref{eq:d1_spt} to match the EFT parameter convention used in the public code CLASS-PT~\cite{Chudaykin:2020aoj}.
We will refer to the new set $b_1,b_2,b_{\nabla^2\delta},b_{\Gamma_3}$ as real space EFT parameters in what follows. 

Eq.~\eqref{eq:naive_eft}
is subject to redshift-space mapping~\cite{Bernardeau:2001qr} and additional renormalization of contact operators~\cite{Senatore:2014vja}, which produces the following expression for the galaxy density field in redshift space $\delta^{(s)}_g$:
\be
\begin{split}
\delta^{(s)}_g({\bf k})= &\sum_{n=1}^3 \Big[ \prod_{j=1}^n\int_{}\frac{d^3{\bf k}_j}{(2\pi)^3} \delta_1({\bf k}_j)\Big]
 Z_n({\bf k}_1,...,{\bf k}_n)\\
 &\times (2\pi)^3\delta^{(3)}_D({\bf k}-{\bf k}_1-...-{\bf k}_n)\\
 &+\sum_{n=0}^2c_{\mu^{2n}} f^{2n} \mu^{2n} \frac{k^2}{k^2_{\rm norm}} \delta_1({\bf k})\\
 &-\frac{b_4}{2} Z_1({\bf k})\frac{k^4}{k_{\rm norm}^4} f^4 \mu^4 \delta_1({\bf k})+\epsilon({\bf k})\,,
 \end{split}
 \label{eq:full_EFT}
\ee
where $Z_n$ are redshift space EFT kernels~\cite{Ivanov:2018gjr,Ivanov:2019pdj}. The parameters $c_{\mu^{2n}}$, $n=1,2$ and $b_4$ are redshift space counterterms.
They capture the response of small scale non-linear redshift space distortions (``FoG''~\cite{Jackson:2008yv}) to the large-scale density field. 
In this sense they describe the \textit{deterministic} component of FoG, which can be interpreted in the halo model as a Gaussian or Lorentzian damping of  galaxy power spectrum~\cite{Ivanov:2025qie}.
The equivalence principle dictates $c_{\mu^4}$ to be the same for galaxies and dark matter.
In contrast, $c_{\mu^2}$ is sensitive to the galaxy population. 
$b_4$ is the higher order counterterm necessary to account for strong FoG~\cite{Ivanov:2019pdj,Chudaykin:2020hbf,Taule:2023izt}.
Note that while some EFT analyses prefer to ignore this term, the evidence for its presence is significant both from simulations and the actual data. 
The reason why this term is often ignored is that it is approximately degenerate with the redshift-space stochasticity counterterm (introduced shortly), if the fit is performed at the power spectrum level. 
In the context of our field level study, however, the two contributions can be clearly separated and measured independently. 
Importantly, both of them are needed to reproduce the field-level results. 
In particular, the other deterministic redshift-space counterterms $c_{\mu^2},c_{\mu^4}$ exhibit a strong scale dependence if $b_4$ is not included in the fit.
We note however that while $b_4$ is significant for the LRGs and ELGs of DESI~\cite{Ivanov:2021fbu,Ivanov:2021zmi}, it is unclear \textit{a priori} if this term is actually necessary for the description of the high-redshift galaxies. 
Our work explicitly addresses this question. 

Taking expectation values of~\eqref{eq:full_EFT} and using the Gaussian distribution for $\delta_1$,
\be 
\langle \delta_1(\k)\delta_1(\k')\rangle =(2\pi)^3\delta_D^{(3)}(\k'+\k)P_{11}(k)\,,
\ee 
(following from linear cosmological
perturbation theory) produces the one-loop EFT power spectrum.
The EFT power spectrum in Eulerian perturbation theory should also be corrected for large-scale bulk flows,
via a procedure known as IR 
resummation~\cite{Senatore:2014via,Baldauf:2015xfa,Vlah:2015zda,Blas:2015qsi,Blas:2016sfa,Ivanov:2018gjr,Vasudevan:2019ewf}. This procedure
is equivalent to resumming the Zel'dovich displacement in Lagrangian perturbation theory. 
The stochastic field $\epsilon$ has the following power spectrum in EFT, 
\be 
\label{eq:err_eft}
P_{\rm err} = \frac{1}{\bar n}\left(
1+\alpha_0+\alpha_1\frac{k^2}{k_S^2} + 
\alpha_2\mu^2 \frac{k^2}{k_S^2}\,,
\right)
\ee 
where $k_S$ is the normalization scale which we take to be $0.45~\hMpc$ following~\cite{Philcox:2021kcw}.
The $\alpha_{0,1,2}$ parameters above are the stochastic EFT parameters. 
Importantly, due to the equivalence principle, the redshift-space contributions to the stochastic power spectrum start with $\mu^2 k^2$, i.e. in the limit $k\to 0$ the error spectrum is white even in redshift space. 
$\alpha_2$ is the redshift-space stochasticity counterterm
that captures the \textit{stochastic} part of FoG, which is uncorrelated with the large-scale density field. 
While the halo model offers an interpretation of this term as produced by virial motion of satellites, in practice it receives significant contributions from the centrals as well~\cite{Maus:2024dzi,Ivanov:2025qie}.

The EFT forward model is based on the promotion of the EFT bias parameters to arbitrary functions of momentum called transfer functions. 
The key objects of the forward model are  ``shifted operators'' defined as
\be 
\label{eq:shift}
\tilde{\mathcal{O}}(\k)
=\int d^3 \q~\mathcal{O}(\q)
e^{-i\k\cdot(\q+\vpsi_1(\q)+f\hat{\bm{z}}(\vpsi(\q)\cdot \hat{\bm{z}}))}~\,,
\ee 
where $\q$ denote Lagrangian space (initial) coordinates, $\mathcal{O}(\q)$ is a Lagrangian bias operator, $\vpsi_1$ is the Zel'dovich displacement, 
\be 
 \bm{\psi}_1(\q,z)=\int d^3q~e^{i\q\k}\frac{i \bm{k} }{k^2}\delta_1(\k,z)\,,
\ee 
where $f$ is the logarithmic growth factor. 
To avoid degeneracies in the forward model description, the shifted operators are orthogonalized following the Gram–Schmidt process ensuring that 
\be 
\langle \tilde{\mathcal{O}}^\perp_m 
\tilde{\mathcal{O}}^\perp_n \rangle  = 0\quad \text{if}
\quad n\neq m\,.
\ee 
Our full EFT forward model inspired by the one-loop EFT power spectrum calculation is given by~\cite{Schmittfull:2020trd,Obuljen:2022cjo}
\be 
\label{eq:eft-field_rsd}
\begin{split}
&\delta^{\rm EFT}_g (\k,\hat{\bm z})= \delta_Z(\k,\hat{\bm z})-\frac{3}{7}\mu^2 f \tilde{\mathcal{G}_2}\\
&\beta_1(k,\mu)\tilde 
\delta_1(\k,\hat{\bm z})
+\beta_2(k,\mu)
(\tilde{\delta}_1^2)^\perp (\k,\hat{\bm z}) \\
& +\beta_{\mathcal{G}_2}(k,\mu)
\tilde{\mathcal{G}_2}^{\perp}(\k,\hat{\bm z})
+\beta_3(k,\mu)
(\tilde{\delta}_1^3)^\perp (\k,\hat{\bm z})\,,
\end{split}
\ee 
where $\delta_Z=\tilde 1$ is the Zel'dovich matter density field, $\mu \equiv (\k\cdot \hat{\bm z})/k$. 
The real space forward model is obtained by setting $f=0$.
This model can be summarized by the transfer functions $\beta_i$
\be 
\label{eq:trfreal}
\beta_i(k,\mu)=\frac{\langle \mathcal{O}^*{}^\perp_i(\k)  \delta_g^{\rm truth}(\k)\rangle'}{\langle
|\mathcal{O}'^\perp_i (\k) |^2 
\rangle' }\,,
\ee 
where $\delta_g^{\rm truth}(\k)$ is the density field from the 
simulation and primes denote the correlation functions with the Dirac deltas removed from them.
The transfer functions $\beta_i$ then can be matched to the EFT predictions at a given order. 
In particular, for the one-loop power spectrum order we use eq.~\eqref{eq:naive_eft}.
This way the EFT parameters are extracted without cosmic variance. 
Note that the EFT model \eqref{eq:eft-field_rsd} includes a cubic operator $\beta_3 \tilde{\delta}^3_1$.
It matches the Eulerian bias operator $\frac{b_3}{3!}\delta^3$ which is important at the level of the one-loop bispectrum~\cite{Eggemeier:2018qae,Philcox:2022frc} and the two-loop power spectrum. 

The performance of the EFT forward model is estimated by the  error spectrum, which by definition captures the part of the density field uncorrelated with the EFT operators in eq.~\eqref{eq:eft-field_rsd},
\be 
P_{\rm err}(k,\mu) = 
\langle 
|\delta^{\rm EFT}(\k,\hat{\bm z}) - \delta_g^{\rm truth}(\k,\hat{\bm z})|^2
\rangle' \,.
\ee 
On large scales the error spectrum matches the EFT  prediction~\eqref{eq:err_eft}, which allows one to extract the stochastic EFT parameters.

\subsection{EFT transfer functions at the field level}
\begin{figure*}
\centering
\includegraphics[width=0.99\textwidth]{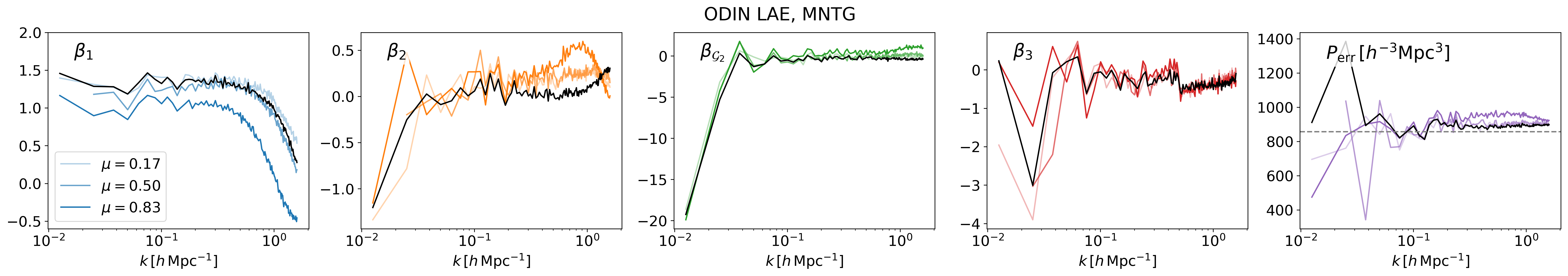}
\includegraphics[width=0.99\textwidth]{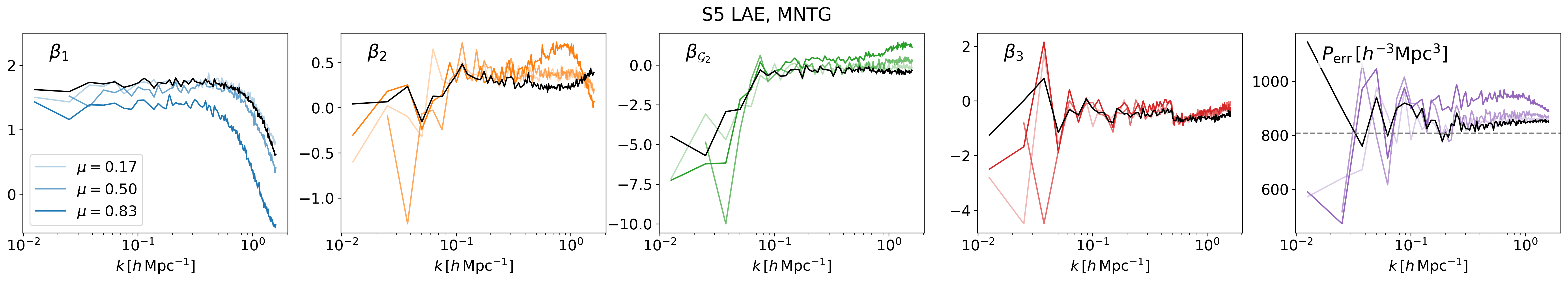}
\includegraphics[width=0.99\textwidth]{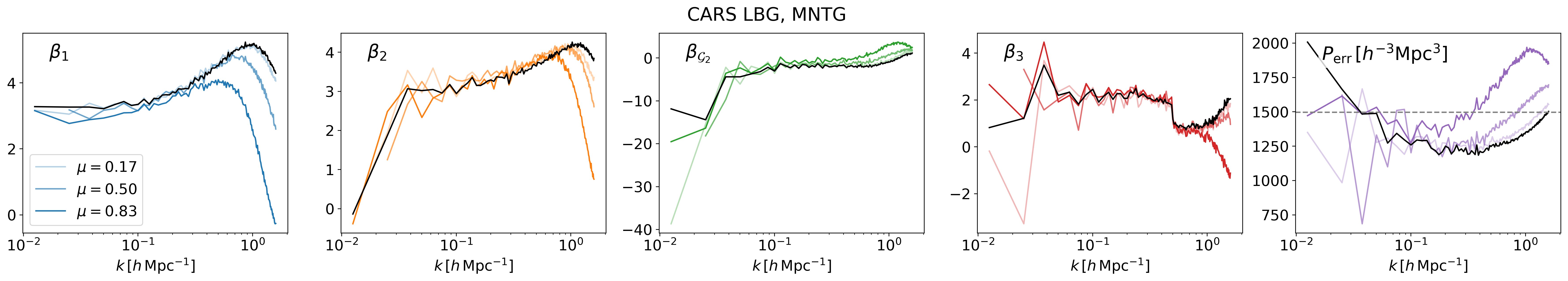}
\includegraphics[width=0.99\textwidth]{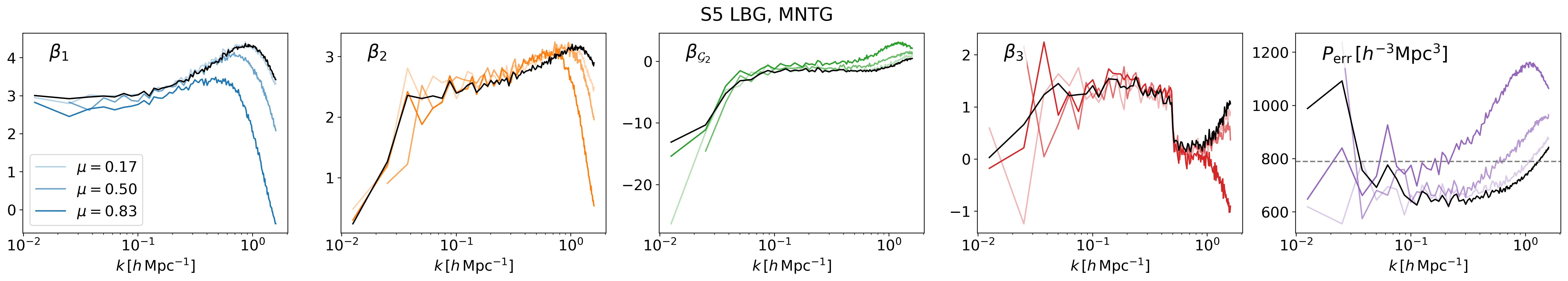}
   \caption{Bias transfer functions extracted form the samples  defined in the MTNG simulations.
   From top to bottom we show each galaxy sample defined in Tab.~\ref{tab:samples_mtng} (ODIN LAEs, S5 LAEs, S5 LBGs, and CARS LBGs).
   From left to right we show the different bias transfer functions ($\beta_1(k,\mu),\beta_2(k,\mu),\beta_{\mathcal{G}_2}(k,\mu),\beta_3(k,\mu)$ and the error power spectrum $P_\mathrm{err}(k)$.
   Increasing darkness of shaded colors represent increasing $\mu$ bins, as indicated in the legend (with the black curve being $\mu=0$).
    } \label{fig:transf_mtng}
\end{figure*}

\begin{figure*}
\centering
\includegraphics[width=0.24\textwidth]{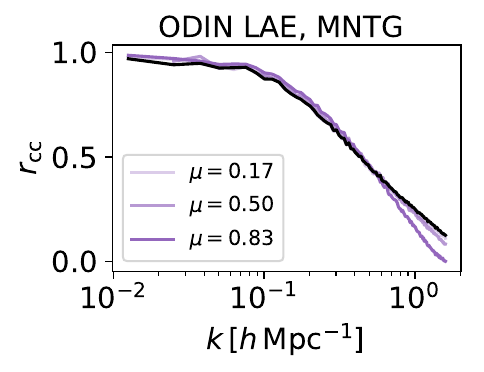}
\includegraphics[width=0.24\textwidth]{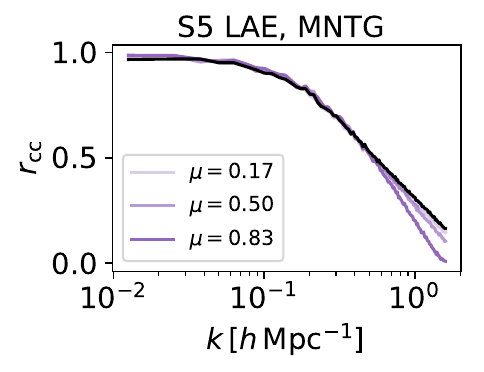}
\includegraphics[width=0.24\textwidth]{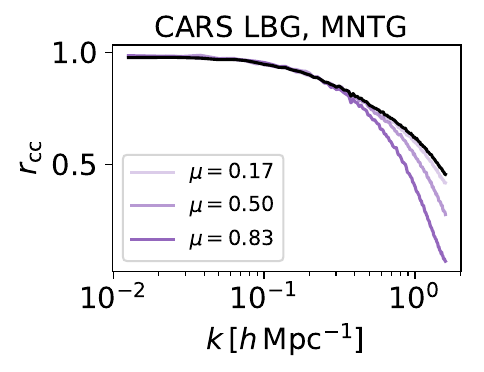}
\includegraphics[width=0.24\textwidth]{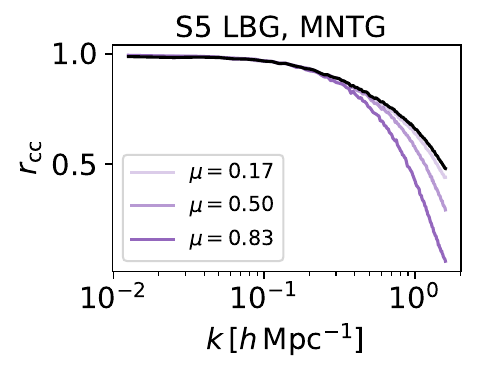}
   \caption{Cross-correlation coefficients 
   between Fourier modes of the EFT model
   and the MTNG galaxy simulation data. 
   Color shades have the same meaning as the rightmost ($P_{\rm{err}}$) column of Fig.~\ref{fig:transf_mtng}.
    } \label{fig:cc_mtng}
\end{figure*}

\begin{figure*}
\centering
\includegraphics[width=0.99\textwidth]{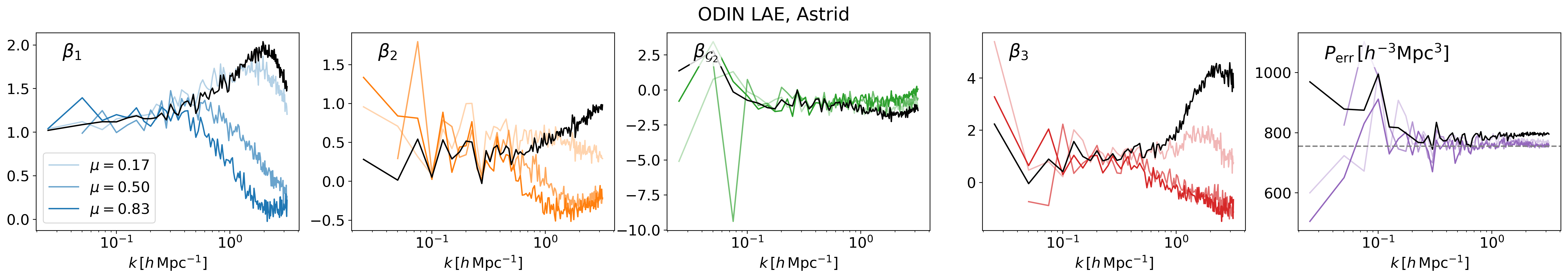}
\includegraphics[width=0.99\textwidth]{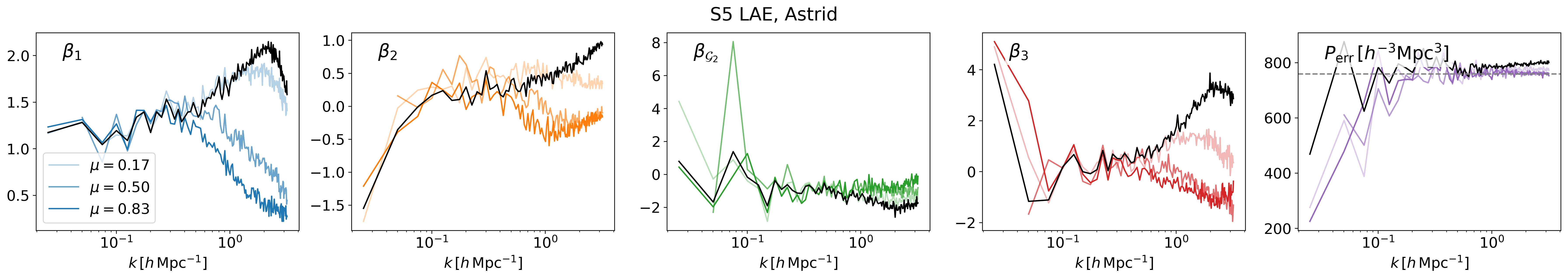}
\includegraphics[width=0.99\textwidth]{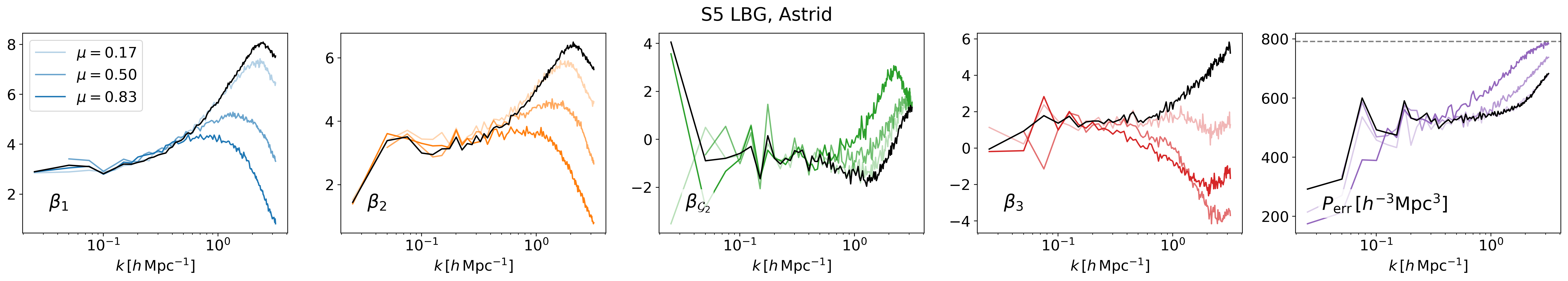}
\includegraphics[width=0.99\textwidth]{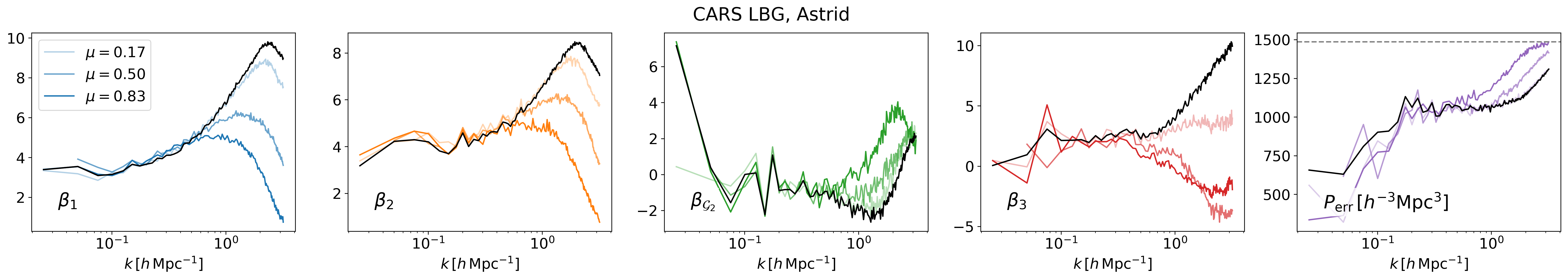}
   \caption{Bias transfer functions extracted form the samples  defined in the Astrid simulations. Panels are organized as in Fig.~\ref{fig:mtng_hod}, where samples are those drawn from Tab.~\ref{tab:samples_astrid} 
    } \label{fig:transf_astrid}
\end{figure*}

\begin{figure*}
\centering
\includegraphics[width=0.24\textwidth]{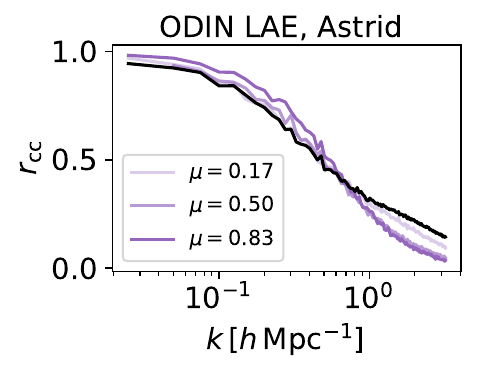}
\includegraphics[width=0.24\textwidth]{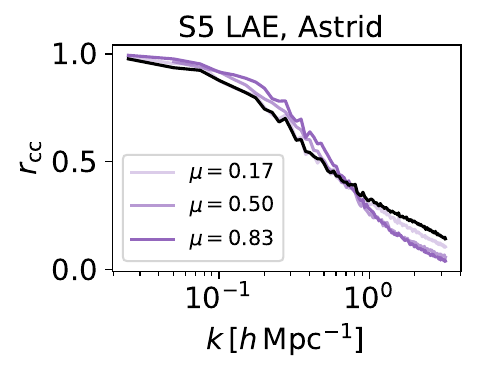}
\includegraphics[width=0.24\textwidth]{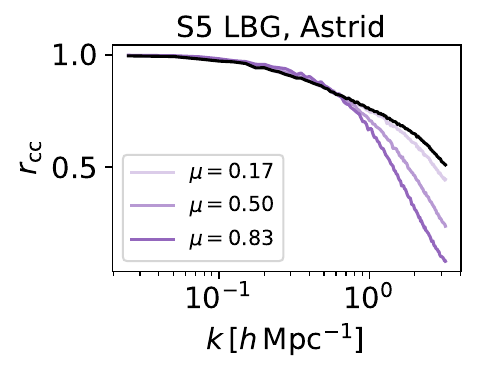}
\includegraphics[width=0.24\textwidth]{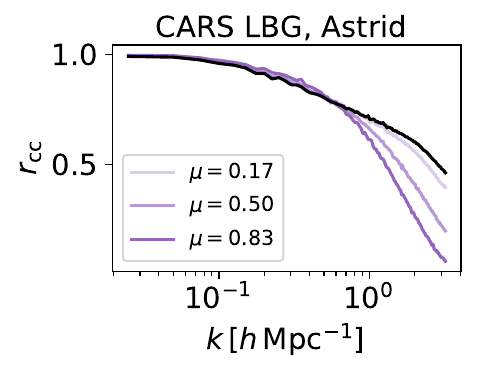}
   \caption{Cross-correlation coefficients 
   between Fourier modes of the EFT model
   and the Astrid simulation data. 
    } \label{fig:cc_astrid}
\end{figure*}

Figures~\ref{fig:transf_mtng},~\ref{fig:cc_mtng} and Figures~\ref{fig:transf_astrid},~\ref{fig:cc_astrid} show the field-level redshift-space bias transfer functions $\beta_i(k,\mu)$
and cross-correlation coefficients
up to cubic order as measured from the MTNG and Astrid samples defined in Section~\ref{sec:selection}, respectively \cite{Schmittfull:2018yuk,Schmittfull:2020trd,Obuljen:2020ypy,Ivanov:2024hgq,Ivanov:2024xgb,Ivanov:2024_mtngeft}.
The final column also shows the error power spectrum.
The bias coefficients are extracted from the large-scale (low $k$) limits of the transfer functions, which are constructed to be $\mu$ independent on large scales.
See Ref.~\cite{Ivanov:2024_mtngeft} for further details on the determination of the bias coefficients and stochastic terms from the field level.

Overall, the transfer functions look very similar
to those of dark matter halos
and HOD galaxies studied in the past~\cite{Schmittfull:2018yuk,Schmittfull:2020trd,Ivanov:2024hgq,Ivanov:2024xgb}.
In particular, the transfer functions are generally flat and $\mu$-independent in the $k\to 0$ limit.\footnote{There is some $\mu$-dependence of the $\beta_1$ transfer functions on very large scales visible for ODIN LAE and S5 LAE MTNG samples. 
We believe that this is due to a statistical fluctuation.
}
Comparing the shape of the transfer functions with those extracted from DESI like LRG and ELG galaxies from MTNG and Astrid~\cite{Ivanov:2024_mtngeft}, we notice that LAEs appear quite similar to ELGs, while LBG's transfer functions behave similarly to those of LRGs. 
As far as the error power spectrum is concerned, it is white and $\mu$-independent in the $k\to 0$ limit in perfect agreement with the EFT expectation. 
In this limit it approaches the Poissonian prediction, shifted by a constant due to the halo exclusion effects~\cite{Baldauf:2013hka}.

We also show the cross-correlation coefficient between the truth density field and the EFT forward model,
\be 
\begin{split}
r_{cc}(k,\mu)&=\frac{\langle \delta_g^{\rm EFT}(\k,\hat{\bm z}) [\delta_g^{\rm truth}(\k,\hat{\bm z})]^*\rangle'}{
\sqrt{\langle |\delta_g^{\rm EFT}(\k,\hat{\bm z})|^2\rangle' 
\langle |\delta_g^{\rm truth}(\k,\hat{\bm z})|^2\rangle'}
}\\
&=\sqrt{1-\frac{P_{\rm err}}{P_{\rm truth}}}
=\sqrt{\frac{P_{\rm EFT}}{P_{\rm truth}}}\,,
\end{split}
\ee 
in Fig.~\ref{fig:cc_mtng} for MTNG and Fig.~\ref{fig:cc_astrid} for Astrid (where the second line applies at the best-fit transfer functions \cite{Schmittfull:2018yuk}).

\textit{MTNG:} 
On large scales, $k<0.1~\hMpc$, we achieve a nearly perfect correlation between the MTNG galaxies and the EFT model for all samples, which illustrates the power of the bias expansion. 
The correlation coefficient is slightly better along the line of sight due to the velocity information encoded in the linear Kaiser redshift space distortions~\cite{Kaiser:1987qv}. 
The EFT model is correlated with the simulated LAEs at the level of about $50\%$ at $k\simeq 0.5~\hMpc$. 
For LBGs in real space, $r_{cc}$ drops below $50\%$ only on very small scales, $k\sim 1~\hMpc$. 
This behavior is similar to that of dark matter halos~\cite{Schmittfull:2018yuk}, suggesting that LBGs trace their parent halos better than LAEs. 
LBGs are significantly affected by FoG because their correlation drops faster for the line-of-sight modes. 
For LAEs, however, the cross-correlation does not exhibit any strong $\mu$-dependence, so the correlation is worsened primarily at the level of the bias expansion (as opposed to at the level of FoG). 

\textit{Astrid:}
The cross-correlation of Astrid galaxies with the EFT model is qualitatively similar to that of MTNG. 
For LAEs, the correlation approaches $100\%$ on very large scales, and drops by about $50\%$ around $k\simeq 0.5~\hMpc$.
For these galaxies the $\mu$-independence of the cross-correlation coefficient is mild.
In contrast, the LBGs is correlated with the EFT model up to $\approx 1~\hMpc$. 
The correlation coefficient 
becomes significantly $\mu$-dependent around this scale, which reflects the effect of fingers of God. 
The real space density field correlation drops below 50$\%$ only for $k\gtrsim 2~\hMpc$, deep in the one-halo regime. 
All in all, we see that the Astrid LAE's correlations with the EFT model are similar to that of MTNG, while Astrid's LBGs are slightly  better correlated with the large-scale fields than MTNG's LBGs.

\subsection{Bias parameters}
\begin{figure*}
\centering
\includegraphics[width=0.99\textwidth]{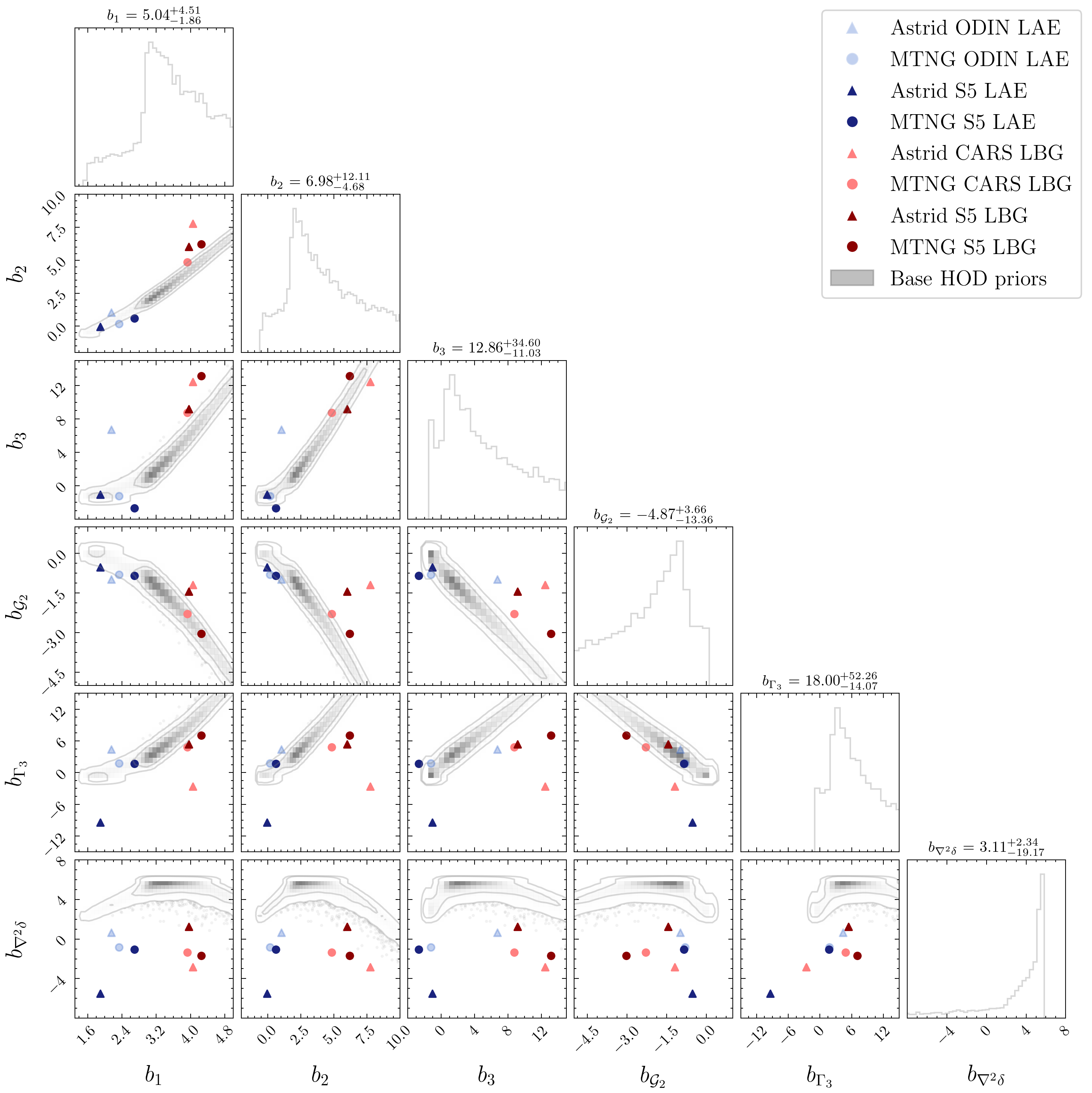}
   \caption{Bias parameters up to third order extracted for the 8 simulated samples we consider in MTNG (red points) and Astrid (blue points).
   We also show samples based on the Zheng07-like 7-parameter HOD model in gray. 
    } \label{fig:bias}
\end{figure*}

Figure~\ref{fig:bias} and Table~\ref{tab:parameters} summarize the main results of this work in terms of the EFT bias parameter values.
Both MTNG and Astrid find low values of the linear bias for LAEs ($b_1\approx2-2.5$, blue points), and high values for LBGs ($b_1 \approx 4$, red points) that are consistent with the bias values roughly estimated from angular clustering (Tabs.~\ref{tab:samples_mtng},\ref{tab:samples_astrid}).
The linear bias values are also consistent with values that have been measured from several existing surveys using the angular correlation function $w(\theta)$.
For LAEs, Ref.~\cite{White:2024_odin} uses the COSMOS field ODIN LAEs ($\approx$1100 galaxies in a $9~\mathrm{deg}^2$ area) at $z=3.1$ to find $b_{1} = 2.0 \pm 0.2$.
The MTNG ODIN sample EFT $b_1$ is within $2\sigma$ of this value while the Astrid sample EFT $b_1$ is within $1\sigma$ of the measured value from Ref.~\cite{White:2024_odin}.
This is the specific sample for which we aimed to match the clustering.
For further context, recent measurements include those of Ref.~\cite{Umeda:2024_SILVERRUSH_LF_ACF}, who use the SILVERRUSH catalog \cite{Kikuta:2023_SILVERRUSH_catalog} of LAEs (with a narrowband filter corresponding to $z=3.3$, 4641 galaxies in a $1.8~\mathrm{deg}^2$ field) to find $b_{1} = 1.59 \pm 0.20$, while older measurements of the linear bias from smaller area surveys are summarized in Fig.~16 of Ref.~\cite{ouchi:2020_AnnRev}\footnote{Though note that the values of $b_1$ there are compared to predictions for dark matter halos under the assumption of 1-1 correspondence between halos and LAEs.}.
For LBGs, the bias values are consistent with the $b_{1} = 4.0 \pm 0.26$ measured from the CARS survey \cite{Hildebrandt:2009_CARS_LBG} (specifically for the $u-$drop galaxies with $r<24.5$ with ``BC\_sim'' photometric redshifts, which corresponds to redshift distribution centered at $z\approx 3$), as expected. 

Figure~\ref{fig:bias} also shows in gray the density contours corresponding to a modified Zheng07 HOD (i.e. with the addition of the central and satellite velocity parameters $\alpha_c,\alpha_s$, as described in Ref.~\cite{Ivanov:2024_mtngeft,Paillas:2023cpk}) with parameters sampled according to the mean halo masses we inferred for $M_{\mathrm{min}}$ and other parameters sampled from the following 
LRG-based priors:
\be 
\label{eq:extHOD_pr}
\begin{split}
& 
\log_{10} M_{\rm cut}\in [11,13]\,,\quad 
\log_{10} M_1\in [12,14]\,,\\
& \log \sigma \in [-3,0]\,,\quad  \alpha \in [0.5,1.5]\,,\quad \\
& \alpha_c \in [0,0.5]\,,
\quad 
\alpha_s \in [0.7,1.3]\,,
\quad  \kappa \in [0.0,1.5]\,,
\end{split}
\ee 
where $M_1$ is the satellite threshold mass, $\sigma$ is the width of the threshold for the centrals, $\alpha$ is the power-law slope for the satellite HOD, $\kappa$ adjust the satellite mass, and $\alpha_{c/s}$ are the velocity bias parameters for centrals and satellites, respectively. 
We compute the EFT parameter predictions for these models using the analytic approach of \cite{Ivanov:2025qie}.
The higher-order bias parameters $b_2,b_{\mathcal{G}_2},b_3,...$ mostly track the HOD contours, though the LBG samples tend to deviate from them more often, and the match for the higher derivative term $b_{\nabla^2 \delta}$ is generally poor.
We stress, however, that the gray LRG-HOD based samples in Fig.~\ref{fig:bias} is only a rough estimate of EFT parameters for high-redshift galaxies. 
The actual shape of the HOD extracted from the MTNG and Astrid data is quite different from the HOD model form (except MTNG's LBGs). 
Assembly bias and baryonic feedback introduce an additional mismatch between the measured EFT parameters and the predictions of the simplistic Zheng07-like model. 
We expect that the use  of more sophisticated ``decorated'' HOD models~\cite{Hearin:2015jnf,Yuan:2021izi,DESI:2023ujh} will produce a better match to  the EFT parameter of the high-redshift galaxies.
Nevertheless, we observe that the measured  bias parameters agree with the 
the simple HOD model predictions used here quite well.

\begin{figure*}
\centering
\includegraphics[width=0.99\textwidth]{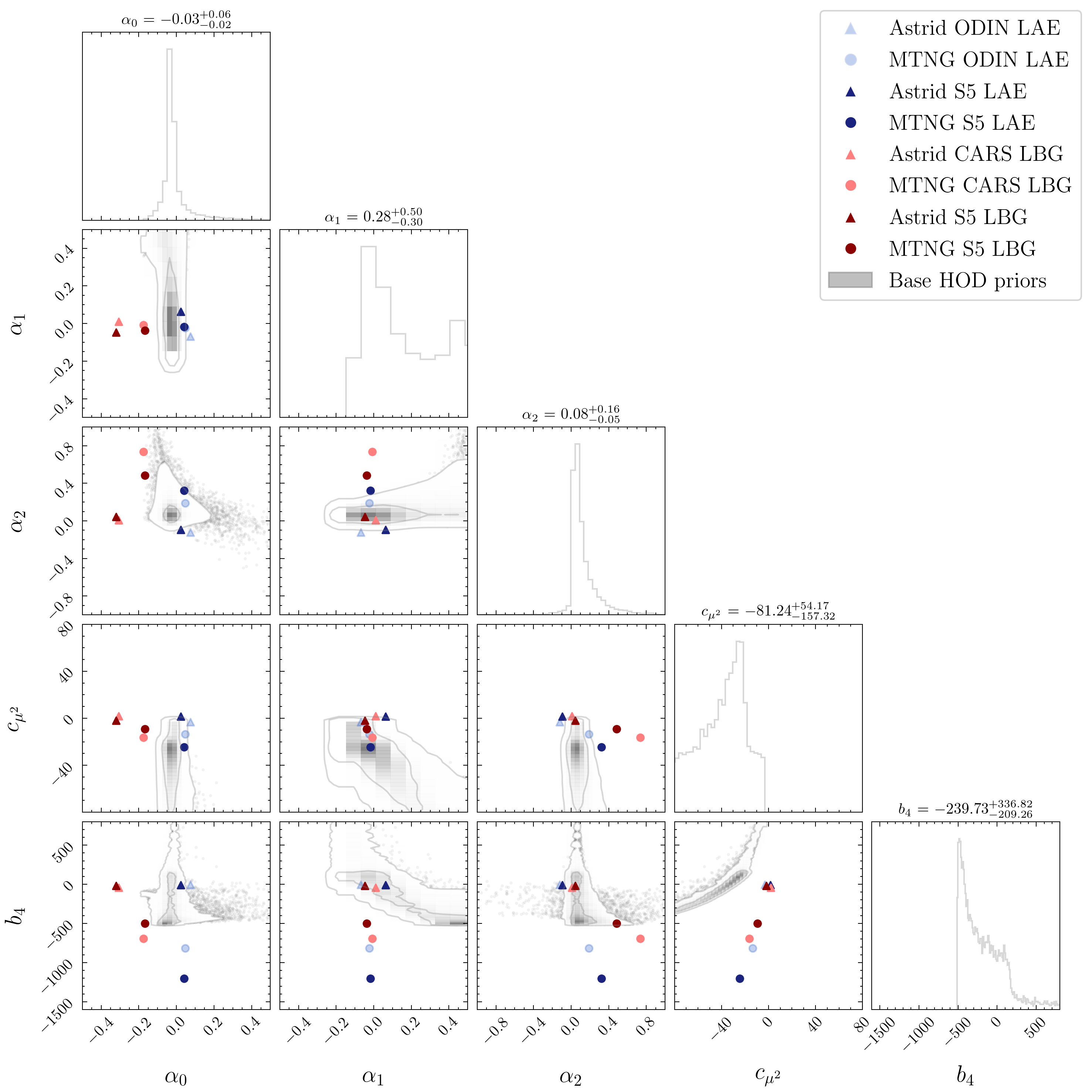}
   \caption{Redshift space distortion and stochastic amplitude EFT parameters. Panels are organized similarly to Figure~\ref{fig:bias} with the same legend meanings.
    } \label{fig:rsd}
\end{figure*}

Figure~\ref{fig:rsd} (and Table~\ref{tab:parameters} ) show the field-level fitted values of the EFT stochastic and counterterm coefficients.
These values are generally consistent with the HOD contours, though the LBGs seem to show a preference for sub-Poisson scale-independent stochasticity $\alpha_0$, and the disagreement between HOD contour and the simulated sample in the $c_{\mu^2}-b_4$ plane is particularly noticeable, and in this case the MTNG galaxies and Astrid galaxies seem to separate regardless of galaxy type.

The large values of the non-linear bias parameters for LBGs suggest that uncertainties in the values of these bias parameters will limit the application of EFT-based methods on quasi-linear scales.
In particular, if the one-loop power spectrum model is used, we do not expect the momentum reach $\kmax$ to be larger than that of the BOSS/DESI LRG galaxies, $\kmax\approx 0.4~\hMpc$ (in real space)~\cite{Chudaykin:2020hbf,Ivanov:2021fbu}. 
In contrast, for LAEs, the small values of the bias parameters indicate that such galaxies will be less dominated by non-linear bias contributions. 
If the real space power spectrum of these galaxies could be observed, it could be fit with EFT down to significantly smaller scales, $\kmax\simeq 1~\hMpc$.

We note, however, that the EFT predictions can be computed to high orders, in which case the actual cutoff $\kmax$ will be limited by the scale-dependence of the shot noise.\footnote{See e.g. for similar discussions in the case of the EFT for dark matter~\cite{Baldauf:2015tla,Baldauf:2015zga}.}
The characteristic non-linear momentum associated with the scale dependence of $P_{\rm err}$ in real space can be estimated as $0.45~\hMpc/\sqrt{|\alpha_1|}$. 
Then, $\kmax$ in real space can be estimated by demanding that the next-to-next-to-leading order (two-loop) correction to the scale-dependence of the noise be $15\%$ of the next-to-leading order ($k^2$),
\be 
\label{eq:kmax_real}
k_{\rm max}^{\rm real}\approx 
0.4
\frac{0.45~\hMpc}{\sqrt{|\alpha_1|}}\,,
\ee 
which implies, for the LAEs:
\be 
\begin{split}
& \text{MTNG LAEs:} \quad \frac{k_{\rm max}^{\rm real}}{\hMpc}=1.2-1.3\,,\\
& \text{Astrid LAEs:} \quad \frac{k_{\rm max}^{\rm real}}{\hMpc}=0.69-0.71\,,\\
\end{split}
\ee 
while for LBGs we obtain:
\be 
\begin{split}
& \text{MTNG LBGs:} \quad \frac{k_{\rm max}^{\rm real}}{\hMpc}=0.94-1.95\,,\\
& \text{Astrid LBGs:} \quad \frac{k_{\rm max}^{\rm real}}{\hMpc}=0.83-1.8\,.
\end{split}
\ee 
Overall, these results are quite similar to those of MTNG galaxies selected to approximate DESI LRGs ($k_{\rm max}^{\rm real}\approx 0.8~\hMpc$) and ELGs ($k_{\rm max}^{\rm real}\approx 1.2~\hMpc$) (see Fig.~6 of Ref.~\cite{Ivanov:2024_mtngeft}).
For comparison, the Astrid figures are $k_{\rm max}^{\rm real}\approx 0.6~\hMpc$ (LRG) and $k_{\rm max}^{\rm real}\approx 0.8~\hMpc$ (ELG).

\begin{table*}[htb!]
\centering
\begin{tabular}{|c|c|c|c|c|c|c|c|c|}
\hline
\tiny{Param.} & \tiny{ODIN LAE MTNG} & \tiny{S5 LAE MTNG} & \tiny{S5 LBG MTNG} & \tiny{CARS LBG MTNG} & \tiny{ODIN LAE Astrid} &  
\tiny{S5 LAE Astrid} & 
\tiny{S5 LBG Astrid} & 
\tiny{CARS LBG Astrid} \\
\hline
$b_1$ & 2.33 & 2.68 & 3.92 & 4.24 & 2.15 & 1.89 & 4.05 & 3.96 \\
\hline
$b_2$ & 0.17 & 0.61 & 4.85 & 6.21 & 1.05 & -0.052 & 7.76 & 6.02 \\
\hline
$b_3$ & -1.23 & -2.68 & 8.75 & 13.14 & 6.73 & -1.05 & 12.46 & 9.17 \\
\hline
$b_{\G}$ & -0.803 & -0.84 & -2.28 & -3.03 & -0.98 & -0.52 & -1.19 & -1.44 \\
\hline
$b_{\GG}$  & 1.76 & 1.71 & 4.83 & 7.04 & 4.37 & -9.44 & -2.62 & 5.38 \\
\hline
$b_{\nabla^2\delta}$  & -0.84 & -1.05 & -1.35 & -1.69 & 0.66 & -5.49 & -2.83 & 1.25 \\
\hline
$\alpha_0$ & 0.046 & 0.041 & -0.17 & -0.16 & 0.074 & 0.023 & -0.307 & -0.32 \\
\hline
$\alpha_1$  & -0.023 & -0.018 & -0.0085 & -0.037 & -0.068 & 0.063 & 0.010 & -0.047 \\
\hline
$\alpha_2$  & 0.18 & 0.32 & 0.73 & 0.48 & -0.12 & -0.093 & 0.0081 & 0.044 \\
\hline
$c_{\mu^2}$ & -13.49 & -24.39 & -16.33 & -9.22 & -3.16 & 1.53 & 1.96 & -1.80 \\
\hline
$b_4$ & -815.02 & -1202.88 & -692.46 & -499.51 & -6.09 & -10.44 & -43.52 & -18.15 \\
\hline
\end{tabular}
\caption{EFT parameters of the main samples of the hydrodynamical galaxies studied in our work.}
\label{tab:parameters}
\end{table*}

\subsection{Fingers of God}
Figure~\ref{fig:transf_mtng} and Figure~\ref{fig:transf_astrid} show, in the right column, the error power spectrum of the field-level model as well as its angular dependence (denoted by color shades).
For LBGs in MTNG, $P_{\mathrm{err}}$ becomes 
strongly $\mu$ dependent around $k\approx0.2-0.3~h\rm{Mpc}^{-1}$, while in Astrid this occurs around $k\approx1.0~h\rm{Mpc}^{-1}$.
It is not immediately clear why this $\mu$ dependence occurs at a smaller scale in Astrid.
The higher stellar mass minimum needed to achieve the same number density in Astrid compared to in MTNG (see Section~\ref{sec:selection}) would naively imply that the more massive Astrid galaxies would have a large FoG effect.
For LAEs, however, in both simulations $P_{\mathrm{err}}$ is (which is essentially completely flat in $k$) has a milder $\mu$ dependence, setting in for MTNG at $k\approx0.2-0.3~h\rm{Mpc}^{-1}$ and for Astrid at $k\approx0.8~h\rm{Mpc}^{-1}$.
Similar qualitative observations about the $\mu$ dependence relevant for FoG strength apply to the cross-correlation coefficients in Figs.~\ref{fig:cc_mtng},\ref{fig:cc_astrid}.
Despite the large satellite fraction of the LAE samples, the $P_\mathrm{err}$ do not suggest the presence of a large FoG effect. 
Since LAE host halos are have low masses, the 
nonlinear velocity dispersion due to virial motions should also be small, potentially leading to the observed low FoG suppression.
The MTNG values of 1-dimensional velocity dispersion $\sigma$ in Table~\ref{tab:samples_mtng} support this interpretation.

As far as the higher order ($\sim k^4 P_{11}$) operator with the $b_4$  coefficient is concerned,  we see that the MTNG and Astrid galaxies give a different answer to the question of whether this term is needed to fit the data. 
MTNG galaxies clearly prefer the presence of this term, while for Astrid galaxies generically it gives a vanishingly small contribution to the transfer functions. 
This can be appreciated from the shape of the transfer function $\beta_1$ for MTNG and Astrid (see Figures~\ref{fig:transf_mtng},~\ref{fig:transf_astrid}) while $\beta_1$ has a clear $\mu$-dependence in MTNG, suggesting significant counterterms, $\beta_1$ of Astrid galaxies is practically $\mu$-independent up to $k\sim 0.5~\hMpc$, implying a very weak FoG effect on the deterministic part of the galaxy power spectrum.

The stochastic fingers of God are the main factor limiting the application of EFT in redshift space. 
When the noise becomes highly scale-dependent, the EFT gradient expansion breaks down and perturbation theory cannot be used anymore. 
To estimate $\kmax$ in redshift space, we use the same logic as in real space. 
Specifically, we require that the two-loop contribution to the scale-dependence of the noise in redshift space be $15\%$ of the next-to-leading order ($k^2\mu^2$) result for $\mu\approx 1$. 
This suggests the estimate 
\be 
\label{eq:kmaxFOG}
k_{\rm max}^{\rm FoG}\approx 
0.4
\frac{0.45~\hMpc}{\sqrt{|\alpha_2|}}\,.
\ee 
For the $\alpha_2$ measurements of MTNG's LRG samples from~\cite{Ivanov:2024_mtngeft} this produces $k^{\rm FoG}_{\rm max}\approx 0.2~\hMpc$, which agrees with the momentum cut used in consistent EFT-based analyses of the LRG samples~\cite{Philcox:2021kcw,Chudaykin:2024wlw}.
For MTNG's ELGs this gives 
$k^{\rm FoG}_{\rm max}\approx 0.25~\hMpc$, in
agreement with the eBOSS ELG data 
analysis~\cite{Ivanov:2021zmi}. 
(For Astrid we have $k^{\rm FoG}_{\rm max}\Big|_{\rm LRG}\simeq 0.26~\hMpc$
and $k^{\rm FoG}_{\rm max}\Big|_{\rm ELG}\simeq 0.7~\hMpc$
~\cite{Ivanov:2024_mtngeft}.)
Applying Eq.~\eqref{eq:kmaxFOG} to our samples we find:
\be 
\begin{split}
& \text{MTNG LAEs:} \quad \frac{k_{\rm max}^{\rm FoG}}{\hMpc}=0.32-0.42\,,\\
& \text{Astrid LAEs:} \quad \frac{k_{\rm max}^{\rm FoG}}{\hMpc}=0.52-0.6\,,
\end{split}
\ee 
\be 
\begin{split}
& \text{MTNG LBGs:} \quad \frac{k_{\rm max}^{\rm FoG}}{\hMpc}=0.21-0.26\,,\\
& \text{Astrid LBGs:} \quad \frac{k_{\rm max}^{\rm FoG}}{\hMpc}=0.63-0.86\,.
\end{split}
\ee 
Importantly, all LAE samples considered here imply $\kmax^{\rm FoG}$ greater than $0.3~\hMpc$, which is a significant improvement over the currently used LRGs. 
As far as LBGs are concerned, MTNG predicts $\kmax$ similar to the current LRG values, while Astrid suggests a significantly larger momentum reach that exceeds even that of the MTNG LAEs.

\section{Conclusions \label{sec:conclusions}}

High-redshift star-forming galaxies will be the targets of leading cosmological survey efforts in the next decade.
Understanding the clustering properties of these galaxies is crucial for extract cosmological information from such surveys.
In this paper, we take the first step toward characterizing the non-linear bias and EFT parameters of these galaxies using sample variance cancellation at the field level.
We reproduce several simulated samples intended to model LBGs and LAEs that match the clustering properties and number densities of existing measurements from smaller surveys.
We find that our simulated samples of LBGs roughly act as high-redshift counterparts of LRGs, having $k_{\rm max}\approx 0.2-0.8~h\rm{Mpc}^{-1}$, 
while the simulated LAEs have quite low bias values and $k_{\rm max}\approx 0.3-0.6~h\rm{Mpc}^{-1}$.
While the selection method is distinct, the qualitative features of the error power spectrum of LAEs is not dissimilar to the case of ELGs \cite{Ivanov:2024_mtngeft}, and exhibits very little scale and $\mu$ dependence.

This suggests that the high-redshift galaxies may be under better overall perturbative control than the currently-used LRGs. 
This offers positive prospects for testing fundamental cosmological physics with these galaxies in the era of future high-precision surveys. 
Our results can be used to revisit the forecasted precision of cosmological parameter measurements with Spec-S5/Megamapper surveys along the lines of~\cite{Chudaykin:2019ock,Sailer:2021yzm,Cabass:2022epm}. 

While this work provides a first step toward characterizing the plausible clustering properties of SFGs in next-generation cosmological surveys, more detailed modeling of LBGs and LAEs and their selection would make the simulated samples more realistic.
A more involved treatment that stops short of full radiative-transfer simulations could involve modeling stellar emission via stellar population synthesis and modeling nebular emission to obtain a forward model for SEDs that includes the impact of dust (and therefore may enrich modeling of $L_{\mathrm{Ly}\alpha}^\mathrm{obs}$, e.g. beyond metallicity dependence).
The resulting observed-frame colors and estimates of equivalent width are how surveys actually select LBGs/LAEs, and would allow for more realistic predictions for the clustering properties of these galaxies.

Our work also indicates that the high-redshift galaxies may feature HODs quite different from the ones used in analyses of LRG/ELG and quasar samples from DESI, see Figs.~\ref{fig:mtng_hod},~\ref{fig:astrid_hod}. 
This motivates further studies of phenomenological parameterizations of the high-redshift galaxy distribution in the HOD framework and beyond it, e.g. with other galaxy formation models.  
Such studies will be important for an efficient production of  simulation-based priors that enhance constraints from EFT-based full-shape analyses~\cite{Ivanov:2024hgq,Ivanov:2024xgb,Akitsu:2024lyt,Zhang:2024thl,Shiferaw:2024_flbias_hydro}. 
Our paper is the first step towards systematic studies of simulation-based priors for high-redshift galaxies.  
We leave the research directions discussed above to  future exploration. 

\section*{Acknowledgments}
We thank Martin White for comments on an early version of this work, useful discussions, and for making his LBG/LAE mock sample code public.
This work is supported by the National Science Foundation under Cooperative Agreement PHY-2019786 (The NSF AI Institute for Artificial Intelligence and Fundamental Interactions, \url{http://iaifi.org/}). 
This material is based upon work supported by the U.S. Department of Energy, Office of Science, Office of High Energy Physics of U.S. Department of Energy under grant Contract Number DE-SC0012567. 
The computations in this paper were run on the FASRC Cannon cluster supported by the FAS Division of Science Research Computing Group at Harvard University.
JMS acknowledges that support for this work was provided by The Brinson Foundation through a Brinson Prize Fellowship grant. 
SB is supported by the UK Research and Innovation (UKRI) Future Leaders Fellowship (grant number MR/V023381/1).
CH-A acknowledges support from the Excellence Cluster ORIGINS which is funded by the Deutsche Forschungsgemeinschaft (DFG, German Research Foundation) under Germany's Excellence Strategy -- EXC-2094 -- 390783311.
RK acknowledges support of the Natural Sciences and Engineering Research Council of Canada (NSERC) through a Discovery Grant and a Discovery Launch Supplement (funding reference numbers RGPIN-2024-06222 and DGECR-2024-00144) and York University's Global Research Excellence Initiative.
LH acknowledges support from the Simons Foundation collaboration Learning the Universe.

\appendix

\section{MTNG stellar mass resolution}
\label{app:mtng_res}

The Astrid simulation has a higher resolution than the MTNG simulations, and is run in a smaller simulation box. 
The two simulations therefore have different stellar particle resolution in each galaxy.
We assess the impact of a minimum stellar particle number cut on our MTNG ODIN LAE results in this Appendix.

\begin{figure}
\centering
\includegraphics[width=0.45\textwidth]{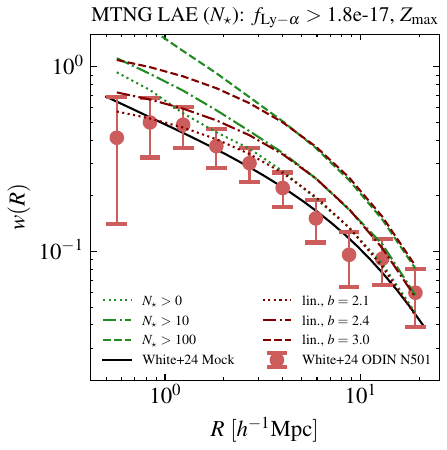}
   \caption{Projected clustering $w(R)$ of simulated LAEs using the ODIN sample selection in MTNG (green curves) for no minimum stellar particle cut (dotted curve, fiducial $\frac{Z}{Z_\odot}<0.04$) a minimum stellar particle cut of 10 star particles (dash-dotted curve, $\frac{Z}{Z_\odot}<0.14$) and for 100 star particles (dashed curve, $\frac{Z}{Z_\odot}<0.69$).
   The ODIN projected clustering data of Ref.~\cite{White:2024_odin} is also plotted (red points) as is the prediction from linear theory for the quoted values of linear bias $b_1$ (dark red curves).} \label{fig:app_mtng_res_wtheta}
\end{figure}

For stellar particle number minima of 10 (100) particles in dash-dotted (dashed) curves in Fig.~\ref{app:mtng_res}, we find LAE samples with linear biases that are consistent with $b_{1} \approx 2.4~(3.0)$, as crudely determined by comparing angular clustering via $w(R)$ (Figure~\ref{fig:app_mtng_res_wtheta}).
These bias values are significantly higher than the linear bias obtained from the fiducial ODIN LAE sample (Table~\ref{tab:samples_mtng}, Table~\ref{tab:parameters}), especially for the 100 star particle cut.
To keep a similar number density as the minimum number of stellar particles increases, we loosen the maximum metallicity cuts used to define the sample to the values described in Fig.~\ref{fig:app_mtng_res_wtheta}.
For the ODIN sample, the application of a stellar particle cut also significantly reduces the sample satellite fraction (by up to 20\%).
This is to be compared with our fiducial sample (with no stellar particle minimum, dotted curves), which is roughly consistent with a linear model using $b_1=2.1$, and passes through the ODIN clustering data.

We thus caution that our results for MTNG ODIN sample are highly sensitive to the minimum stellar particle number cut used for the LAE sample, as the LAE sample produced by our selection largely consists of low-mass galaxies, many of which are satellites.
However, we find the qualitative agreement of our results between MTNG and the higher-resolution Astrid to be reassuring.
Therefore, in the main text, we do not cut by any stellar particle minimum.

\section{Downsampled LAE cuts \label{app:downsampling}}

In this Appendix we briefly discuss the impact of allowing more permissive cuts than those we allowed in the main text under the assumption that only a fixed fraction of physical LAEs actually enter the observed sample.
This procedure can be interpreted as varying the mean LAE effective escape fraction while keeping the number density fixed.
Though, as discussed in the main text, the effective escape fraction is also naturally implemented through any of the cuts applied without random downsampling (as the mean occupation of those samples is much less than 1, see Figs.~\ref{fig:mtng_hod},\ref{fig:astrid_hod}), here we explore the more straightforward procedure of constructing samples by randomly downsampling all galaxies satisfying the flux cut.

Table~\ref{tab:samples_mtng_lae_downsample} shows the halo summary properties for these samples.
As the downsampling fraction increases (the cut becomes more permissive) the approximate clustering bias of the resulting $w(\theta)$ grows, getting further from the data-preferred value of $2.0\pm0.2$ of Ref.~\cite{White:2024_odin}.
Except when compared to the fiducial sample, the satellite fraction does not change appreciably, while the mean halo masses and 1D velocity dispersions get slightly larger.
This suggests that the clustering bias of Ref.~\cite{White:2024_odin} prefers the low bias samples selected by our most aggressive cut in the main text fiducial samples.
As discussed in the main text, a lower bias sample could also be achieved (at the same number density) by a higher value of the flux conversion $f_{\rm{Ly}-\alpha}(Z)$, which would allow more low mass (and low linear bias) galaxies (at fixed metallicity) to exceed the flux threshold.

\begin{table}[] 
\begin{tabular}{|cccccc|} 
\hline 
 $f_{\mathrm{DS}}$ & $\frac{Z_{\mathrm{max}}}{Z_\odot}$ & $ \log_{10} \left( \frac{\langle M_{h}\rangle}{h^{-1}M_{\odot}}\right) $ & $b_{1}^{w(\theta)}$ & $f_{\mathrm{sat}}$ & $\langle \frac{\sigma}{\mathrm{km}/\mathrm{s}} \rangle$\\
\hline
  $1$    & 0.04  &   11.40     &  2.1  &    $30\%$  &   60\\
  $0.5$    & 0.09  &   11.36   &  2.2  &    $23\%$  &   65\\
  $0.25$   & 0.17  & 11.37     &  2.2  &  $21\%$    &  71\\
  $0.1$    & 0.29  & 11.67     &  2.2  &  $21\%$    &  74\\
  $0.05$   & 0.55  & 11.67     &  2.4  &  $21\%$    & 90\\
\hline
\end{tabular}
\caption{Several alternate ODIN LAE samples that match the fiducial number density (top row) but with different maximum metallicity cuts $Z_{\rm{max}}$ and concomitant downsampling factors $f_{\rm{DS}}$. 
Columns are similar to Table~\ref{tab:samples_mtng}. \label{tab:samples_mtng_lae_downsample}}
\end{table}

\bibliography{short.bib}

\begin{thebibliography}{112}%
\makeatletter
\providecommand \@ifxundefined [1]{%
 \@ifx{#1\undefined}
}%
\providecommand \@ifnum [1]{%
 \ifnum #1\expandafter \@firstoftwo
 \else \expandafter \@secondoftwo
 \fi
}%
\providecommand \@ifx [1]{%
 \ifx #1\expandafter \@firstoftwo
 \else \expandafter \@secondoftwo
 \fi
}%
\providecommand \natexlab [1]{#1}%
\providecommand \enquote  [1]{``#1''}%
\providecommand \bibnamefont  [1]{#1}%
\providecommand \bibfnamefont [1]{#1}%
\providecommand \citenamefont [1]{#1}%
\providecommand \href@noop [0]{\@secondoftwo}%
\providecommand \href [0]{\begingroup \@sanitize@url \@href}%
\providecommand \@href[1]{\@@startlink{#1}\@@href}%
\providecommand \@@href[1]{\endgroup#1\@@endlink}%
\providecommand \@sanitize@url [0]{\catcode `\\12\catcode `\$12\catcode `\&12\catcode `\#12\catcode `\^12\catcode `\_12\catcode `\%12\relax}%
\providecommand \@@startlink[1]{}%
\providecommand \@@endlink[0]{}%
\providecommand \url  [0]{\begingroup\@sanitize@url \@url }%
\providecommand \@url [1]{\endgroup\@href {#1}{\urlprefix }}%
\providecommand \urlprefix  [0]{URL }%
\providecommand \Eprint [0]{\href }%
\providecommand \doibase [0]{http://dx.doi.org/}%
\providecommand \selectlanguage [0]{\@gobble}%
\providecommand \bibinfo  [0]{\@secondoftwo}%
\providecommand \bibfield  [0]{\@secondoftwo}%
\providecommand \translation [1]{[#1]}%
\providecommand \BibitemOpen [0]{}%
\providecommand \bibitemStop [0]{}%
\providecommand \bibitemNoStop [0]{.\EOS\space}%
\providecommand \EOS [0]{\spacefactor3000\relax}%
\providecommand \BibitemShut  [1]{\csname bibitem#1\endcsname}%
\let\auto@bib@innerbib\@empty
\bibitem [{\citenamefont {{Schlegel}}\ \emph {et~al.}(2022)\citenamefont {{Schlegel}}, \citenamefont {{Ferraro}}, \citenamefont {{Aldering}}, \citenamefont {{Baltay}}, \citenamefont {{BenZvi}}, \citenamefont {{Besuner}}, \citenamefont {{Blanc}}, \citenamefont {{Bolton}}, \citenamefont {{Bonaca}}, \citenamefont {{Brooks}}, \citenamefont {{Buckley-Geer}}, \citenamefont {{Cai}}, \citenamefont {{DeRose}}, \citenamefont {{Dey}}, \citenamefont {{Doel}}, \citenamefont {{Drlica-Wagner}}, \citenamefont {{Fan}}, \citenamefont {{Gutierrez}}, \citenamefont {{Green}}, \citenamefont {{Guy}}, \citenamefont {{Huterer}}, \citenamefont {{Infante}}, \citenamefont {{Jelinsky}}, \citenamefont {{Karagiannis}}, \citenamefont {{Kent}}, \citenamefont {{Kim}}, \citenamefont {{Kneib}}, \citenamefont {{Kollmeier}}, \citenamefont {{Kremin}}, \citenamefont {{Lahav}}, \citenamefont {{Landriau}}, \citenamefont {{Lang}}, \citenamefont {{Leauthaud}}, \citenamefont {{Levi}}, \citenamefont {{Linder}}, \citenamefont {{Magneville}}, \citenamefont
  {{Martini}}, \citenamefont {{McDonald}}, \citenamefont {{Miller}}, \citenamefont {{Myers}}, \citenamefont {{Newman}}, \citenamefont {{Nugent}}, \citenamefont {{Palanque-Delabrouille}}, \citenamefont {{Padmanabhan}}, \citenamefont {{Palmese}}, \citenamefont {{Poppett}}, \citenamefont {{Prochaska}}, \citenamefont {{Raichoor}}, \citenamefont {{Ramirez}}, \citenamefont {{Sailer}}, \citenamefont {{Schaan}}, \citenamefont {{Schubnell}}, \citenamefont {{Seljak}}, \citenamefont {{Seo}}, \citenamefont {{Silber}}, \citenamefont {{Simon}}, \citenamefont {{Slepian}}, \citenamefont {{Soares-Santos}}, \citenamefont {{Tarle}}, \citenamefont {{Valluri}}, \citenamefont {{Weaverdyck}}, \citenamefont {{Wechsler}}, \citenamefont {{White}}, \citenamefont {{Yeche}},\ and\ \citenamefont {{Zhou}}}]{Schlegel:2022_spec_roadmap}%
  \BibitemOpen
  \bibfield  {author} {\bibinfo {author} {\bibfnamefont {D.~J.}\ \bibnamefont {{Schlegel}}}, \bibinfo {author} {\bibfnamefont {S.}~\bibnamefont {{Ferraro}}}, \bibinfo {author} {\bibfnamefont {G.}~\bibnamefont {{Aldering}}}, \bibinfo {author} {\bibfnamefont {C.}~\bibnamefont {{Baltay}}}, \bibinfo {author} {\bibfnamefont {S.}~\bibnamefont {{BenZvi}}}, \bibinfo {author} {\bibfnamefont {R.}~\bibnamefont {{Besuner}}}, \bibinfo {author} {\bibfnamefont {G.~A.}\ \bibnamefont {{Blanc}}}, \bibinfo {author} {\bibfnamefont {A.~S.}\ \bibnamefont {{Bolton}}}, \bibinfo {author} {\bibfnamefont {A.}~\bibnamefont {{Bonaca}}}, \bibinfo {author} {\bibfnamefont {D.}~\bibnamefont {{Brooks}}}, \bibinfo {author} {\bibfnamefont {E.}~\bibnamefont {{Buckley-Geer}}}, \bibinfo {author} {\bibfnamefont {Z.}~\bibnamefont {{Cai}}}, \bibinfo {author} {\bibfnamefont {J.}~\bibnamefont {{DeRose}}}, \bibinfo {author} {\bibfnamefont {A.}~\bibnamefont {{Dey}}}, \bibinfo {author} {\bibfnamefont {P.}~\bibnamefont {{Doel}}}, \bibinfo {author}
  {\bibfnamefont {A.}~\bibnamefont {{Drlica-Wagner}}}, \bibinfo {author} {\bibfnamefont {X.}~\bibnamefont {{Fan}}}, \bibinfo {author} {\bibfnamefont {G.}~\bibnamefont {{Gutierrez}}}, \bibinfo {author} {\bibfnamefont {D.}~\bibnamefont {{Green}}}, \bibinfo {author} {\bibfnamefont {J.}~\bibnamefont {{Guy}}}, \bibinfo {author} {\bibfnamefont {D.}~\bibnamefont {{Huterer}}}, \bibinfo {author} {\bibfnamefont {L.}~\bibnamefont {{Infante}}}, \bibinfo {author} {\bibfnamefont {P.}~\bibnamefont {{Jelinsky}}}, \bibinfo {author} {\bibfnamefont {D.}~\bibnamefont {{Karagiannis}}}, \bibinfo {author} {\bibfnamefont {S.~M.}\ \bibnamefont {{Kent}}}, \bibinfo {author} {\bibfnamefont {A.~G.}\ \bibnamefont {{Kim}}}, \bibinfo {author} {\bibfnamefont {J.-P.}\ \bibnamefont {{Kneib}}}, \bibinfo {author} {\bibfnamefont {J.~A.}\ \bibnamefont {{Kollmeier}}}, \bibinfo {author} {\bibfnamefont {A.}~\bibnamefont {{Kremin}}}, \bibinfo {author} {\bibfnamefont {O.}~\bibnamefont {{Lahav}}}, \bibinfo {author} {\bibfnamefont {M.}~\bibnamefont
  {{Landriau}}}, \bibinfo {author} {\bibfnamefont {D.}~\bibnamefont {{Lang}}}, \bibinfo {author} {\bibfnamefont {A.}~\bibnamefont {{Leauthaud}}}, \bibinfo {author} {\bibfnamefont {M.~E.}\ \bibnamefont {{Levi}}}, \bibinfo {author} {\bibfnamefont {E.~V.}\ \bibnamefont {{Linder}}}, \bibinfo {author} {\bibfnamefont {C.}~\bibnamefont {{Magneville}}}, \bibinfo {author} {\bibfnamefont {P.}~\bibnamefont {{Martini}}}, \bibinfo {author} {\bibfnamefont {P.}~\bibnamefont {{McDonald}}}, \bibinfo {author} {\bibfnamefont {C.~J.}\ \bibnamefont {{Miller}}}, \bibinfo {author} {\bibfnamefont {A.~D.}\ \bibnamefont {{Myers}}}, \bibinfo {author} {\bibfnamefont {J.~A.}\ \bibnamefont {{Newman}}}, \bibinfo {author} {\bibfnamefont {P.~E.}\ \bibnamefont {{Nugent}}}, \bibinfo {author} {\bibfnamefont {N.}~\bibnamefont {{Palanque-Delabrouille}}}, \bibinfo {author} {\bibfnamefont {N.}~\bibnamefont {{Padmanabhan}}}, \bibinfo {author} {\bibfnamefont {A.}~\bibnamefont {{Palmese}}}, \bibinfo {author} {\bibfnamefont {C.}~\bibnamefont
  {{Poppett}}}, \bibinfo {author} {\bibfnamefont {J.~X.}\ \bibnamefont {{Prochaska}}}, \bibinfo {author} {\bibfnamefont {A.}~\bibnamefont {{Raichoor}}}, \bibinfo {author} {\bibfnamefont {S.}~\bibnamefont {{Ramirez}}}, \bibinfo {author} {\bibfnamefont {N.}~\bibnamefont {{Sailer}}}, \bibinfo {author} {\bibfnamefont {E.}~\bibnamefont {{Schaan}}}, \bibinfo {author} {\bibfnamefont {M.}~\bibnamefont {{Schubnell}}}, \bibinfo {author} {\bibfnamefont {U.}~\bibnamefont {{Seljak}}}, \bibinfo {author} {\bibfnamefont {H.-J.}\ \bibnamefont {{Seo}}}, \bibinfo {author} {\bibfnamefont {J.}~\bibnamefont {{Silber}}}, \bibinfo {author} {\bibfnamefont {J.~D.}\ \bibnamefont {{Simon}}}, \bibinfo {author} {\bibfnamefont {Z.}~\bibnamefont {{Slepian}}}, \bibinfo {author} {\bibfnamefont {M.}~\bibnamefont {{Soares-Santos}}}, \bibinfo {author} {\bibfnamefont {G.}~\bibnamefont {{Tarle}}}, \bibinfo {author} {\bibfnamefont {M.}~\bibnamefont {{Valluri}}}, \bibinfo {author} {\bibfnamefont {N.~J.}\ \bibnamefont {{Weaverdyck}}}, \bibinfo
  {author} {\bibfnamefont {R.~H.}\ \bibnamefont {{Wechsler}}}, \bibinfo {author} {\bibfnamefont {M.}~\bibnamefont {{White}}}, \bibinfo {author} {\bibfnamefont {C.}~\bibnamefont {{Yeche}}}, \ and\ \bibinfo {author} {\bibfnamefont {R.}~\bibnamefont {{Zhou}}},\ }\href {\doibase 10.48550/arXiv.2209.03585} {\bibfield  {journal} {\bibinfo  {journal} {arXiv e-prints}\ ,\ \bibinfo {eid} {arXiv:2209.03585}} (\bibinfo {year} {2022})},\ \Eprint {http://arxiv.org/abs/2209.03585} {arXiv:2209.03585 [astro-ph.CO]} \BibitemShut {NoStop}%
\bibitem [{\citenamefont {{Besuner}}\ \emph {et~al.}(2025)\citenamefont {{Besuner}}, \citenamefont {{Dey}}, \citenamefont {{Drlica-Wagner}}, \citenamefont {{Ebina}}, \citenamefont {{Fernandez Moroni}}, \citenamefont {{Ferraro}}, \citenamefont {{Forero-Romero}}, \citenamefont {{Honscheid}}, \citenamefont {{Jelinsky}}, \citenamefont {{Lang}}, \citenamefont {{Levi}}, \citenamefont {{Martini}}, \citenamefont {{Myers}}, \citenamefont {{Palanque-Delabrouille}}, \citenamefont {{Panda}}, \citenamefont {{Poppett}}, \citenamefont {{Sailer}}, \citenamefont {{Schlegel}}, \citenamefont {{Shafieloo}}, \citenamefont {{Silber}}, \citenamefont {{White}}, \citenamefont {{Abbott}}, \citenamefont {{Allen}}, \citenamefont {{Avila}}, \citenamefont {{Bailey}}, \citenamefont {{Bault}}, \citenamefont {{Bouri}}, \citenamefont {{Boutsia}}, \citenamefont {{Burtin}}, \citenamefont {{Chierchie}}, \citenamefont {{Coulton}}, \citenamefont {{Dawson}}, \citenamefont {{Dey}}, \citenamefont {{Dunlop}}, \citenamefont {{Eisenstein}},
  \citenamefont {{Emanuele}}, \citenamefont {{Escoffier}}, \citenamefont {{Estrada}}, \citenamefont {{Fagrelius}}, \citenamefont {{Fanning}}, \citenamefont {{Fanning}}, \citenamefont {{Font-Ribera}}, \citenamefont {{Frieman}}, \citenamefont {{Galal}}, \citenamefont {{Gluscevic}}, \citenamefont {{Gontcho}}, \citenamefont {{Green}}, \citenamefont {{Gutierrez}}, \citenamefont {{Guy}}, \citenamefont {{Hashemi}}, \citenamefont {{Heathcote}}, \citenamefont {{Holland}}, \citenamefont {{Hou}}, \citenamefont {{Huterer}}, \citenamefont {{Irigoyen Gimenez}}, \citenamefont {{Ivanov}}, \citenamefont {{Joyce}}, \citenamefont {{Jullo}}, \citenamefont {{Juneau}}, \citenamefont {{Juramy}}, \citenamefont {{Karcher}}, \citenamefont {{Kent}}, \citenamefont {{Kirkby}}, \citenamefont {{Kneib}}, \citenamefont {{Krause}}, \citenamefont {{Krolewski}}, \citenamefont {{Lahav}}, \citenamefont {{Lapi}}, \citenamefont {{Leauthaud}}, \citenamefont {{Lewandowski}}, \citenamefont {{Li}}, \citenamefont {{Lin}}, \citenamefont {{Loverde}},
  \citenamefont {{MacBride}}, \citenamefont {{Magneville}}, \citenamefont {{Marshall}}, \citenamefont {{McDonald}}, \citenamefont {{Miller}}, \citenamefont {{Moustakas}}, \citenamefont {{M{\"u}nchmeyer}}, \citenamefont {{Najita}}, \citenamefont {{Newman}}, \citenamefont {{Percival}}, \citenamefont {{Philcox}}, \citenamefont {{Pires}}, \citenamefont {{Raichoor}}, \citenamefont {{Roach}}, \citenamefont {{Rockosi}}, \citenamefont {{Rombach}}, \citenamefont {{Ross}}, \citenamefont {{Sanchez}}, \citenamefont {{Schmidt}}, \citenamefont {{Schubnell}}, \citenamefont {{Sebok}}, \citenamefont {{Seljak}}, \citenamefont {{Silverstein}}, \citenamefont {{Slepian}}, \citenamefont {{Stupak}}, \citenamefont {{Tarl{\'e}}}, \citenamefont {{Tyas}}, \citenamefont {{Vargas-Maga{\~n}a}}, \citenamefont {{Walker}}, \citenamefont {{Wenner}}, \citenamefont {{Y{\`e}che}}, \citenamefont {{Zhang}},\ and\ \citenamefont {{Zhou}}}]{2025:spec_s5}%
  \BibitemOpen
  \bibfield  {author} {\bibinfo {author} {\bibfnamefont {R.}~\bibnamefont {{Besuner}}}, \bibinfo {author} {\bibfnamefont {A.}~\bibnamefont {{Dey}}}, \bibinfo {author} {\bibfnamefont {A.}~\bibnamefont {{Drlica-Wagner}}}, \bibinfo {author} {\bibfnamefont {H.}~\bibnamefont {{Ebina}}}, \bibinfo {author} {\bibfnamefont {G.}~\bibnamefont {{Fernandez Moroni}}}, \bibinfo {author} {\bibfnamefont {S.}~\bibnamefont {{Ferraro}}}, \bibinfo {author} {\bibfnamefont {J.}~\bibnamefont {{Forero-Romero}}}, \bibinfo {author} {\bibfnamefont {K.}~\bibnamefont {{Honscheid}}}, \bibinfo {author} {\bibfnamefont {P.}~\bibnamefont {{Jelinsky}}}, \bibinfo {author} {\bibfnamefont {D.}~\bibnamefont {{Lang}}}, \bibinfo {author} {\bibfnamefont {M.}~\bibnamefont {{Levi}}}, \bibinfo {author} {\bibfnamefont {P.}~\bibnamefont {{Martini}}}, \bibinfo {author} {\bibfnamefont {A.}~\bibnamefont {{Myers}}}, \bibinfo {author} {\bibfnamefont {N.}~\bibnamefont {{Palanque-Delabrouille}}}, \bibinfo {author} {\bibfnamefont {S.}~\bibnamefont {{Panda}}},
  \bibinfo {author} {\bibfnamefont {C.}~\bibnamefont {{Poppett}}}, \bibinfo {author} {\bibfnamefont {N.}~\bibnamefont {{Sailer}}}, \bibinfo {author} {\bibfnamefont {D.}~\bibnamefont {{Schlegel}}}, \bibinfo {author} {\bibfnamefont {A.}~\bibnamefont {{Shafieloo}}}, \bibinfo {author} {\bibfnamefont {J.}~\bibnamefont {{Silber}}}, \bibinfo {author} {\bibfnamefont {M.}~\bibnamefont {{White}}}, \bibinfo {author} {\bibfnamefont {T.}~\bibnamefont {{Abbott}}}, \bibinfo {author} {\bibfnamefont {L.}~\bibnamefont {{Allen}}}, \bibinfo {author} {\bibfnamefont {S.}~\bibnamefont {{Avila}}}, \bibinfo {author} {\bibfnamefont {S.}~\bibnamefont {{Bailey}}}, \bibinfo {author} {\bibfnamefont {A.}~\bibnamefont {{Bault}}}, \bibinfo {author} {\bibfnamefont {M.}~\bibnamefont {{Bouri}}}, \bibinfo {author} {\bibfnamefont {K.}~\bibnamefont {{Boutsia}}}, \bibinfo {author} {\bibfnamefont {E.}~\bibnamefont {{Burtin}}}, \bibinfo {author} {\bibfnamefont {F.}~\bibnamefont {{Chierchie}}}, \bibinfo {author} {\bibfnamefont {W.}~\bibnamefont
  {{Coulton}}}, \bibinfo {author} {\bibfnamefont {K.}~\bibnamefont {{Dawson}}}, \bibinfo {author} {\bibfnamefont {B.}~\bibnamefont {{Dey}}}, \bibinfo {author} {\bibfnamefont {P.}~\bibnamefont {{Dunlop}}}, \bibinfo {author} {\bibfnamefont {D.}~\bibnamefont {{Eisenstein}}}, \bibinfo {author} {\bibfnamefont {C.}~\bibnamefont {{Emanuele}}}, \bibinfo {author} {\bibfnamefont {S.}~\bibnamefont {{Escoffier}}}, \bibinfo {author} {\bibfnamefont {J.}~\bibnamefont {{Estrada}}}, \bibinfo {author} {\bibfnamefont {P.}~\bibnamefont {{Fagrelius}}}, \bibinfo {author} {\bibfnamefont {K.}~\bibnamefont {{Fanning}}}, \bibinfo {author} {\bibfnamefont {T.}~\bibnamefont {{Fanning}}}, \bibinfo {author} {\bibfnamefont {A.}~\bibnamefont {{Font-Ribera}}}, \bibinfo {author} {\bibfnamefont {J.}~\bibnamefont {{Frieman}}}, \bibinfo {author} {\bibfnamefont {M.}~\bibnamefont {{Galal}}}, \bibinfo {author} {\bibfnamefont {V.}~\bibnamefont {{Gluscevic}}}, \bibinfo {author} {\bibfnamefont {S.~G.~A.}\ \bibnamefont {{Gontcho}}}, \bibinfo {author}
  {\bibfnamefont {D.}~\bibnamefont {{Green}}}, \bibinfo {author} {\bibfnamefont {G.}~\bibnamefont {{Gutierrez}}}, \bibinfo {author} {\bibfnamefont {J.}~\bibnamefont {{Guy}}}, \bibinfo {author} {\bibfnamefont {K.}~\bibnamefont {{Hashemi}}}, \bibinfo {author} {\bibfnamefont {S.}~\bibnamefont {{Heathcote}}}, \bibinfo {author} {\bibfnamefont {S.}~\bibnamefont {{Holland}}}, \bibinfo {author} {\bibfnamefont {J.}~\bibnamefont {{Hou}}}, \bibinfo {author} {\bibfnamefont {D.}~\bibnamefont {{Huterer}}}, \bibinfo {author} {\bibfnamefont {B.}~\bibnamefont {{Irigoyen Gimenez}}}, \bibinfo {author} {\bibfnamefont {M.}~\bibnamefont {{Ivanov}}}, \bibinfo {author} {\bibfnamefont {R.}~\bibnamefont {{Joyce}}}, \bibinfo {author} {\bibfnamefont {E.}~\bibnamefont {{Jullo}}}, \bibinfo {author} {\bibfnamefont {S.}~\bibnamefont {{Juneau}}}, \bibinfo {author} {\bibfnamefont {C.}~\bibnamefont {{Juramy}}}, \bibinfo {author} {\bibfnamefont {A.}~\bibnamefont {{Karcher}}}, \bibinfo {author} {\bibfnamefont {S.}~\bibnamefont {{Kent}}},
  \bibinfo {author} {\bibfnamefont {D.}~\bibnamefont {{Kirkby}}}, \bibinfo {author} {\bibfnamefont {J.-P.}\ \bibnamefont {{Kneib}}}, \bibinfo {author} {\bibfnamefont {E.}~\bibnamefont {{Krause}}}, \bibinfo {author} {\bibfnamefont {A.}~\bibnamefont {{Krolewski}}}, \bibinfo {author} {\bibfnamefont {O.}~\bibnamefont {{Lahav}}}, \bibinfo {author} {\bibfnamefont {A.}~\bibnamefont {{Lapi}}}, \bibinfo {author} {\bibfnamefont {A.}~\bibnamefont {{Leauthaud}}}, \bibinfo {author} {\bibfnamefont {M.}~\bibnamefont {{Lewandowski}}}, \bibinfo {author} {\bibfnamefont {T.}~\bibnamefont {{Li}}}, \bibinfo {author} {\bibfnamefont {K.}~\bibnamefont {{Lin}}}, \bibinfo {author} {\bibfnamefont {M.}~\bibnamefont {{Loverde}}}, \bibinfo {author} {\bibfnamefont {S.}~\bibnamefont {{MacBride}}}, \bibinfo {author} {\bibfnamefont {C.}~\bibnamefont {{Magneville}}}, \bibinfo {author} {\bibfnamefont {J.}~\bibnamefont {{Marshall}}}, \bibinfo {author} {\bibfnamefont {P.}~\bibnamefont {{McDonald}}}, \bibinfo {author} {\bibfnamefont
  {T.}~\bibnamefont {{Miller}}}, \bibinfo {author} {\bibfnamefont {J.}~\bibnamefont {{Moustakas}}}, \bibinfo {author} {\bibfnamefont {M.}~\bibnamefont {{M{\"u}nchmeyer}}}, \bibinfo {author} {\bibfnamefont {J.}~\bibnamefont {{Najita}}}, \bibinfo {author} {\bibfnamefont {J.}~\bibnamefont {{Newman}}}, \bibinfo {author} {\bibfnamefont {W.}~\bibnamefont {{Percival}}}, \bibinfo {author} {\bibfnamefont {O.}~\bibnamefont {{Philcox}}}, \bibinfo {author} {\bibfnamefont {P.}~\bibnamefont {{Pires}}}, \bibinfo {author} {\bibfnamefont {A.}~\bibnamefont {{Raichoor}}}, \bibinfo {author} {\bibfnamefont {B.}~\bibnamefont {{Roach}}}, \bibinfo {author} {\bibfnamefont {C.}~\bibnamefont {{Rockosi}}}, \bibinfo {author} {\bibfnamefont {M.}~\bibnamefont {{Rombach}}}, \bibinfo {author} {\bibfnamefont {A.}~\bibnamefont {{Ross}}}, \bibinfo {author} {\bibfnamefont {E.}~\bibnamefont {{Sanchez}}}, \bibinfo {author} {\bibfnamefont {L.}~\bibnamefont {{Schmidt}}}, \bibinfo {author} {\bibfnamefont {M.}~\bibnamefont {{Schubnell}}}, \bibinfo
  {author} {\bibfnamefont {R.}~\bibnamefont {{Sebok}}}, \bibinfo {author} {\bibfnamefont {U.}~\bibnamefont {{Seljak}}}, \bibinfo {author} {\bibfnamefont {E.}~\bibnamefont {{Silverstein}}}, \bibinfo {author} {\bibfnamefont {Z.}~\bibnamefont {{Slepian}}}, \bibinfo {author} {\bibfnamefont {R.}~\bibnamefont {{Stupak}}}, \bibinfo {author} {\bibfnamefont {G.}~\bibnamefont {{Tarl{\'e}}}}, \bibinfo {author} {\bibfnamefont {L.}~\bibnamefont {{Tyas}}}, \bibinfo {author} {\bibfnamefont {M.}~\bibnamefont {{Vargas-Maga{\~n}a}}}, \bibinfo {author} {\bibfnamefont {A.}~\bibnamefont {{Walker}}}, \bibinfo {author} {\bibfnamefont {N.}~\bibnamefont {{Wenner}}}, \bibinfo {author} {\bibfnamefont {C.}~\bibnamefont {{Y{\`e}che}}}, \bibinfo {author} {\bibfnamefont {Y.}~\bibnamefont {{Zhang}}}, \ and\ \bibinfo {author} {\bibfnamefont {R.}~\bibnamefont {{Zhou}}},\ }\href {\doibase 10.48550/arXiv.2503.07923} {\bibfield  {journal} {\bibinfo  {journal} {arXiv e-prints}\ ,\ \bibinfo {eid} {arXiv:2503.07923}} (\bibinfo {year} {2025})},\
  \Eprint {http://arxiv.org/abs/2503.07923} {arXiv:2503.07923 [astro-ph.CO]} \BibitemShut {NoStop}%
\bibitem [{\citenamefont {Baumann}\ \emph {et~al.}(2012)\citenamefont {Baumann}, \citenamefont {Nicolis}, \citenamefont {Senatore},\ and\ \citenamefont {Zaldarriaga}}]{Baumann:2010tm}%
  \BibitemOpen
  \bibfield  {author} {\bibinfo {author} {\bibfnamefont {D.}~\bibnamefont {Baumann}}, \bibinfo {author} {\bibfnamefont {A.}~\bibnamefont {Nicolis}}, \bibinfo {author} {\bibfnamefont {L.}~\bibnamefont {Senatore}}, \ and\ \bibinfo {author} {\bibfnamefont {M.}~\bibnamefont {Zaldarriaga}},\ }\href {\doibase 10.1088/1475-7516/2012/07/051} {\bibfield  {journal} {\bibinfo  {journal} {JCAP}\ }\textbf {\bibinfo {volume} {1207}},\ \bibinfo {pages} {051} (\bibinfo {year} {2012})},\ \Eprint {http://arxiv.org/abs/1004.2488} {arXiv:1004.2488 [astro-ph.CO]} \BibitemShut {NoStop}%
\bibitem [{\citenamefont {Carrasco}\ \emph {et~al.}(2012)\citenamefont {Carrasco}, \citenamefont {Hertzberg},\ and\ \citenamefont {Senatore}}]{Carrasco:2012cv}%
  \BibitemOpen
  \bibfield  {author} {\bibinfo {author} {\bibfnamefont {J.~J.~M.}\ \bibnamefont {Carrasco}}, \bibinfo {author} {\bibfnamefont {M.~P.}\ \bibnamefont {Hertzberg}}, \ and\ \bibinfo {author} {\bibfnamefont {L.}~\bibnamefont {Senatore}},\ }\href {\doibase 10.1007/JHEP09(2012)082} {\bibfield  {journal} {\bibinfo  {journal} {JHEP}\ }\textbf {\bibinfo {volume} {09}},\ \bibinfo {pages} {082} (\bibinfo {year} {2012})},\ \Eprint {http://arxiv.org/abs/1206.2926} {arXiv:1206.2926 [astro-ph.CO]} \BibitemShut {NoStop}%
\bibitem [{\citenamefont {Ivanov}(2022)}]{Ivanov:2022mrd}%
  \BibitemOpen
  \bibfield  {author} {\bibinfo {author} {\bibfnamefont {M.~M.}\ \bibnamefont {Ivanov}},\ }\href@noop {} {\  (\bibinfo {year} {2022})},\ \Eprint {http://arxiv.org/abs/2212.08488} {arXiv:2212.08488 [astro-ph.CO]} \BibitemShut {NoStop}%
\bibitem [{\citenamefont {Desjacques}\ \emph {et~al.}(2018)\citenamefont {Desjacques}, \citenamefont {Jeong},\ and\ \citenamefont {Schmidt}}]{Desjacques:2016bnm}%
  \BibitemOpen
  \bibfield  {author} {\bibinfo {author} {\bibfnamefont {V.}~\bibnamefont {Desjacques}}, \bibinfo {author} {\bibfnamefont {D.}~\bibnamefont {Jeong}}, \ and\ \bibinfo {author} {\bibfnamefont {F.}~\bibnamefont {Schmidt}},\ }\href {\doibase 10.1016/j.physrep.2017.12.002} {\bibfield  {journal} {\bibinfo  {journal} {Phys. Rept.}\ }\textbf {\bibinfo {volume} {733}},\ \bibinfo {pages} {1} (\bibinfo {year} {2018})},\ \Eprint {http://arxiv.org/abs/1611.09787} {arXiv:1611.09787 [astro-ph.CO]} \BibitemShut {NoStop}%
\bibitem [{\citenamefont {Bardeen}\ \emph {et~al.}(1986)\citenamefont {Bardeen}, \citenamefont {Bond}, \citenamefont {Kaiser},\ and\ \citenamefont {Szalay}}]{Bardeen:1985tr}%
  \BibitemOpen
  \bibfield  {author} {\bibinfo {author} {\bibfnamefont {J.~M.}\ \bibnamefont {Bardeen}}, \bibinfo {author} {\bibfnamefont {J.~R.}\ \bibnamefont {Bond}}, \bibinfo {author} {\bibfnamefont {N.}~\bibnamefont {Kaiser}}, \ and\ \bibinfo {author} {\bibfnamefont {A.~S.}\ \bibnamefont {Szalay}},\ }\href {\doibase 10.1086/164143} {\bibfield  {journal} {\bibinfo  {journal} {Astrophys. J.}\ }\textbf {\bibinfo {volume} {304}},\ \bibinfo {pages} {15} (\bibinfo {year} {1986})}\BibitemShut {NoStop}%
\bibitem [{\citenamefont {{Kaiser}}(1984)}]{Kaiser:1984}%
  \BibitemOpen
  \bibfield  {author} {\bibinfo {author} {\bibfnamefont {N.}~\bibnamefont {{Kaiser}}},\ }\href {\doibase 10.1086/184341} {\bibfield  {journal} {\bibinfo  {journal} {\apjl}\ }\textbf {\bibinfo {volume} {284}},\ \bibinfo {pages} {L9} (\bibinfo {year} {1984})}\BibitemShut {NoStop}%
\bibitem [{\citenamefont {{Eisenstein}}\ \emph {et~al.}(2001)\citenamefont {{Eisenstein}}, \citenamefont {{Annis}}, \citenamefont {{Gunn}}, \citenamefont {{Szalay}}, \citenamefont {{Connolly}}, \citenamefont {{Nichol}}, \citenamefont {{Bahcall}}, \citenamefont {{Bernardi}}, \citenamefont {{Burles}}, \citenamefont {{Castander}}, \citenamefont {{Fukugita}}, \citenamefont {{Hogg}}, \citenamefont {{Ivezi{\'c}}}, \citenamefont {{Knapp}}, \citenamefont {{Lupton}}, \citenamefont {{Narayanan}}, \citenamefont {{Postman}}, \citenamefont {{Reichart}}, \citenamefont {{Richmond}}, \citenamefont {{Schneider}}, \citenamefont {{Schlegel}}, \citenamefont {{Strauss}}, \citenamefont {{SubbaRao}}, \citenamefont {{Tucker}}, \citenamefont {{Vanden Berk}}, \citenamefont {{Vogeley}}, \citenamefont {{Weinberg}},\ and\ \citenamefont {{Yanny}}}]{2001:eisenstein_lrgs}%
  \BibitemOpen
  \bibfield  {author} {\bibinfo {author} {\bibfnamefont {D.~J.}\ \bibnamefont {{Eisenstein}}}, \bibinfo {author} {\bibfnamefont {J.}~\bibnamefont {{Annis}}}, \bibinfo {author} {\bibfnamefont {J.~E.}\ \bibnamefont {{Gunn}}}, \bibinfo {author} {\bibfnamefont {A.~S.}\ \bibnamefont {{Szalay}}}, \bibinfo {author} {\bibfnamefont {A.~J.}\ \bibnamefont {{Connolly}}}, \bibinfo {author} {\bibfnamefont {R.~C.}\ \bibnamefont {{Nichol}}}, \bibinfo {author} {\bibfnamefont {N.~A.}\ \bibnamefont {{Bahcall}}}, \bibinfo {author} {\bibfnamefont {M.}~\bibnamefont {{Bernardi}}}, \bibinfo {author} {\bibfnamefont {S.}~\bibnamefont {{Burles}}}, \bibinfo {author} {\bibfnamefont {F.~J.}\ \bibnamefont {{Castander}}}, \bibinfo {author} {\bibfnamefont {M.}~\bibnamefont {{Fukugita}}}, \bibinfo {author} {\bibfnamefont {D.~W.}\ \bibnamefont {{Hogg}}}, \bibinfo {author} {\bibfnamefont {{\v{Z}}.}~\bibnamefont {{Ivezi{\'c}}}}, \bibinfo {author} {\bibfnamefont {G.~R.}\ \bibnamefont {{Knapp}}}, \bibinfo {author} {\bibfnamefont {R.~H.}\
  \bibnamefont {{Lupton}}}, \bibinfo {author} {\bibfnamefont {V.}~\bibnamefont {{Narayanan}}}, \bibinfo {author} {\bibfnamefont {M.}~\bibnamefont {{Postman}}}, \bibinfo {author} {\bibfnamefont {D.~E.}\ \bibnamefont {{Reichart}}}, \bibinfo {author} {\bibfnamefont {M.}~\bibnamefont {{Richmond}}}, \bibinfo {author} {\bibfnamefont {D.~P.}\ \bibnamefont {{Schneider}}}, \bibinfo {author} {\bibfnamefont {D.~J.}\ \bibnamefont {{Schlegel}}}, \bibinfo {author} {\bibfnamefont {M.~A.}\ \bibnamefont {{Strauss}}}, \bibinfo {author} {\bibfnamefont {M.}~\bibnamefont {{SubbaRao}}}, \bibinfo {author} {\bibfnamefont {D.~L.}\ \bibnamefont {{Tucker}}}, \bibinfo {author} {\bibfnamefont {D.}~\bibnamefont {{Vanden Berk}}}, \bibinfo {author} {\bibfnamefont {M.~S.}\ \bibnamefont {{Vogeley}}}, \bibinfo {author} {\bibfnamefont {D.~H.}\ \bibnamefont {{Weinberg}}}, \ and\ \bibinfo {author} {\bibfnamefont {B.}~\bibnamefont {{Yanny}}},\ }\href {\doibase 10.1086/323717} {\bibfield  {journal} {\bibinfo  {journal} {\aj}\ }\textbf {\bibinfo
  {volume} {122}},\ \bibinfo {pages} {2267} (\bibinfo {year} {2001})},\ \Eprint {http://arxiv.org/abs/astro-ph/0108153} {arXiv:astro-ph/0108153 [astro-ph]} \BibitemShut {NoStop}%
\bibitem [{\citenamefont {Aghamousa}\ \emph {et~al.}(2016)\citenamefont {Aghamousa} \emph {et~al.}}]{Aghamousa:2016zmz}%
  \BibitemOpen
  \bibfield  {author} {\bibinfo {author} {\bibfnamefont {A.}~\bibnamefont {Aghamousa}} \emph {et~al.} (\bibinfo {collaboration} {DESI}),\ }\href@noop {} {\  (\bibinfo {year} {2016})},\ \Eprint {http://arxiv.org/abs/1611.00036} {arXiv:1611.00036 [astro-ph.IM]} \BibitemShut {NoStop}%
\bibitem [{\citenamefont {{Raichoor}}\ \emph {et~al.}(2023)\citenamefont {{Raichoor}}, \citenamefont {{Moustakas}}, \citenamefont {{Newman}}, \citenamefont {{Karim}}, \citenamefont {{Ahlen}}, \citenamefont {{Alam}}, \citenamefont {{Bailey}}, \citenamefont {{Brooks}}, \citenamefont {{Dawson}}, \citenamefont {{de la Macorra}}, \citenamefont {{de Mattia}}, \citenamefont {{Dey}}, \citenamefont {{Dey}}, \citenamefont {{Dhungana}}, \citenamefont {{Eftekharzadeh}}, \citenamefont {{Eisenstein}}, \citenamefont {{Fanning}}, \citenamefont {{Font-Ribera}}, \citenamefont {{Garc{\'\i}a-Bellido}}, \citenamefont {{Gazta{\~n}aga}}, \citenamefont {{A Gontcho}}, \citenamefont {{Guy}}, \citenamefont {{Honscheid}}, \citenamefont {{Ishak}}, \citenamefont {{Kehoe}}, \citenamefont {{Kisner}}, \citenamefont {{Kremin}}, \citenamefont {{Lan}}, \citenamefont {{Landriau}}, \citenamefont {{Le Guillou}}, \citenamefont {{Levi}}, \citenamefont {{Magneville}}, \citenamefont {{Manera}}, \citenamefont {{Martini}}, \citenamefont {{Meisner}},
  \citenamefont {{Myers}}, \citenamefont {{Nie}}, \citenamefont {{Palanque-Delabrouille}}, \citenamefont {{Percival}}, \citenamefont {{Poppett}}, \citenamefont {{Prada}}, \citenamefont {{Ross}}, \citenamefont {{Ruhlmann-Kleider}}, \citenamefont {{Sabiu}}, \citenamefont {{Schlafly}}, \citenamefont {{Schlegel}}, \citenamefont {{Tarl{\'e}}}, \citenamefont {{Weaver}}, \citenamefont {{Y{\`e}che}}, \citenamefont {{Zhou}}, \citenamefont {{Zhou}},\ and\ \citenamefont {{Zou}}}]{Raichoor:2023_elg}%
  \BibitemOpen
  \bibfield  {author} {\bibinfo {author} {\bibfnamefont {A.}~\bibnamefont {{Raichoor}}}, \bibinfo {author} {\bibfnamefont {J.}~\bibnamefont {{Moustakas}}}, \bibinfo {author} {\bibfnamefont {J.~A.}\ \bibnamefont {{Newman}}}, \bibinfo {author} {\bibfnamefont {T.}~\bibnamefont {{Karim}}}, \bibinfo {author} {\bibfnamefont {S.}~\bibnamefont {{Ahlen}}}, \bibinfo {author} {\bibfnamefont {S.}~\bibnamefont {{Alam}}}, \bibinfo {author} {\bibfnamefont {S.}~\bibnamefont {{Bailey}}}, \bibinfo {author} {\bibfnamefont {D.}~\bibnamefont {{Brooks}}}, \bibinfo {author} {\bibfnamefont {K.}~\bibnamefont {{Dawson}}}, \bibinfo {author} {\bibfnamefont {A.}~\bibnamefont {{de la Macorra}}}, \bibinfo {author} {\bibfnamefont {A.}~\bibnamefont {{de Mattia}}}, \bibinfo {author} {\bibfnamefont {A.}~\bibnamefont {{Dey}}}, \bibinfo {author} {\bibfnamefont {B.}~\bibnamefont {{Dey}}}, \bibinfo {author} {\bibfnamefont {G.}~\bibnamefont {{Dhungana}}}, \bibinfo {author} {\bibfnamefont {S.}~\bibnamefont {{Eftekharzadeh}}}, \bibinfo {author}
  {\bibfnamefont {D.~J.}\ \bibnamefont {{Eisenstein}}}, \bibinfo {author} {\bibfnamefont {K.}~\bibnamefont {{Fanning}}}, \bibinfo {author} {\bibfnamefont {A.}~\bibnamefont {{Font-Ribera}}}, \bibinfo {author} {\bibfnamefont {J.}~\bibnamefont {{Garc{\'\i}a-Bellido}}}, \bibinfo {author} {\bibfnamefont {E.}~\bibnamefont {{Gazta{\~n}aga}}}, \bibinfo {author} {\bibfnamefont {S.~G.}\ \bibnamefont {{A Gontcho}}}, \bibinfo {author} {\bibfnamefont {J.}~\bibnamefont {{Guy}}}, \bibinfo {author} {\bibfnamefont {K.}~\bibnamefont {{Honscheid}}}, \bibinfo {author} {\bibfnamefont {M.}~\bibnamefont {{Ishak}}}, \bibinfo {author} {\bibfnamefont {R.}~\bibnamefont {{Kehoe}}}, \bibinfo {author} {\bibfnamefont {T.}~\bibnamefont {{Kisner}}}, \bibinfo {author} {\bibfnamefont {A.}~\bibnamefont {{Kremin}}}, \bibinfo {author} {\bibfnamefont {T.-W.}\ \bibnamefont {{Lan}}}, \bibinfo {author} {\bibfnamefont {M.}~\bibnamefont {{Landriau}}}, \bibinfo {author} {\bibfnamefont {L.}~\bibnamefont {{Le Guillou}}}, \bibinfo {author} {\bibfnamefont
  {M.~E.}\ \bibnamefont {{Levi}}}, \bibinfo {author} {\bibfnamefont {C.}~\bibnamefont {{Magneville}}}, \bibinfo {author} {\bibfnamefont {M.}~\bibnamefont {{Manera}}}, \bibinfo {author} {\bibfnamefont {P.}~\bibnamefont {{Martini}}}, \bibinfo {author} {\bibfnamefont {A.~M.}\ \bibnamefont {{Meisner}}}, \bibinfo {author} {\bibfnamefont {A.~D.}\ \bibnamefont {{Myers}}}, \bibinfo {author} {\bibfnamefont {J.}~\bibnamefont {{Nie}}}, \bibinfo {author} {\bibfnamefont {N.}~\bibnamefont {{Palanque-Delabrouille}}}, \bibinfo {author} {\bibfnamefont {W.~J.}\ \bibnamefont {{Percival}}}, \bibinfo {author} {\bibfnamefont {C.}~\bibnamefont {{Poppett}}}, \bibinfo {author} {\bibfnamefont {F.}~\bibnamefont {{Prada}}}, \bibinfo {author} {\bibfnamefont {A.~J.}\ \bibnamefont {{Ross}}}, \bibinfo {author} {\bibfnamefont {V.}~\bibnamefont {{Ruhlmann-Kleider}}}, \bibinfo {author} {\bibfnamefont {C.~G.}\ \bibnamefont {{Sabiu}}}, \bibinfo {author} {\bibfnamefont {E.~F.}\ \bibnamefont {{Schlafly}}}, \bibinfo {author} {\bibfnamefont
  {D.}~\bibnamefont {{Schlegel}}}, \bibinfo {author} {\bibfnamefont {G.}~\bibnamefont {{Tarl{\'e}}}}, \bibinfo {author} {\bibfnamefont {B.~A.}\ \bibnamefont {{Weaver}}}, \bibinfo {author} {\bibfnamefont {C.}~\bibnamefont {{Y{\`e}che}}}, \bibinfo {author} {\bibfnamefont {R.}~\bibnamefont {{Zhou}}}, \bibinfo {author} {\bibfnamefont {Z.}~\bibnamefont {{Zhou}}}, \ and\ \bibinfo {author} {\bibfnamefont {H.}~\bibnamefont {{Zou}}},\ }\href {\doibase 10.3847/1538-3881/acb213} {\bibfield  {journal} {\bibinfo  {journal} {\aj}\ }\textbf {\bibinfo {volume} {165}},\ \bibinfo {eid} {126} (\bibinfo {year} {2023})},\ \Eprint {http://arxiv.org/abs/2208.08513} {arXiv:2208.08513 [astro-ph.CO]} \BibitemShut {NoStop}%
\bibitem [{\citenamefont {{Steidel}}\ and\ \citenamefont {{Hamilton}}(1992)}]{Steidel1992:lyman_break_technique}%
  \BibitemOpen
  \bibfield  {author} {\bibinfo {author} {\bibfnamefont {C.~C.}\ \bibnamefont {{Steidel}}}\ and\ \bibinfo {author} {\bibfnamefont {D.}~\bibnamefont {{Hamilton}}},\ }\href {\doibase 10.1086/116287} {\bibfield  {journal} {\bibinfo  {journal} {\aj}\ }\textbf {\bibinfo {volume} {104}},\ \bibinfo {pages} {941} (\bibinfo {year} {1992})}\BibitemShut {NoStop}%
\bibitem [{\citenamefont {{Shapley}}\ \emph {et~al.}(2003)\citenamefont {{Shapley}}, \citenamefont {{Steidel}}, \citenamefont {{Pettini}},\ and\ \citenamefont {{Adelberger}}}]{Shapley:2003_lbg}%
  \BibitemOpen
  \bibfield  {author} {\bibinfo {author} {\bibfnamefont {A.~E.}\ \bibnamefont {{Shapley}}}, \bibinfo {author} {\bibfnamefont {C.~C.}\ \bibnamefont {{Steidel}}}, \bibinfo {author} {\bibfnamefont {M.}~\bibnamefont {{Pettini}}}, \ and\ \bibinfo {author} {\bibfnamefont {K.~L.}\ \bibnamefont {{Adelberger}}},\ }\href {\doibase 10.1086/373922} {\bibfield  {journal} {\bibinfo  {journal} {\apj}\ }\textbf {\bibinfo {volume} {588}},\ \bibinfo {pages} {65} (\bibinfo {year} {2003})},\ \Eprint {http://arxiv.org/abs/astro-ph/0301230} {arXiv:astro-ph/0301230 [astro-ph]} \BibitemShut {NoStop}%
\bibitem [{\citenamefont {{Moscardini}}\ \emph {et~al.}(1998)\citenamefont {{Moscardini}}, \citenamefont {{Coles}}, \citenamefont {{Lucchin}},\ and\ \citenamefont {{Matarrese}}}]{Moscardini:1998_early_lbg_acf}%
  \BibitemOpen
  \bibfield  {author} {\bibinfo {author} {\bibfnamefont {L.}~\bibnamefont {{Moscardini}}}, \bibinfo {author} {\bibfnamefont {P.}~\bibnamefont {{Coles}}}, \bibinfo {author} {\bibfnamefont {F.}~\bibnamefont {{Lucchin}}}, \ and\ \bibinfo {author} {\bibfnamefont {S.}~\bibnamefont {{Matarrese}}},\ }\href {\doibase 10.1046/j.1365-8711.1998.01728.x} {\bibfield  {journal} {\bibinfo  {journal} {\mnras}\ }\textbf {\bibinfo {volume} {299}},\ \bibinfo {pages} {95} (\bibinfo {year} {1998})},\ \Eprint {http://arxiv.org/abs/astro-ph/9712184} {arXiv:astro-ph/9712184 [astro-ph]} \BibitemShut {NoStop}%
\bibitem [{\citenamefont {{Wilson}}\ and\ \citenamefont {{White}}(2019)}]{Wilson_White:2019_lbg}%
  \BibitemOpen
  \bibfield  {author} {\bibinfo {author} {\bibfnamefont {M.~J.}\ \bibnamefont {{Wilson}}}\ and\ \bibinfo {author} {\bibfnamefont {M.}~\bibnamefont {{White}}},\ }\href {\doibase 10.1088/1475-7516/2019/10/015} {\bibfield  {journal} {\bibinfo  {journal} {\jcap}\ }\textbf {\bibinfo {volume} {2019}},\ \bibinfo {eid} {015} (\bibinfo {year} {2019})},\ \Eprint {http://arxiv.org/abs/1904.13378} {arXiv:1904.13378 [astro-ph.CO]} \BibitemShut {NoStop}%
\bibitem [{\citenamefont {{Ebina}}\ and\ \citenamefont {{White}}(2024)}]{Ebina_White:2024_forecast}%
  \BibitemOpen
  \bibfield  {author} {\bibinfo {author} {\bibfnamefont {H.}~\bibnamefont {{Ebina}}}\ and\ \bibinfo {author} {\bibfnamefont {M.}~\bibnamefont {{White}}},\ }\href {\doibase 10.1088/1475-7516/2024/06/052} {\bibfield  {journal} {\bibinfo  {journal} {\jcap}\ }\textbf {\bibinfo {volume} {2024}},\ \bibinfo {eid} {052} (\bibinfo {year} {2024})},\ \Eprint {http://arxiv.org/abs/2401.13166} {arXiv:2401.13166 [astro-ph.CO]} \BibitemShut {NoStop}%
\bibitem [{\citenamefont {{Park}}\ \emph {et~al.}(2016)\citenamefont {{Park}}, \citenamefont {{Kim}}, \citenamefont {{Wyithe}}, \citenamefont {{Lacey}}, \citenamefont {{Baugh}}, \citenamefont {{Barone-Nugent}}, \citenamefont {{Trenti}},\ and\ \citenamefont {{Bouwens}}}]{Park:2016_lbg_hod_sam_small}%
  \BibitemOpen
  \bibfield  {author} {\bibinfo {author} {\bibfnamefont {J.}~\bibnamefont {{Park}}}, \bibinfo {author} {\bibfnamefont {H.-S.}\ \bibnamefont {{Kim}}}, \bibinfo {author} {\bibfnamefont {J.~S.~B.}\ \bibnamefont {{Wyithe}}}, \bibinfo {author} {\bibfnamefont {C.~G.}\ \bibnamefont {{Lacey}}}, \bibinfo {author} {\bibfnamefont {C.~M.}\ \bibnamefont {{Baugh}}}, \bibinfo {author} {\bibfnamefont {R.~L.}\ \bibnamefont {{Barone-Nugent}}}, \bibinfo {author} {\bibfnamefont {M.}~\bibnamefont {{Trenti}}}, \ and\ \bibinfo {author} {\bibfnamefont {R.~J.}\ \bibnamefont {{Bouwens}}},\ }\href {\doibase 10.1093/mnras/stw1316} {\bibfield  {journal} {\bibinfo  {journal} {\mnras}\ }\textbf {\bibinfo {volume} {461}},\ \bibinfo {pages} {176} (\bibinfo {year} {2016})},\ \Eprint {http://arxiv.org/abs/1511.01983} {arXiv:1511.01983 [astro-ph.GA]} \BibitemShut {NoStop}%
\bibitem [{\citenamefont {{Ruhlmann-Kleider}}\ \emph {et~al.}(2024)\citenamefont {{Ruhlmann-Kleider}}, \citenamefont {{Y{\`e}che}}, \citenamefont {{Magneville}}, \citenamefont {{Coquinot}}, \citenamefont {{Armengaud}}, \citenamefont {{Palanque-Delabrouille}}, \citenamefont {{Raichoor}}, \citenamefont {{Aguilar}}, \citenamefont {{Ahlen}}, \citenamefont {{Arnouts}}, \citenamefont {{Brooks}}, \citenamefont {{Chaussidon}}, \citenamefont {{Claybaugh}}, \citenamefont {{Dawson}}, \citenamefont {{de la Macorra}}, \citenamefont {{Dey}}, \citenamefont {{Dey}}, \citenamefont {{Doel}}, \citenamefont {{Fanning}}, \citenamefont {{Ferraro}}, \citenamefont {{Forero-Romero}}, \citenamefont {{Gontcho A Gontcho}}, \citenamefont {{Gutierrez}}, \citenamefont {{Gwyn}}, \citenamefont {{Honscheid}}, \citenamefont {{Juneau}}, \citenamefont {{Kehoe}}, \citenamefont {{Kisner}}, \citenamefont {{Kremin}}, \citenamefont {{Lambert}}, \citenamefont {{Landriau}}, \citenamefont {{Le Guillou}}, \citenamefont {{Levi}}, \citenamefont {{Manera}},
  \citenamefont {{Martini}}, \citenamefont {{Meisner}}, \citenamefont {{Miquel}}, \citenamefont {{Moustakas}}, \citenamefont {{Mueller}}, \citenamefont {{Mu{\~n}oz-Guti{\'e}rrez}}, \citenamefont {{Newman}}, \citenamefont {{Nie}}, \citenamefont {{Niz}}, \citenamefont {{Payerne}}, \citenamefont {{Picouet}}, \citenamefont {{Ravoux}}, \citenamefont {{Rezaie}}, \citenamefont {{Rossi}}, \citenamefont {{Sanchez}}, \citenamefont {{Sawicki}}, \citenamefont {{Schlafly}}, \citenamefont {{Schlegel}}, \citenamefont {{Schubnell}}, \citenamefont {{Seo}}, \citenamefont {{Silber}}, \citenamefont {{Sprayberry}}, \citenamefont {{Taran}}, \citenamefont {{Tarl{\'e}}}, \citenamefont {{Weaver}}, \citenamefont {{White}}, \citenamefont {{Wilson}}, \citenamefont {{Zhou}},\ and\ \citenamefont {{Zou}}}]{Ruhlmann-Kleider:2024_desi_lbg}%
  \BibitemOpen
  \bibfield  {author} {\bibinfo {author} {\bibfnamefont {V.}~\bibnamefont {{Ruhlmann-Kleider}}}, \bibinfo {author} {\bibfnamefont {C.}~\bibnamefont {{Y{\`e}che}}}, \bibinfo {author} {\bibfnamefont {C.}~\bibnamefont {{Magneville}}}, \bibinfo {author} {\bibfnamefont {H.}~\bibnamefont {{Coquinot}}}, \bibinfo {author} {\bibfnamefont {E.}~\bibnamefont {{Armengaud}}}, \bibinfo {author} {\bibfnamefont {N.}~\bibnamefont {{Palanque-Delabrouille}}}, \bibinfo {author} {\bibfnamefont {A.}~\bibnamefont {{Raichoor}}}, \bibinfo {author} {\bibfnamefont {J.~N.}\ \bibnamefont {{Aguilar}}}, \bibinfo {author} {\bibfnamefont {S.}~\bibnamefont {{Ahlen}}}, \bibinfo {author} {\bibfnamefont {S.}~\bibnamefont {{Arnouts}}}, \bibinfo {author} {\bibfnamefont {D.}~\bibnamefont {{Brooks}}}, \bibinfo {author} {\bibfnamefont {E.}~\bibnamefont {{Chaussidon}}}, \bibinfo {author} {\bibfnamefont {T.}~\bibnamefont {{Claybaugh}}}, \bibinfo {author} {\bibfnamefont {K.}~\bibnamefont {{Dawson}}}, \bibinfo {author} {\bibfnamefont {A.}~\bibnamefont
  {{de la Macorra}}}, \bibinfo {author} {\bibfnamefont {A.}~\bibnamefont {{Dey}}}, \bibinfo {author} {\bibfnamefont {B.}~\bibnamefont {{Dey}}}, \bibinfo {author} {\bibfnamefont {P.}~\bibnamefont {{Doel}}}, \bibinfo {author} {\bibfnamefont {K.}~\bibnamefont {{Fanning}}}, \bibinfo {author} {\bibfnamefont {S.}~\bibnamefont {{Ferraro}}}, \bibinfo {author} {\bibfnamefont {J.~E.}\ \bibnamefont {{Forero-Romero}}}, \bibinfo {author} {\bibfnamefont {S.}~\bibnamefont {{Gontcho A Gontcho}}}, \bibinfo {author} {\bibfnamefont {G.}~\bibnamefont {{Gutierrez}}}, \bibinfo {author} {\bibfnamefont {S.}~\bibnamefont {{Gwyn}}}, \bibinfo {author} {\bibfnamefont {K.}~\bibnamefont {{Honscheid}}}, \bibinfo {author} {\bibfnamefont {S.}~\bibnamefont {{Juneau}}}, \bibinfo {author} {\bibfnamefont {R.}~\bibnamefont {{Kehoe}}}, \bibinfo {author} {\bibfnamefont {T.}~\bibnamefont {{Kisner}}}, \bibinfo {author} {\bibfnamefont {A.}~\bibnamefont {{Kremin}}}, \bibinfo {author} {\bibfnamefont {A.}~\bibnamefont {{Lambert}}}, \bibinfo {author}
  {\bibfnamefont {M.}~\bibnamefont {{Landriau}}}, \bibinfo {author} {\bibfnamefont {L.}~\bibnamefont {{Le Guillou}}}, \bibinfo {author} {\bibfnamefont {M.~E.}\ \bibnamefont {{Levi}}}, \bibinfo {author} {\bibfnamefont {M.}~\bibnamefont {{Manera}}}, \bibinfo {author} {\bibfnamefont {P.}~\bibnamefont {{Martini}}}, \bibinfo {author} {\bibfnamefont {A.}~\bibnamefont {{Meisner}}}, \bibinfo {author} {\bibfnamefont {R.}~\bibnamefont {{Miquel}}}, \bibinfo {author} {\bibfnamefont {J.}~\bibnamefont {{Moustakas}}}, \bibinfo {author} {\bibfnamefont {E.-M.}\ \bibnamefont {{Mueller}}}, \bibinfo {author} {\bibfnamefont {A.}~\bibnamefont {{Mu{\~n}oz-Guti{\'e}rrez}}}, \bibinfo {author} {\bibfnamefont {J.~A.}\ \bibnamefont {{Newman}}}, \bibinfo {author} {\bibfnamefont {J.}~\bibnamefont {{Nie}}}, \bibinfo {author} {\bibfnamefont {G.}~\bibnamefont {{Niz}}}, \bibinfo {author} {\bibfnamefont {C.}~\bibnamefont {{Payerne}}}, \bibinfo {author} {\bibfnamefont {V.}~\bibnamefont {{Picouet}}}, \bibinfo {author} {\bibfnamefont
  {C.}~\bibnamefont {{Ravoux}}}, \bibinfo {author} {\bibfnamefont {M.}~\bibnamefont {{Rezaie}}}, \bibinfo {author} {\bibfnamefont {G.}~\bibnamefont {{Rossi}}}, \bibinfo {author} {\bibfnamefont {E.}~\bibnamefont {{Sanchez}}}, \bibinfo {author} {\bibfnamefont {M.}~\bibnamefont {{Sawicki}}}, \bibinfo {author} {\bibfnamefont {E.~F.}\ \bibnamefont {{Schlafly}}}, \bibinfo {author} {\bibfnamefont {D.}~\bibnamefont {{Schlegel}}}, \bibinfo {author} {\bibfnamefont {M.}~\bibnamefont {{Schubnell}}}, \bibinfo {author} {\bibfnamefont {H.-J.}\ \bibnamefont {{Seo}}}, \bibinfo {author} {\bibfnamefont {J.}~\bibnamefont {{Silber}}}, \bibinfo {author} {\bibfnamefont {D.}~\bibnamefont {{Sprayberry}}}, \bibinfo {author} {\bibfnamefont {J.}~\bibnamefont {{Taran}}}, \bibinfo {author} {\bibfnamefont {G.}~\bibnamefont {{Tarl{\'e}}}}, \bibinfo {author} {\bibfnamefont {B.~A.}\ \bibnamefont {{Weaver}}}, \bibinfo {author} {\bibfnamefont {M.}~\bibnamefont {{White}}}, \bibinfo {author} {\bibfnamefont {M.~J.}\ \bibnamefont {{Wilson}}},
  \bibinfo {author} {\bibfnamefont {Z.}~\bibnamefont {{Zhou}}}, \ and\ \bibinfo {author} {\bibfnamefont {H.}~\bibnamefont {{Zou}}},\ }\href {\doibase 10.1088/1475-7516/2024/08/059} {\bibfield  {journal} {\bibinfo  {journal} {\jcap}\ }\textbf {\bibinfo {volume} {2024}},\ \bibinfo {eid} {059} (\bibinfo {year} {2024})},\ \Eprint {http://arxiv.org/abs/2404.03569} {arXiv:2404.03569 [astro-ph.CO]} \BibitemShut {NoStop}%
\bibitem [{\citenamefont {{Firestone}}\ \emph {et~al.}(2025)\citenamefont {{Firestone}}, \citenamefont {{Gawiser}}, \citenamefont {{Iyer}}, \citenamefont {{Lee}}, \citenamefont {{Ramakrishnan}}, \citenamefont {{Valdes}}, \citenamefont {{Park}}, \citenamefont {{Yang}}, \citenamefont {{Alavi}}, \citenamefont {{Ciardullo}}, \citenamefont {{Grogin}}, \citenamefont {{Gronwall}}, \citenamefont {{Guaita}}, \citenamefont {{Hong}}, \citenamefont {{Hwang}}, \citenamefont {{Im}}, \citenamefont {{Jeong}}, \citenamefont {{Kim}}, \citenamefont {{Koekemoer}}, \citenamefont {{Kumar}}, \citenamefont {{Lee}}, \citenamefont {{Mehta}}, \citenamefont {{Nagaraj}}, \citenamefont {{Nantais}}, \citenamefont {{Prichard}}, \citenamefont {{Rafelski}}, \citenamefont {{Song}}, \citenamefont {{Sunnquist}}, \citenamefont {{Teplitz}},\ and\ \citenamefont {{Wang}}}]{Firestone:2025_ODIN_SFH}%
  \BibitemOpen
  \bibfield  {author} {\bibinfo {author} {\bibfnamefont {N.~M.}\ \bibnamefont {{Firestone}}}, \bibinfo {author} {\bibfnamefont {E.}~\bibnamefont {{Gawiser}}}, \bibinfo {author} {\bibfnamefont {K.~G.}\ \bibnamefont {{Iyer}}}, \bibinfo {author} {\bibfnamefont {K.-S.}\ \bibnamefont {{Lee}}}, \bibinfo {author} {\bibfnamefont {V.}~\bibnamefont {{Ramakrishnan}}}, \bibinfo {author} {\bibfnamefont {F.}~\bibnamefont {{Valdes}}}, \bibinfo {author} {\bibfnamefont {C.}~\bibnamefont {{Park}}}, \bibinfo {author} {\bibfnamefont {Y.}~\bibnamefont {{Yang}}}, \bibinfo {author} {\bibfnamefont {A.}~\bibnamefont {{Alavi}}}, \bibinfo {author} {\bibfnamefont {R.}~\bibnamefont {{Ciardullo}}}, \bibinfo {author} {\bibfnamefont {N.}~\bibnamefont {{Grogin}}}, \bibinfo {author} {\bibfnamefont {C.}~\bibnamefont {{Gronwall}}}, \bibinfo {author} {\bibfnamefont {L.}~\bibnamefont {{Guaita}}}, \bibinfo {author} {\bibfnamefont {S.}~\bibnamefont {{Hong}}}, \bibinfo {author} {\bibfnamefont {H.~S.}\ \bibnamefont {{Hwang}}}, \bibinfo {author}
  {\bibfnamefont {S.~H.}\ \bibnamefont {{Im}}}, \bibinfo {author} {\bibfnamefont {W.-S.}\ \bibnamefont {{Jeong}}}, \bibinfo {author} {\bibfnamefont {S.}~\bibnamefont {{Kim}}}, \bibinfo {author} {\bibfnamefont {A.~M.}\ \bibnamefont {{Koekemoer}}}, \bibinfo {author} {\bibfnamefont {A.}~\bibnamefont {{Kumar}}}, \bibinfo {author} {\bibfnamefont {J.}~\bibnamefont {{Lee}}}, \bibinfo {author} {\bibfnamefont {V.}~\bibnamefont {{Mehta}}}, \bibinfo {author} {\bibfnamefont {G.}~\bibnamefont {{Nagaraj}}}, \bibinfo {author} {\bibfnamefont {J.}~\bibnamefont {{Nantais}}}, \bibinfo {author} {\bibfnamefont {L.}~\bibnamefont {{Prichard}}}, \bibinfo {author} {\bibfnamefont {M.}~\bibnamefont {{Rafelski}}}, \bibinfo {author} {\bibfnamefont {H.}~\bibnamefont {{Song}}}, \bibinfo {author} {\bibfnamefont {B.}~\bibnamefont {{Sunnquist}}}, \bibinfo {author} {\bibfnamefont {H.~I.}\ \bibnamefont {{Teplitz}}}, \ and\ \bibinfo {author} {\bibfnamefont {X.}~\bibnamefont {{Wang}}},\ }\href@noop {} {\bibfield  {journal} {\bibinfo  {journal}
  {arXiv e-prints}\ ,\ \bibinfo {eid} {arXiv:2501.08568}} (\bibinfo {year} {2025})},\ \Eprint {http://arxiv.org/abs/2501.08568} {arXiv:2501.08568 [astro-ph.GA]} \BibitemShut {NoStop}%
\bibitem [{\citenamefont {{Partridge}}\ and\ \citenamefont {{Peebles}}(1967)}]{Partridge:1967_peebles}%
  \BibitemOpen
  \bibfield  {author} {\bibinfo {author} {\bibfnamefont {R.~B.}\ \bibnamefont {{Partridge}}}\ and\ \bibinfo {author} {\bibfnamefont {P.~J.~E.}\ \bibnamefont {{Peebles}}},\ }\href {\doibase 10.1086/149079} {\bibfield  {journal} {\bibinfo  {journal} {\apj}\ }\textbf {\bibinfo {volume} {147}},\ \bibinfo {pages} {868} (\bibinfo {year} {1967})}\BibitemShut {NoStop}%
\bibitem [{\citenamefont {{Furlanetto}}\ \emph {et~al.}(2005)\citenamefont {{Furlanetto}}, \citenamefont {{Schaye}}, \citenamefont {{Springel}},\ and\ \citenamefont {{Hernquist}}}]{Furlanetto:2005_lae_igm}%
  \BibitemOpen
  \bibfield  {author} {\bibinfo {author} {\bibfnamefont {S.~R.}\ \bibnamefont {{Furlanetto}}}, \bibinfo {author} {\bibfnamefont {J.}~\bibnamefont {{Schaye}}}, \bibinfo {author} {\bibfnamefont {V.}~\bibnamefont {{Springel}}}, \ and\ \bibinfo {author} {\bibfnamefont {L.}~\bibnamefont {{Hernquist}}},\ }\href {\doibase 10.1086/426808} {\bibfield  {journal} {\bibinfo  {journal} {\apj}\ }\textbf {\bibinfo {volume} {622}},\ \bibinfo {pages} {7} (\bibinfo {year} {2005})},\ \Eprint {http://arxiv.org/abs/astro-ph/0409736} {arXiv:astro-ph/0409736 [astro-ph]} \BibitemShut {NoStop}%
\bibitem [{\citenamefont {{Ouchi}}\ \emph {et~al.}(2020)\citenamefont {{Ouchi}}, \citenamefont {{Ono}},\ and\ \citenamefont {{Shibuya}}}]{ouchi:2020_AnnRev}%
  \BibitemOpen
  \bibfield  {author} {\bibinfo {author} {\bibfnamefont {M.}~\bibnamefont {{Ouchi}}}, \bibinfo {author} {\bibfnamefont {Y.}~\bibnamefont {{Ono}}}, \ and\ \bibinfo {author} {\bibfnamefont {T.}~\bibnamefont {{Shibuya}}},\ }\href {\doibase 10.1146/annurev-astro-032620-021859} {\bibfield  {journal} {\bibinfo  {journal} {\araa}\ }\textbf {\bibinfo {volume} {58}},\ \bibinfo {pages} {617} (\bibinfo {year} {2020})},\ \Eprint {http://arxiv.org/abs/2012.07960} {arXiv:2012.07960 [astro-ph.GA]} \BibitemShut {NoStop}%
\bibitem [{\citenamefont {{Lee}}\ \emph {et~al.}(2024)\citenamefont {{Lee}}, \citenamefont {{Gawiser}}, \citenamefont {{Park}}, \citenamefont {{Yang}}, \citenamefont {{Valdes}}, \citenamefont {{Lang}}, \citenamefont {{Ramakrishnan}}, \citenamefont {{Moon}}, \citenamefont {{Firestone}}, \citenamefont {{Appleby}}, \citenamefont {{Artale}}, \citenamefont {{Andrews}}, \citenamefont {{Bauer}}, \citenamefont {{Benda}}, \citenamefont {{Broussard}}, \citenamefont {{Chiang}}, \citenamefont {{Ciardullo}}, \citenamefont {{Dey}}, \citenamefont {{Farooq}}, \citenamefont {{Gronwall}}, \citenamefont {{Guaita}}, \citenamefont {{Huang}}, \citenamefont {{Hwang}}, \citenamefont {{Im}}, \citenamefont {{Jeong}}, \citenamefont {{Karthikeyan}}, \citenamefont {{Kim}}, \citenamefont {{Kim}}, \citenamefont {{Kumar}}, \citenamefont {{Nagaraj}}, \citenamefont {{Nantais}}, \citenamefont {{Padilla}}, \citenamefont {{Park}}, \citenamefont {{Pope}}, \citenamefont {{Popescu}}, \citenamefont {{Schlegel}}, \citenamefont {{Seo}}, \citenamefont
  {{Singh}}, \citenamefont {{Song}}, \citenamefont {{Troncoso}}, \citenamefont {{Vivas}}, \citenamefont {{Zabludoff}},\ and\ \citenamefont {{Zenteno}}}]{Lee:2024_ODIN_initial}%
  \BibitemOpen
  \bibfield  {author} {\bibinfo {author} {\bibfnamefont {K.-S.}\ \bibnamefont {{Lee}}}, \bibinfo {author} {\bibfnamefont {E.}~\bibnamefont {{Gawiser}}}, \bibinfo {author} {\bibfnamefont {C.}~\bibnamefont {{Park}}}, \bibinfo {author} {\bibfnamefont {Y.}~\bibnamefont {{Yang}}}, \bibinfo {author} {\bibfnamefont {F.}~\bibnamefont {{Valdes}}}, \bibinfo {author} {\bibfnamefont {D.}~\bibnamefont {{Lang}}}, \bibinfo {author} {\bibfnamefont {V.}~\bibnamefont {{Ramakrishnan}}}, \bibinfo {author} {\bibfnamefont {B.}~\bibnamefont {{Moon}}}, \bibinfo {author} {\bibfnamefont {N.}~\bibnamefont {{Firestone}}}, \bibinfo {author} {\bibfnamefont {S.}~\bibnamefont {{Appleby}}}, \bibinfo {author} {\bibfnamefont {M.~C.}\ \bibnamefont {{Artale}}}, \bibinfo {author} {\bibfnamefont {M.}~\bibnamefont {{Andrews}}}, \bibinfo {author} {\bibfnamefont {F.}~\bibnamefont {{Bauer}}}, \bibinfo {author} {\bibfnamefont {B.}~\bibnamefont {{Benda}}}, \bibinfo {author} {\bibfnamefont {A.}~\bibnamefont {{Broussard}}}, \bibinfo {author}
  {\bibfnamefont {Y.-K.}\ \bibnamefont {{Chiang}}}, \bibinfo {author} {\bibfnamefont {R.}~\bibnamefont {{Ciardullo}}}, \bibinfo {author} {\bibfnamefont {A.}~\bibnamefont {{Dey}}}, \bibinfo {author} {\bibfnamefont {R.}~\bibnamefont {{Farooq}}}, \bibinfo {author} {\bibfnamefont {C.}~\bibnamefont {{Gronwall}}}, \bibinfo {author} {\bibfnamefont {L.}~\bibnamefont {{Guaita}}}, \bibinfo {author} {\bibfnamefont {Y.}~\bibnamefont {{Huang}}}, \bibinfo {author} {\bibfnamefont {H.~S.}\ \bibnamefont {{Hwang}}}, \bibinfo {author} {\bibfnamefont {S.~H.}\ \bibnamefont {{Im}}}, \bibinfo {author} {\bibfnamefont {W.-S.}\ \bibnamefont {{Jeong}}}, \bibinfo {author} {\bibfnamefont {S.}~\bibnamefont {{Karthikeyan}}}, \bibinfo {author} {\bibfnamefont {H.}~\bibnamefont {{Kim}}}, \bibinfo {author} {\bibfnamefont {S.}~\bibnamefont {{Kim}}}, \bibinfo {author} {\bibfnamefont {A.}~\bibnamefont {{Kumar}}}, \bibinfo {author} {\bibfnamefont {G.~R.}\ \bibnamefont {{Nagaraj}}}, \bibinfo {author} {\bibfnamefont {J.}~\bibnamefont {{Nantais}}},
  \bibinfo {author} {\bibfnamefont {N.}~\bibnamefont {{Padilla}}}, \bibinfo {author} {\bibfnamefont {J.}~\bibnamefont {{Park}}}, \bibinfo {author} {\bibfnamefont {A.}~\bibnamefont {{Pope}}}, \bibinfo {author} {\bibfnamefont {R.}~\bibnamefont {{Popescu}}}, \bibinfo {author} {\bibfnamefont {D.}~\bibnamefont {{Schlegel}}}, \bibinfo {author} {\bibfnamefont {E.}~\bibnamefont {{Seo}}}, \bibinfo {author} {\bibfnamefont {A.}~\bibnamefont {{Singh}}}, \bibinfo {author} {\bibfnamefont {H.}~\bibnamefont {{Song}}}, \bibinfo {author} {\bibfnamefont {P.}~\bibnamefont {{Troncoso}}}, \bibinfo {author} {\bibfnamefont {A.~K.}\ \bibnamefont {{Vivas}}}, \bibinfo {author} {\bibfnamefont {A.}~\bibnamefont {{Zabludoff}}}, \ and\ \bibinfo {author} {\bibfnamefont {A.}~\bibnamefont {{Zenteno}}},\ }\href {\doibase 10.3847/1538-4357/ad165e} {\bibfield  {journal} {\bibinfo  {journal} {\apj}\ }\textbf {\bibinfo {volume} {962}},\ \bibinfo {eid} {36} (\bibinfo {year} {2024})},\ \Eprint {http://arxiv.org/abs/2309.10191} {arXiv:2309.10191
  [astro-ph.GA]} \BibitemShut {NoStop}%
\bibitem [{\citenamefont {{Kikuta}}\ \emph {et~al.}(2023)\citenamefont {{Kikuta}}, \citenamefont {{Ouchi}}, \citenamefont {{Shibuya}}, \citenamefont {{Liang}}, \citenamefont {{Umeda}}, \citenamefont {{Matsumoto}}, \citenamefont {{Shimasaku}}, \citenamefont {{Harikane}}, \citenamefont {{Ono}}, \citenamefont {{Inoue}}, \citenamefont {{Yamanaka}}, \citenamefont {{Kusakabe}}, \citenamefont {{Momose}}, \citenamefont {{Kashikawa}}, \citenamefont {{Matsuda}},\ and\ \citenamefont {{Lee}}}]{Kikuta:2023_SILVERRUSH_catalog}%
  \BibitemOpen
  \bibfield  {author} {\bibinfo {author} {\bibfnamefont {S.}~\bibnamefont {{Kikuta}}}, \bibinfo {author} {\bibfnamefont {M.}~\bibnamefont {{Ouchi}}}, \bibinfo {author} {\bibfnamefont {T.}~\bibnamefont {{Shibuya}}}, \bibinfo {author} {\bibfnamefont {Y.}~\bibnamefont {{Liang}}}, \bibinfo {author} {\bibfnamefont {H.}~\bibnamefont {{Umeda}}}, \bibinfo {author} {\bibfnamefont {A.}~\bibnamefont {{Matsumoto}}}, \bibinfo {author} {\bibfnamefont {K.}~\bibnamefont {{Shimasaku}}}, \bibinfo {author} {\bibfnamefont {Y.}~\bibnamefont {{Harikane}}}, \bibinfo {author} {\bibfnamefont {Y.}~\bibnamefont {{Ono}}}, \bibinfo {author} {\bibfnamefont {A.~K.}\ \bibnamefont {{Inoue}}}, \bibinfo {author} {\bibfnamefont {S.}~\bibnamefont {{Yamanaka}}}, \bibinfo {author} {\bibfnamefont {H.}~\bibnamefont {{Kusakabe}}}, \bibinfo {author} {\bibfnamefont {R.}~\bibnamefont {{Momose}}}, \bibinfo {author} {\bibfnamefont {N.}~\bibnamefont {{Kashikawa}}}, \bibinfo {author} {\bibfnamefont {Y.}~\bibnamefont {{Matsuda}}}, \ and\ \bibinfo {author}
  {\bibfnamefont {C.-H.}\ \bibnamefont {{Lee}}},\ }\href {\doibase 10.3847/1538-4365/ace4cb} {\bibfield  {journal} {\bibinfo  {journal} {\apjs}\ }\textbf {\bibinfo {volume} {268}},\ \bibinfo {eid} {24} (\bibinfo {year} {2023})},\ \Eprint {http://arxiv.org/abs/2305.08921} {arXiv:2305.08921 [astro-ph.GA]} \BibitemShut {NoStop}%
\bibitem [{\citenamefont {{Davis}}\ \emph {et~al.}(2023)\citenamefont {{Davis}}, \citenamefont {{Gebhardt}}, \citenamefont {{Cooper}}, \citenamefont {{Ciardullo}}, \citenamefont {{Fabricius}}, \citenamefont {{Farrow}}, \citenamefont {{Feldmeier}}, \citenamefont {{Finkelstein}}, \citenamefont {{Gawiser}}, \citenamefont {{Gronwall}}, \citenamefont {{Hill}}, \citenamefont {{Hopp}}, \citenamefont {{House}}, \citenamefont {{Jeong}}, \citenamefont {{Kollatschny}}, \citenamefont {{Komatsu}}, \citenamefont {{Landriau}}, \citenamefont {{Liu}}, \citenamefont {{Saito}}, \citenamefont {{Tuttle}}, \citenamefont {{Wold}}, \citenamefont {{Zeimann}},\ and\ \citenamefont {{Zhang}}}]{Davis:2023_hetdex_classif}%
  \BibitemOpen
  \bibfield  {author} {\bibinfo {author} {\bibfnamefont {D.}~\bibnamefont {{Davis}}}, \bibinfo {author} {\bibfnamefont {K.}~\bibnamefont {{Gebhardt}}}, \bibinfo {author} {\bibfnamefont {E.~M.}\ \bibnamefont {{Cooper}}}, \bibinfo {author} {\bibfnamefont {R.}~\bibnamefont {{Ciardullo}}}, \bibinfo {author} {\bibfnamefont {M.}~\bibnamefont {{Fabricius}}}, \bibinfo {author} {\bibfnamefont {D.~J.}\ \bibnamefont {{Farrow}}}, \bibinfo {author} {\bibfnamefont {J.~J.}\ \bibnamefont {{Feldmeier}}}, \bibinfo {author} {\bibfnamefont {S.~L.}\ \bibnamefont {{Finkelstein}}}, \bibinfo {author} {\bibfnamefont {E.}~\bibnamefont {{Gawiser}}}, \bibinfo {author} {\bibfnamefont {C.}~\bibnamefont {{Gronwall}}}, \bibinfo {author} {\bibfnamefont {G.~J.}\ \bibnamefont {{Hill}}}, \bibinfo {author} {\bibfnamefont {U.}~\bibnamefont {{Hopp}}}, \bibinfo {author} {\bibfnamefont {L.~R.}\ \bibnamefont {{House}}}, \bibinfo {author} {\bibfnamefont {D.}~\bibnamefont {{Jeong}}}, \bibinfo {author} {\bibfnamefont {W.}~\bibnamefont {{Kollatschny}}},
  \bibinfo {author} {\bibfnamefont {E.}~\bibnamefont {{Komatsu}}}, \bibinfo {author} {\bibfnamefont {M.}~\bibnamefont {{Landriau}}}, \bibinfo {author} {\bibfnamefont {C.}~\bibnamefont {{Liu}}}, \bibinfo {author} {\bibfnamefont {S.}~\bibnamefont {{Saito}}}, \bibinfo {author} {\bibfnamefont {S.}~\bibnamefont {{Tuttle}}}, \bibinfo {author} {\bibfnamefont {I.~G.~B.}\ \bibnamefont {{Wold}}}, \bibinfo {author} {\bibfnamefont {G.~R.}\ \bibnamefont {{Zeimann}}}, \ and\ \bibinfo {author} {\bibfnamefont {Y.}~\bibnamefont {{Zhang}}},\ }\href {\doibase 10.3847/1538-4357/acb0ca} {\bibfield  {journal} {\bibinfo  {journal} {\apj}\ }\textbf {\bibinfo {volume} {946}},\ \bibinfo {eid} {86} (\bibinfo {year} {2023})},\ \Eprint {http://arxiv.org/abs/2301.01799} {arXiv:2301.01799 [astro-ph.GA]} \BibitemShut {NoStop}%
\bibitem [{\citenamefont {{Landriau}}\ \emph {et~al.}(2025)\citenamefont {{Landriau}}, \citenamefont {{Mentuch Cooper}}, \citenamefont {{Davis}}, \citenamefont {{Gebhardt}}, \citenamefont {{Ciardullo}}, \citenamefont {{Armengaud}}, \citenamefont {{Dey}}, \citenamefont {{Raichoor}}, \citenamefont {{Schlegel}}, \citenamefont {{Wilson}}, \citenamefont {{Aguilar}}, \citenamefont {{Ahlen}}, \citenamefont {{Bianchi}}, \citenamefont {{Brooks}}, \citenamefont {{Claybaugh}}, \citenamefont {{de la Macorra}}, \citenamefont {{Ferraro}}, \citenamefont {{Forero-Romero}}, \citenamefont {{Gazta{\~n}aga}}, \citenamefont {{Gontcho}}, \citenamefont {{Gutierrez}}, \citenamefont {{Hahn}}, \citenamefont {{Honscheid}}, \citenamefont {{Howlett}}, \citenamefont {{Ishak}}, \citenamefont {{Juneau}}, \citenamefont {{Kehoe}}, \citenamefont {{Kisner}}, \citenamefont {{Kremin}}, \citenamefont {{Le Guillou}}, \citenamefont {{Levi}}, \citenamefont {{Manera}}, \citenamefont {{Meisner}}, \citenamefont {{Miquel}}, \citenamefont {{Moustakas}},
  \citenamefont {{Nadathur}}, \citenamefont {{P{\'e}rez-R{\`a}fols}}, \citenamefont {{Poppett}}, \citenamefont {{Prada}}, \citenamefont {{Rossi}}, \citenamefont {{Sanchez}}, \citenamefont {{Schubnell}}, \citenamefont {{Sprayberry}}, \citenamefont {{Tarl{\'e}}}, \citenamefont {{Weaver}}, \citenamefont {{Zhou}}, \citenamefont {{Zou}}, \citenamefont {{Farrow}}, \citenamefont {{Hill}}, \citenamefont {{Jeong}}, \citenamefont {{Liu}}, \citenamefont {{Saito}},\ and\ \citenamefont {{Schneider}}}]{Landriau:2025_desi_hetdex_line}%
  \BibitemOpen
  \bibfield  {author} {\bibinfo {author} {\bibfnamefont {M.}~\bibnamefont {{Landriau}}}, \bibinfo {author} {\bibfnamefont {E.}~\bibnamefont {{Mentuch Cooper}}}, \bibinfo {author} {\bibfnamefont {D.}~\bibnamefont {{Davis}}}, \bibinfo {author} {\bibfnamefont {K.}~\bibnamefont {{Gebhardt}}}, \bibinfo {author} {\bibfnamefont {R.}~\bibnamefont {{Ciardullo}}}, \bibinfo {author} {\bibfnamefont {{\'E}.}~\bibnamefont {{Armengaud}}}, \bibinfo {author} {\bibfnamefont {A.}~\bibnamefont {{Dey}}}, \bibinfo {author} {\bibfnamefont {A.}~\bibnamefont {{Raichoor}}}, \bibinfo {author} {\bibfnamefont {D.~J.}\ \bibnamefont {{Schlegel}}}, \bibinfo {author} {\bibfnamefont {M.}~\bibnamefont {{Wilson}}}, \bibinfo {author} {\bibfnamefont {J.}~\bibnamefont {{Aguilar}}}, \bibinfo {author} {\bibfnamefont {S.}~\bibnamefont {{Ahlen}}}, \bibinfo {author} {\bibfnamefont {D.}~\bibnamefont {{Bianchi}}}, \bibinfo {author} {\bibfnamefont {D.}~\bibnamefont {{Brooks}}}, \bibinfo {author} {\bibfnamefont {T.}~\bibnamefont {{Claybaugh}}}, \bibinfo
  {author} {\bibfnamefont {A.}~\bibnamefont {{de la Macorra}}}, \bibinfo {author} {\bibfnamefont {S.}~\bibnamefont {{Ferraro}}}, \bibinfo {author} {\bibfnamefont {J.~E.}\ \bibnamefont {{Forero-Romero}}}, \bibinfo {author} {\bibfnamefont {E.}~\bibnamefont {{Gazta{\~n}aga}}}, \bibinfo {author} {\bibfnamefont {S.~G.~A.}\ \bibnamefont {{Gontcho}}}, \bibinfo {author} {\bibfnamefont {G.}~\bibnamefont {{Gutierrez}}}, \bibinfo {author} {\bibfnamefont {C.}~\bibnamefont {{Hahn}}}, \bibinfo {author} {\bibfnamefont {K.}~\bibnamefont {{Honscheid}}}, \bibinfo {author} {\bibfnamefont {C.}~\bibnamefont {{Howlett}}}, \bibinfo {author} {\bibfnamefont {M.}~\bibnamefont {{Ishak}}}, \bibinfo {author} {\bibfnamefont {S.}~\bibnamefont {{Juneau}}}, \bibinfo {author} {\bibfnamefont {R.}~\bibnamefont {{Kehoe}}}, \bibinfo {author} {\bibfnamefont {T.}~\bibnamefont {{Kisner}}}, \bibinfo {author} {\bibfnamefont {A.}~\bibnamefont {{Kremin}}}, \bibinfo {author} {\bibfnamefont {L.}~\bibnamefont {{Le Guillou}}}, \bibinfo {author}
  {\bibfnamefont {M.~E.}\ \bibnamefont {{Levi}}}, \bibinfo {author} {\bibfnamefont {M.}~\bibnamefont {{Manera}}}, \bibinfo {author} {\bibfnamefont {A.}~\bibnamefont {{Meisner}}}, \bibinfo {author} {\bibfnamefont {R.}~\bibnamefont {{Miquel}}}, \bibinfo {author} {\bibfnamefont {J.}~\bibnamefont {{Moustakas}}}, \bibinfo {author} {\bibfnamefont {S.}~\bibnamefont {{Nadathur}}}, \bibinfo {author} {\bibfnamefont {I.}~\bibnamefont {{P{\'e}rez-R{\`a}fols}}}, \bibinfo {author} {\bibfnamefont {C.}~\bibnamefont {{Poppett}}}, \bibinfo {author} {\bibfnamefont {F.}~\bibnamefont {{Prada}}}, \bibinfo {author} {\bibfnamefont {G.}~\bibnamefont {{Rossi}}}, \bibinfo {author} {\bibfnamefont {E.}~\bibnamefont {{Sanchez}}}, \bibinfo {author} {\bibfnamefont {M.}~\bibnamefont {{Schubnell}}}, \bibinfo {author} {\bibfnamefont {D.}~\bibnamefont {{Sprayberry}}}, \bibinfo {author} {\bibfnamefont {G.}~\bibnamefont {{Tarl{\'e}}}}, \bibinfo {author} {\bibfnamefont {B.~A.}\ \bibnamefont {{Weaver}}}, \bibinfo {author} {\bibfnamefont
  {R.}~\bibnamefont {{Zhou}}}, \bibinfo {author} {\bibfnamefont {H.}~\bibnamefont {{Zou}}}, \bibinfo {author} {\bibfnamefont {D.~J.}\ \bibnamefont {{Farrow}}}, \bibinfo {author} {\bibfnamefont {G.~J.}\ \bibnamefont {{Hill}}}, \bibinfo {author} {\bibfnamefont {D.}~\bibnamefont {{Jeong}}}, \bibinfo {author} {\bibfnamefont {C.}~\bibnamefont {{Liu}}}, \bibinfo {author} {\bibfnamefont {S.}~\bibnamefont {{Saito}}}, \ and\ \bibinfo {author} {\bibfnamefont {D.~P.}\ \bibnamefont {{Schneider}}},\ }\href {\doibase 10.48550/arXiv.2503.02229} {\bibfield  {journal} {\bibinfo  {journal} {arXiv e-prints}\ ,\ \bibinfo {eid} {arXiv:2503.02229}} (\bibinfo {year} {2025})},\ \Eprint {http://arxiv.org/abs/2503.02229} {arXiv:2503.02229 [astro-ph.CO]} \BibitemShut {NoStop}%
\bibitem [{\citenamefont {{Firestone}}\ \emph {et~al.}(2024)\citenamefont {{Firestone}}, \citenamefont {{Gawiser}}, \citenamefont {{Ramakrishnan}}, \citenamefont {{Lee}}, \citenamefont {{Valdes}}, \citenamefont {{Park}}, \citenamefont {{Yang}}, \citenamefont {{Ciardullo}}, \citenamefont {{Artale}}, \citenamefont {{Benda}}, \citenamefont {{Broussard}}, \citenamefont {{Eid}}, \citenamefont {{Farooq}}, \citenamefont {{Gronwall}}, \citenamefont {{Guaita}}, \citenamefont {{Gwyn}}, \citenamefont {{Hwang}}, \citenamefont {{Im}}, \citenamefont {{Jeong}}, \citenamefont {{Karthikeyan}}, \citenamefont {{Lang}}, \citenamefont {{Moon}}, \citenamefont {{Padilla}}, \citenamefont {{Sawicki}}, \citenamefont {{Seo}}, \citenamefont {{Singh}}, \citenamefont {{Song}},\ and\ \citenamefont {{Troncoso Iribarren}}}]{Firestone:2024_odin_selection}%
  \BibitemOpen
  \bibfield  {author} {\bibinfo {author} {\bibfnamefont {N.~M.}\ \bibnamefont {{Firestone}}}, \bibinfo {author} {\bibfnamefont {E.}~\bibnamefont {{Gawiser}}}, \bibinfo {author} {\bibfnamefont {V.}~\bibnamefont {{Ramakrishnan}}}, \bibinfo {author} {\bibfnamefont {K.-S.}\ \bibnamefont {{Lee}}}, \bibinfo {author} {\bibfnamefont {F.}~\bibnamefont {{Valdes}}}, \bibinfo {author} {\bibfnamefont {C.}~\bibnamefont {{Park}}}, \bibinfo {author} {\bibfnamefont {Y.}~\bibnamefont {{Yang}}}, \bibinfo {author} {\bibfnamefont {R.}~\bibnamefont {{Ciardullo}}}, \bibinfo {author} {\bibfnamefont {M.~C.}\ \bibnamefont {{Artale}}}, \bibinfo {author} {\bibfnamefont {B.}~\bibnamefont {{Benda}}}, \bibinfo {author} {\bibfnamefont {A.}~\bibnamefont {{Broussard}}}, \bibinfo {author} {\bibfnamefont {L.}~\bibnamefont {{Eid}}}, \bibinfo {author} {\bibfnamefont {R.}~\bibnamefont {{Farooq}}}, \bibinfo {author} {\bibfnamefont {C.}~\bibnamefont {{Gronwall}}}, \bibinfo {author} {\bibfnamefont {L.}~\bibnamefont {{Guaita}}}, \bibinfo {author}
  {\bibfnamefont {S.}~\bibnamefont {{Gwyn}}}, \bibinfo {author} {\bibfnamefont {H.~S.}\ \bibnamefont {{Hwang}}}, \bibinfo {author} {\bibfnamefont {S.~H.}\ \bibnamefont {{Im}}}, \bibinfo {author} {\bibfnamefont {W.-S.}\ \bibnamefont {{Jeong}}}, \bibinfo {author} {\bibfnamefont {S.}~\bibnamefont {{Karthikeyan}}}, \bibinfo {author} {\bibfnamefont {D.}~\bibnamefont {{Lang}}}, \bibinfo {author} {\bibfnamefont {B.}~\bibnamefont {{Moon}}}, \bibinfo {author} {\bibfnamefont {N.}~\bibnamefont {{Padilla}}}, \bibinfo {author} {\bibfnamefont {M.}~\bibnamefont {{Sawicki}}}, \bibinfo {author} {\bibfnamefont {E.}~\bibnamefont {{Seo}}}, \bibinfo {author} {\bibfnamefont {A.}~\bibnamefont {{Singh}}}, \bibinfo {author} {\bibfnamefont {H.}~\bibnamefont {{Song}}}, \ and\ \bibinfo {author} {\bibfnamefont {P.}~\bibnamefont {{Troncoso Iribarren}}},\ }\href {\doibase 10.3847/1538-4357/ad71c9} {\bibfield  {journal} {\bibinfo  {journal} {\apj}\ }\textbf {\bibinfo {volume} {974}},\ \bibinfo {eid} {217} (\bibinfo {year} {2024})},\ \Eprint
  {http://arxiv.org/abs/2312.16075} {arXiv:2312.16075 [astro-ph.GA]} \BibitemShut {NoStop}%
\bibitem [{\citenamefont {{Ivanov}}\ \emph {et~al.}(2024)\citenamefont {{Ivanov}}, \citenamefont {{Cuesta-Lazaro}}, \citenamefont {{Obuljen}}, \citenamefont {{Toomey}}, \citenamefont {{Ni}}, \citenamefont {{Bose}}, \citenamefont {{Hadzhiyska}}, \citenamefont {{Hern{\'a}ndez-Aguayo}}, \citenamefont {{Hernquist}}, \citenamefont {{Kannan}},\ and\ \citenamefont {{Springel}}}]{Ivanov:2024_mtngeft}%
  \BibitemOpen
  \bibfield  {author} {\bibinfo {author} {\bibfnamefont {M.~M.}\ \bibnamefont {{Ivanov}}}, \bibinfo {author} {\bibfnamefont {C.}~\bibnamefont {{Cuesta-Lazaro}}}, \bibinfo {author} {\bibfnamefont {A.}~\bibnamefont {{Obuljen}}}, \bibinfo {author} {\bibfnamefont {M.~W.}\ \bibnamefont {{Toomey}}}, \bibinfo {author} {\bibfnamefont {Y.}~\bibnamefont {{Ni}}}, \bibinfo {author} {\bibfnamefont {S.}~\bibnamefont {{Bose}}}, \bibinfo {author} {\bibfnamefont {B.}~\bibnamefont {{Hadzhiyska}}}, \bibinfo {author} {\bibfnamefont {C.}~\bibnamefont {{Hern{\'a}ndez-Aguayo}}}, \bibinfo {author} {\bibfnamefont {L.}~\bibnamefont {{Hernquist}}}, \bibinfo {author} {\bibfnamefont {R.}~\bibnamefont {{Kannan}}}, \ and\ \bibinfo {author} {\bibfnamefont {V.}~\bibnamefont {{Springel}}},\ }\href {\doibase 10.48550/arXiv.2412.01888} {\bibfield  {journal} {\bibinfo  {journal} {arXiv e-prints}\ ,\ \bibinfo {eid} {arXiv:2412.01888}} (\bibinfo {year} {2024})},\ \Eprint {http://arxiv.org/abs/2412.01888} {arXiv:2412.01888 [astro-ph.CO]}
  \BibitemShut {NoStop}%
\bibitem [{\citenamefont {Pakmor}\ \emph {et~al.}(2023)\citenamefont {Pakmor} \emph {et~al.}}]{Pakmor:2022yyn}%
  \BibitemOpen
  \bibfield  {author} {\bibinfo {author} {\bibfnamefont {R.}~\bibnamefont {Pakmor}} \emph {et~al.},\ }\href {\doibase 10.1093/mnras/stac3620} {\bibfield  {journal} {\bibinfo  {journal} {Mon. Not. Roy. Astron. Soc.}\ }\textbf {\bibinfo {volume} {524}},\ \bibinfo {pages} {2539} (\bibinfo {year} {2023})},\ \Eprint {http://arxiv.org/abs/2210.10060} {arXiv:2210.10060 [astro-ph.CO]} \BibitemShut {NoStop}%
\bibitem [{\citenamefont {Bird}\ \emph {et~al.}(2022)\citenamefont {Bird}, \citenamefont {Ni}, \citenamefont {Di~Matteo}, \citenamefont {Croft}, \citenamefont {Feng},\ and\ \citenamefont {Chen}}]{Bird:2022ulj}%
  \BibitemOpen
  \bibfield  {author} {\bibinfo {author} {\bibfnamefont {S.}~\bibnamefont {Bird}}, \bibinfo {author} {\bibfnamefont {Y.}~\bibnamefont {Ni}}, \bibinfo {author} {\bibfnamefont {T.}~\bibnamefont {Di~Matteo}}, \bibinfo {author} {\bibfnamefont {R.}~\bibnamefont {Croft}}, \bibinfo {author} {\bibfnamefont {Y.}~\bibnamefont {Feng}}, \ and\ \bibinfo {author} {\bibfnamefont {N.}~\bibnamefont {Chen}},\ }\href {\doibase 10.1093/mnras/stac648} {\bibfield  {journal} {\bibinfo  {journal} {Mon. Not. Roy. Astron. Soc.}\ }\textbf {\bibinfo {volume} {512}},\ \bibinfo {pages} {3703} (\bibinfo {year} {2022})},\ \Eprint {http://arxiv.org/abs/2111.01160} {arXiv:2111.01160} \BibitemShut {NoStop}%
\bibitem [{\citenamefont {{Ni}}\ \emph {et~al.}(2022)\citenamefont {{Ni}}, \citenamefont {{Di Matteo}}, \citenamefont {{Bird}}, \citenamefont {{Croft}}, \citenamefont {{Feng}}, \citenamefont {{Chen}}, \citenamefont {{Tremmel}}, \citenamefont {{DeGraf}},\ and\ \citenamefont {{Li}}}]{Ni2022}%
  \BibitemOpen
  \bibfield  {author} {\bibinfo {author} {\bibfnamefont {Y.}~\bibnamefont {{Ni}}}, \bibinfo {author} {\bibfnamefont {T.}~\bibnamefont {{Di Matteo}}}, \bibinfo {author} {\bibfnamefont {S.}~\bibnamefont {{Bird}}}, \bibinfo {author} {\bibfnamefont {R.}~\bibnamefont {{Croft}}}, \bibinfo {author} {\bibfnamefont {Y.}~\bibnamefont {{Feng}}}, \bibinfo {author} {\bibfnamefont {N.}~\bibnamefont {{Chen}}}, \bibinfo {author} {\bibfnamefont {M.}~\bibnamefont {{Tremmel}}}, \bibinfo {author} {\bibfnamefont {C.}~\bibnamefont {{DeGraf}}}, \ and\ \bibinfo {author} {\bibfnamefont {Y.}~\bibnamefont {{Li}}},\ }\href {\doibase 10.1093/mnras/stac351} {\bibfield  {journal} {\bibinfo  {journal} {\mnras}\ }\textbf {\bibinfo {volume} {513}},\ \bibinfo {pages} {670} (\bibinfo {year} {2022})},\ \Eprint {http://arxiv.org/abs/2110.14154} {arXiv:2110.14154 [astro-ph.GA]} \BibitemShut {NoStop}%
\bibitem [{\citenamefont {Hern\'andez-Aguayo}\ \emph {et~al.}(2023)\citenamefont {Hern\'andez-Aguayo} \emph {et~al.}}]{Hernandez-Aguayo:2022xcl}%
  \BibitemOpen
  \bibfield  {author} {\bibinfo {author} {\bibfnamefont {C.}~\bibnamefont {Hern\'andez-Aguayo}} \emph {et~al.},\ }\href {\doibase 10.1093/mnras/stad1657} {\bibfield  {journal} {\bibinfo  {journal} {Mon. Not. Roy. Astron. Soc.}\ }\textbf {\bibinfo {volume} {524}},\ \bibinfo {pages} {2556} (\bibinfo {year} {2023})},\ \Eprint {http://arxiv.org/abs/2210.10059} {arXiv:2210.10059 [astro-ph.CO]} \BibitemShut {NoStop}%
\bibitem [{\citenamefont {{Ni}}\ \emph {et~al.}(2024)\citenamefont {{Ni}}, \citenamefont {{Chen}}, \citenamefont {{Zhou}}, \citenamefont {{Park}}, \citenamefont {{Yang}}, \citenamefont {{DiMatteo}}, \citenamefont {{Bird}},\ and\ \citenamefont {{Croft}}}]{Ni2024:Astrid2}%
  \BibitemOpen
  \bibfield  {author} {\bibinfo {author} {\bibfnamefont {Y.}~\bibnamefont {{Ni}}}, \bibinfo {author} {\bibfnamefont {N.}~\bibnamefont {{Chen}}}, \bibinfo {author} {\bibfnamefont {Y.}~\bibnamefont {{Zhou}}}, \bibinfo {author} {\bibfnamefont {M.}~\bibnamefont {{Park}}}, \bibinfo {author} {\bibfnamefont {Y.}~\bibnamefont {{Yang}}}, \bibinfo {author} {\bibfnamefont {T.}~\bibnamefont {{DiMatteo}}}, \bibinfo {author} {\bibfnamefont {S.}~\bibnamefont {{Bird}}}, \ and\ \bibinfo {author} {\bibfnamefont {R.}~\bibnamefont {{Croft}}},\ }\href {\doibase 10.48550/arXiv.2409.10666} {\bibfield  {journal} {\bibinfo  {journal} {arXiv e-prints}\ ,\ \bibinfo {eid} {arXiv:2409.10666}} (\bibinfo {year} {2024})},\ \Eprint {http://arxiv.org/abs/2409.10666} {arXiv:2409.10666 [astro-ph.GA]} \BibitemShut {NoStop}%
\bibitem [{\citenamefont {{McQuinn}}\ \emph {et~al.}(2007)\citenamefont {{McQuinn}}, \citenamefont {{Hernquist}}, \citenamefont {{Zaldarriaga}},\ and\ \citenamefont {{Dutta}}}]{mcquinn_lae_07}%
  \BibitemOpen
  \bibfield  {author} {\bibinfo {author} {\bibfnamefont {M.}~\bibnamefont {{McQuinn}}}, \bibinfo {author} {\bibfnamefont {L.}~\bibnamefont {{Hernquist}}}, \bibinfo {author} {\bibfnamefont {M.}~\bibnamefont {{Zaldarriaga}}}, \ and\ \bibinfo {author} {\bibfnamefont {S.}~\bibnamefont {{Dutta}}},\ }\href {\doibase 10.1111/j.1365-2966.2007.12085.x} {\bibfield  {journal} {\bibinfo  {journal} {\mnras}\ }\textbf {\bibinfo {volume} {381}},\ \bibinfo {pages} {75} (\bibinfo {year} {2007})},\ \Eprint {http://arxiv.org/abs/0704.2239} {arXiv:0704.2239 [astro-ph]} \BibitemShut {NoStop}%
\bibitem [{\citenamefont {{Ouchi}}\ \emph {et~al.}(2010)\citenamefont {{Ouchi}}, \citenamefont {{Shimasaku}}, \citenamefont {{Furusawa}}, \citenamefont {{Saito}}, \citenamefont {{Yoshida}}, \citenamefont {{Akiyama}}, \citenamefont {{Ono}}, \citenamefont {{Yamada}}, \citenamefont {{Ota}}, \citenamefont {{Kashikawa}}, \citenamefont {{Iye}}, \citenamefont {{Kodama}}, \citenamefont {{Okamura}}, \citenamefont {{Simpson}},\ and\ \citenamefont {{Yoshida}}}]{Ouchi2010:highzlae_7}%
  \BibitemOpen
  \bibfield  {author} {\bibinfo {author} {\bibfnamefont {M.}~\bibnamefont {{Ouchi}}}, \bibinfo {author} {\bibfnamefont {K.}~\bibnamefont {{Shimasaku}}}, \bibinfo {author} {\bibfnamefont {H.}~\bibnamefont {{Furusawa}}}, \bibinfo {author} {\bibfnamefont {T.}~\bibnamefont {{Saito}}}, \bibinfo {author} {\bibfnamefont {M.}~\bibnamefont {{Yoshida}}}, \bibinfo {author} {\bibfnamefont {M.}~\bibnamefont {{Akiyama}}}, \bibinfo {author} {\bibfnamefont {Y.}~\bibnamefont {{Ono}}}, \bibinfo {author} {\bibfnamefont {T.}~\bibnamefont {{Yamada}}}, \bibinfo {author} {\bibfnamefont {K.}~\bibnamefont {{Ota}}}, \bibinfo {author} {\bibfnamefont {N.}~\bibnamefont {{Kashikawa}}}, \bibinfo {author} {\bibfnamefont {M.}~\bibnamefont {{Iye}}}, \bibinfo {author} {\bibfnamefont {T.}~\bibnamefont {{Kodama}}}, \bibinfo {author} {\bibfnamefont {S.}~\bibnamefont {{Okamura}}}, \bibinfo {author} {\bibfnamefont {C.}~\bibnamefont {{Simpson}}}, \ and\ \bibinfo {author} {\bibfnamefont {M.}~\bibnamefont {{Yoshida}}},\ }\href {\doibase
  10.1088/0004-637X/723/1/869} {\bibfield  {journal} {\bibinfo  {journal} {\apj}\ }\textbf {\bibinfo {volume} {723}},\ \bibinfo {pages} {869} (\bibinfo {year} {2010})},\ \Eprint {http://arxiv.org/abs/1007.2961} {arXiv:1007.2961 [astro-ph.CO]} \BibitemShut {NoStop}%
\bibitem [{\citenamefont {{Konno}}\ \emph {et~al.}(2014)\citenamefont {{Konno}}, \citenamefont {{Ouchi}}, \citenamefont {{Ono}}, \citenamefont {{Shimasaku}}, \citenamefont {{Shibuya}}, \citenamefont {{Furusawa}}, \citenamefont {{Nakajima}}, \citenamefont {{Naito}}, \citenamefont {{Momose}}, \citenamefont {{Yuma}},\ and\ \citenamefont {{Iye}}}]{Konno2014:highz_lae}%
  \BibitemOpen
  \bibfield  {author} {\bibinfo {author} {\bibfnamefont {A.}~\bibnamefont {{Konno}}}, \bibinfo {author} {\bibfnamefont {M.}~\bibnamefont {{Ouchi}}}, \bibinfo {author} {\bibfnamefont {Y.}~\bibnamefont {{Ono}}}, \bibinfo {author} {\bibfnamefont {K.}~\bibnamefont {{Shimasaku}}}, \bibinfo {author} {\bibfnamefont {T.}~\bibnamefont {{Shibuya}}}, \bibinfo {author} {\bibfnamefont {H.}~\bibnamefont {{Furusawa}}}, \bibinfo {author} {\bibfnamefont {K.}~\bibnamefont {{Nakajima}}}, \bibinfo {author} {\bibfnamefont {Y.}~\bibnamefont {{Naito}}}, \bibinfo {author} {\bibfnamefont {R.}~\bibnamefont {{Momose}}}, \bibinfo {author} {\bibfnamefont {S.}~\bibnamefont {{Yuma}}}, \ and\ \bibinfo {author} {\bibfnamefont {M.}~\bibnamefont {{Iye}}},\ }\href {\doibase 10.1088/0004-637X/797/1/16} {\bibfield  {journal} {\bibinfo  {journal} {\apj}\ }\textbf {\bibinfo {volume} {797}},\ \bibinfo {eid} {16} (\bibinfo {year} {2014})},\ \Eprint {http://arxiv.org/abs/1404.6066} {arXiv:1404.6066 [astro-ph.CO]} \BibitemShut {NoStop}%
\bibitem [{\citenamefont {{Ouchi}}\ \emph {et~al.}(2018)\citenamefont {{Ouchi}}, \citenamefont {{Harikane}}, \citenamefont {{Shibuya}}, \citenamefont {{Shimasaku}}, \citenamefont {{Taniguchi}}, \citenamefont {{Konno}}, \citenamefont {{Kobayashi}}, \citenamefont {{Kajisawa}}, \citenamefont {{Nagao}}, \citenamefont {{Ono}}, \citenamefont {{Inoue}}, \citenamefont {{Umemura}}, \citenamefont {{Mori}}, \citenamefont {{Hasegawa}}, \citenamefont {{Higuchi}}, \citenamefont {{Komiyama}}, \citenamefont {{Matsuda}}, \citenamefont {{Nakajima}}, \citenamefont {{Saito}},\ and\ \citenamefont {{Wang}}}]{Ouchi2018:silverrush_highz}%
  \BibitemOpen
  \bibfield  {author} {\bibinfo {author} {\bibfnamefont {M.}~\bibnamefont {{Ouchi}}}, \bibinfo {author} {\bibfnamefont {Y.}~\bibnamefont {{Harikane}}}, \bibinfo {author} {\bibfnamefont {T.}~\bibnamefont {{Shibuya}}}, \bibinfo {author} {\bibfnamefont {K.}~\bibnamefont {{Shimasaku}}}, \bibinfo {author} {\bibfnamefont {Y.}~\bibnamefont {{Taniguchi}}}, \bibinfo {author} {\bibfnamefont {A.}~\bibnamefont {{Konno}}}, \bibinfo {author} {\bibfnamefont {M.}~\bibnamefont {{Kobayashi}}}, \bibinfo {author} {\bibfnamefont {M.}~\bibnamefont {{Kajisawa}}}, \bibinfo {author} {\bibfnamefont {T.}~\bibnamefont {{Nagao}}}, \bibinfo {author} {\bibfnamefont {Y.}~\bibnamefont {{Ono}}}, \bibinfo {author} {\bibfnamefont {A.~K.}\ \bibnamefont {{Inoue}}}, \bibinfo {author} {\bibfnamefont {M.}~\bibnamefont {{Umemura}}}, \bibinfo {author} {\bibfnamefont {M.}~\bibnamefont {{Mori}}}, \bibinfo {author} {\bibfnamefont {K.}~\bibnamefont {{Hasegawa}}}, \bibinfo {author} {\bibfnamefont {R.}~\bibnamefont {{Higuchi}}}, \bibinfo {author}
  {\bibfnamefont {Y.}~\bibnamefont {{Komiyama}}}, \bibinfo {author} {\bibfnamefont {Y.}~\bibnamefont {{Matsuda}}}, \bibinfo {author} {\bibfnamefont {K.}~\bibnamefont {{Nakajima}}}, \bibinfo {author} {\bibfnamefont {T.}~\bibnamefont {{Saito}}}, \ and\ \bibinfo {author} {\bibfnamefont {S.-Y.}\ \bibnamefont {{Wang}}},\ }\href {\doibase 10.1093/pasj/psx074} {\bibfield  {journal} {\bibinfo  {journal} {\pasj}\ }\textbf {\bibinfo {volume} {70}},\ \bibinfo {eid} {S13} (\bibinfo {year} {2018})},\ \Eprint {http://arxiv.org/abs/1704.07455} {arXiv:1704.07455 [astro-ph.GA]} \BibitemShut {NoStop}%
\bibitem [{\citenamefont {{Inoue}}\ \emph {et~al.}(2018)\citenamefont {{Inoue}}, \citenamefont {{Hasegawa}}, \citenamefont {{Ishiyama}}, \citenamefont {{Yajima}}, \citenamefont {{Shimizu}}, \citenamefont {{Umemura}}, \citenamefont {{Konno}}, \citenamefont {{Harikane}}, \citenamefont {{Shibuya}}, \citenamefont {{Ouchi}}, \citenamefont {{Shimasaku}}, \citenamefont {{Ono}}, \citenamefont {{Kusakabe}}, \citenamefont {{Higuchi}},\ and\ \citenamefont {{Lee}}}]{Inoue2018:silverrush_sims_highz}%
  \BibitemOpen
  \bibfield  {author} {\bibinfo {author} {\bibfnamefont {A.~K.}\ \bibnamefont {{Inoue}}}, \bibinfo {author} {\bibfnamefont {K.}~\bibnamefont {{Hasegawa}}}, \bibinfo {author} {\bibfnamefont {T.}~\bibnamefont {{Ishiyama}}}, \bibinfo {author} {\bibfnamefont {H.}~\bibnamefont {{Yajima}}}, \bibinfo {author} {\bibfnamefont {I.}~\bibnamefont {{Shimizu}}}, \bibinfo {author} {\bibfnamefont {M.}~\bibnamefont {{Umemura}}}, \bibinfo {author} {\bibfnamefont {A.}~\bibnamefont {{Konno}}}, \bibinfo {author} {\bibfnamefont {Y.}~\bibnamefont {{Harikane}}}, \bibinfo {author} {\bibfnamefont {T.}~\bibnamefont {{Shibuya}}}, \bibinfo {author} {\bibfnamefont {M.}~\bibnamefont {{Ouchi}}}, \bibinfo {author} {\bibfnamefont {K.}~\bibnamefont {{Shimasaku}}}, \bibinfo {author} {\bibfnamefont {Y.}~\bibnamefont {{Ono}}}, \bibinfo {author} {\bibfnamefont {H.}~\bibnamefont {{Kusakabe}}}, \bibinfo {author} {\bibfnamefont {R.}~\bibnamefont {{Higuchi}}}, \ and\ \bibinfo {author} {\bibfnamefont {C.-H.}\ \bibnamefont {{Lee}}},\ }\href {\doibase
  10.1093/pasj/psy048} {\bibfield  {journal} {\bibinfo  {journal} {\pasj}\ }\textbf {\bibinfo {volume} {70}},\ \bibinfo {eid} {55} (\bibinfo {year} {2018})},\ \Eprint {http://arxiv.org/abs/1801.00067} {arXiv:1801.00067 [astro-ph.GA]} \BibitemShut {NoStop}%
\bibitem [{\citenamefont {{Ning}}\ \emph {et~al.}(2022)\citenamefont {{Ning}}, \citenamefont {{Jiang}}, \citenamefont {{Zheng}},\ and\ \citenamefont {{Wu}}}]{Ning2022:magellan_highz_lae}%
  \BibitemOpen
  \bibfield  {author} {\bibinfo {author} {\bibfnamefont {Y.}~\bibnamefont {{Ning}}}, \bibinfo {author} {\bibfnamefont {L.}~\bibnamefont {{Jiang}}}, \bibinfo {author} {\bibfnamefont {Z.-Y.}\ \bibnamefont {{Zheng}}}, \ and\ \bibinfo {author} {\bibfnamefont {J.}~\bibnamefont {{Wu}}},\ }\href {\doibase 10.3847/1538-4357/ac4268} {\bibfield  {journal} {\bibinfo  {journal} {\apj}\ }\textbf {\bibinfo {volume} {926}},\ \bibinfo {eid} {230} (\bibinfo {year} {2022})},\ \Eprint {http://arxiv.org/abs/2112.07800} {arXiv:2112.07800 [astro-ph.GA]} \BibitemShut {NoStop}%
\bibitem [{\citenamefont {{Umeda}}\ \emph {et~al.}(2025)\citenamefont {{Umeda}}, \citenamefont {{Ouchi}}, \citenamefont {{Kageura}}, \citenamefont {{Harikane}}, \citenamefont {{Nakane}}, \citenamefont {{Thai}},\ and\ \citenamefont {{Nakajima}}}]{Umeda2025highzlae}%
  \BibitemOpen
  \bibfield  {author} {\bibinfo {author} {\bibfnamefont {H.}~\bibnamefont {{Umeda}}}, \bibinfo {author} {\bibfnamefont {M.}~\bibnamefont {{Ouchi}}}, \bibinfo {author} {\bibfnamefont {Y.}~\bibnamefont {{Kageura}}}, \bibinfo {author} {\bibfnamefont {Y.}~\bibnamefont {{Harikane}}}, \bibinfo {author} {\bibfnamefont {M.}~\bibnamefont {{Nakane}}}, \bibinfo {author} {\bibfnamefont {T.~T.}\ \bibnamefont {{Thai}}}, \ and\ \bibinfo {author} {\bibfnamefont {K.}~\bibnamefont {{Nakajima}}},\ }\href {\doibase 10.48550/arXiv.2504.04683} {\bibfield  {journal} {\bibinfo  {journal} {arXiv e-prints}\ ,\ \bibinfo {eid} {arXiv:2504.04683}} (\bibinfo {year} {2025})},\ \Eprint {http://arxiv.org/abs/2504.04683} {arXiv:2504.04683 [astro-ph.GA]} \BibitemShut {NoStop}%
\bibitem [{\citenamefont {{Runnholm}}\ \emph {et~al.}(2025)\citenamefont {{Runnholm}}, \citenamefont {{Hayes}}, \citenamefont {{Mehta}}, \citenamefont {{Malkan}}, \citenamefont {{Scarlata}}, \citenamefont {{Nedkova}}, \citenamefont {{Rafelski}}, \citenamefont {{Vulcani}}, \citenamefont {{Huberty}}, \citenamefont {{Herenz}}, \citenamefont {{Hutter}}, \citenamefont {{Bruton}}, \citenamefont {{Acharyya}}, \citenamefont {{Atek}}, \citenamefont {{Baronchelli}}, \citenamefont {{Battisti}}, \citenamefont {{Brada{\v{c}}}}, \citenamefont {{Bunker}}, \citenamefont {{Dai}}, \citenamefont {{Hannahs}}, \citenamefont {{Hasan}}, \citenamefont {{Kim}}, \citenamefont {{Leethochawalit}}, \citenamefont {{Lin}}, \citenamefont {{Rutkowski}}, \citenamefont {{Saldana-Lopez}}, \citenamefont {{Sattari}},\ and\ \citenamefont {{Wang}}}]{Runnholm2025:jwst_highz_lae}%
  \BibitemOpen
  \bibfield  {author} {\bibinfo {author} {\bibfnamefont {A.}~\bibnamefont {{Runnholm}}}, \bibinfo {author} {\bibfnamefont {M.~J.}\ \bibnamefont {{Hayes}}}, \bibinfo {author} {\bibfnamefont {V.}~\bibnamefont {{Mehta}}}, \bibinfo {author} {\bibfnamefont {M.~A.}\ \bibnamefont {{Malkan}}}, \bibinfo {author} {\bibfnamefont {C.}~\bibnamefont {{Scarlata}}}, \bibinfo {author} {\bibfnamefont {K.~V.}\ \bibnamefont {{Nedkova}}}, \bibinfo {author} {\bibfnamefont {M.}~\bibnamefont {{Rafelski}}}, \bibinfo {author} {\bibfnamefont {B.}~\bibnamefont {{Vulcani}}}, \bibinfo {author} {\bibfnamefont {M.}~\bibnamefont {{Huberty}}}, \bibinfo {author} {\bibfnamefont {E.~C.}\ \bibnamefont {{Herenz}}}, \bibinfo {author} {\bibfnamefont {A.}~\bibnamefont {{Hutter}}}, \bibinfo {author} {\bibfnamefont {S.}~\bibnamefont {{Bruton}}}, \bibinfo {author} {\bibfnamefont {A.}~\bibnamefont {{Acharyya}}}, \bibinfo {author} {\bibfnamefont {H.}~\bibnamefont {{Atek}}}, \bibinfo {author} {\bibfnamefont {I.}~\bibnamefont {{Baronchelli}}}, \bibinfo
  {author} {\bibfnamefont {A.~J.}\ \bibnamefont {{Battisti}}}, \bibinfo {author} {\bibfnamefont {M.}~\bibnamefont {{Brada{\v{c}}}}}, \bibinfo {author} {\bibfnamefont {A.~J.}\ \bibnamefont {{Bunker}}}, \bibinfo {author} {\bibfnamefont {Y.~S.}\ \bibnamefont {{Dai}}}, \bibinfo {author} {\bibfnamefont {C.}~\bibnamefont {{Hannahs}}}, \bibinfo {author} {\bibfnamefont {F.}~\bibnamefont {{Hasan}}}, \bibinfo {author} {\bibfnamefont {K.~J.}\ \bibnamefont {{Kim}}}, \bibinfo {author} {\bibfnamefont {N.}~\bibnamefont {{Leethochawalit}}}, \bibinfo {author} {\bibfnamefont {Y.-H.}\ \bibnamefont {{Lin}}}, \bibinfo {author} {\bibfnamefont {M.~J.}\ \bibnamefont {{Rutkowski}}}, \bibinfo {author} {\bibfnamefont {A.}~\bibnamefont {{Saldana-Lopez}}}, \bibinfo {author} {\bibfnamefont {Z.}~\bibnamefont {{Sattari}}}, \ and\ \bibinfo {author} {\bibfnamefont {X.}~\bibnamefont {{Wang}}},\ }\href {\doibase 10.48550/arXiv.2502.19174} {\bibfield  {journal} {\bibinfo  {journal} {arXiv e-prints}\ ,\ \bibinfo {eid} {arXiv:2502.19174}}
  (\bibinfo {year} {2025})},\ \Eprint {http://arxiv.org/abs/2502.19174} {arXiv:2502.19174 [astro-ph.GA]} \BibitemShut {NoStop}%
\bibitem [{\citenamefont {{Sobacchi}}\ and\ \citenamefont {{Mesinger}}(2015)}]{2015Sobacchi:highz_lae}%
  \BibitemOpen
  \bibfield  {author} {\bibinfo {author} {\bibfnamefont {E.}~\bibnamefont {{Sobacchi}}}\ and\ \bibinfo {author} {\bibfnamefont {A.}~\bibnamefont {{Mesinger}}},\ }\href {\doibase 10.1093/mnras/stv1751} {\bibfield  {journal} {\bibinfo  {journal} {\mnras}\ }\textbf {\bibinfo {volume} {453}},\ \bibinfo {pages} {1843} (\bibinfo {year} {2015})},\ \Eprint {http://arxiv.org/abs/1505.02787} {arXiv:1505.02787 [astro-ph.CO]} \BibitemShut {NoStop}%
\bibitem [{\citenamefont {{Nagamine}}\ \emph {et~al.}(2010)\citenamefont {{Nagamine}}, \citenamefont {{Ouchi}}, \citenamefont {{Springel}},\ and\ \citenamefont {{Hernquist}}}]{Nagamine:2010_sim_sfgal}%
  \BibitemOpen
  \bibfield  {author} {\bibinfo {author} {\bibfnamefont {K.}~\bibnamefont {{Nagamine}}}, \bibinfo {author} {\bibfnamefont {M.}~\bibnamefont {{Ouchi}}}, \bibinfo {author} {\bibfnamefont {V.}~\bibnamefont {{Springel}}}, \ and\ \bibinfo {author} {\bibfnamefont {L.}~\bibnamefont {{Hernquist}}},\ }\href {\doibase 10.1093/pasj/62.6.1455} {\bibfield  {journal} {\bibinfo  {journal} {\pasj}\ }\textbf {\bibinfo {volume} {62}},\ \bibinfo {pages} {1455} (\bibinfo {year} {2010})},\ \Eprint {http://arxiv.org/abs/0802.0228} {arXiv:0802.0228 [astro-ph]} \BibitemShut {NoStop}%
\bibitem [{\citenamefont {{Nagamine}}\ \emph {et~al.}(2004)\citenamefont {{Nagamine}}, \citenamefont {{Springel}}, \citenamefont {{Hernquist}},\ and\ \citenamefont {{Machacek}}}]{Nagamine:2004_LBG}%
  \BibitemOpen
  \bibfield  {author} {\bibinfo {author} {\bibfnamefont {K.}~\bibnamefont {{Nagamine}}}, \bibinfo {author} {\bibfnamefont {V.}~\bibnamefont {{Springel}}}, \bibinfo {author} {\bibfnamefont {L.}~\bibnamefont {{Hernquist}}}, \ and\ \bibinfo {author} {\bibfnamefont {M.}~\bibnamefont {{Machacek}}},\ }\href {\doibase 10.1111/j.1365-2966.2004.07664.x} {\bibfield  {journal} {\bibinfo  {journal} {\mnras}\ }\textbf {\bibinfo {volume} {350}},\ \bibinfo {pages} {385} (\bibinfo {year} {2004})},\ \Eprint {http://arxiv.org/abs/astro-ph/0311295} {arXiv:astro-ph/0311295 [astro-ph]} \BibitemShut {NoStop}%
\bibitem [{\citenamefont {{Night}}\ \emph {et~al.}(2006)\citenamefont {{Night}}, \citenamefont {{Nagamine}}, \citenamefont {{Springel}},\ and\ \citenamefont {{Hernquist}}}]{Night:2006_LBG}%
  \BibitemOpen
  \bibfield  {author} {\bibinfo {author} {\bibfnamefont {C.}~\bibnamefont {{Night}}}, \bibinfo {author} {\bibfnamefont {K.}~\bibnamefont {{Nagamine}}}, \bibinfo {author} {\bibfnamefont {V.}~\bibnamefont {{Springel}}}, \ and\ \bibinfo {author} {\bibfnamefont {L.}~\bibnamefont {{Hernquist}}},\ }\href {\doibase 10.1111/j.1365-2966.2005.09730.x} {\bibfield  {journal} {\bibinfo  {journal} {\mnras}\ }\textbf {\bibinfo {volume} {366}},\ \bibinfo {pages} {705} (\bibinfo {year} {2006})},\ \Eprint {http://arxiv.org/abs/astro-ph/0503631} {arXiv:astro-ph/0503631 [astro-ph]} \BibitemShut {NoStop}%
\bibitem [{\citenamefont {{Im}}\ \emph {et~al.}(2024)\citenamefont {{Im}}, \citenamefont {{Hwang}}, \citenamefont {{Park}}, \citenamefont {{Lee}}, \citenamefont {{Song}}, \citenamefont {{Appleby}}, \citenamefont {{Dubois}}, \citenamefont {{Few}}, \citenamefont {{Gibson}}, \citenamefont {{Kim}}, \citenamefont {{Kim}}, \citenamefont {{Park}}, \citenamefont {{Pichon}}, \citenamefont {{Shin}}, \citenamefont {{Snaith}}, \citenamefont {{Artale}}, \citenamefont {{Gawiser}}, \citenamefont {{Guaita}}, \citenamefont {{Jeong}}, \citenamefont {{Lee}}, \citenamefont {{Padilla}}, \citenamefont {{Ramakrishnan}}, \citenamefont {{Troncoso}},\ and\ \citenamefont {{Yang}}}]{Im:2024_sim_lbg_lae}%
  \BibitemOpen
  \bibfield  {author} {\bibinfo {author} {\bibfnamefont {S.~H.}\ \bibnamefont {{Im}}}, \bibinfo {author} {\bibfnamefont {H.~S.}\ \bibnamefont {{Hwang}}}, \bibinfo {author} {\bibfnamefont {J.}~\bibnamefont {{Park}}}, \bibinfo {author} {\bibfnamefont {J.}~\bibnamefont {{Lee}}}, \bibinfo {author} {\bibfnamefont {H.}~\bibnamefont {{Song}}}, \bibinfo {author} {\bibfnamefont {S.}~\bibnamefont {{Appleby}}}, \bibinfo {author} {\bibfnamefont {Y.}~\bibnamefont {{Dubois}}}, \bibinfo {author} {\bibfnamefont {C.~G.}\ \bibnamefont {{Few}}}, \bibinfo {author} {\bibfnamefont {B.~K.}\ \bibnamefont {{Gibson}}}, \bibinfo {author} {\bibfnamefont {J.}~\bibnamefont {{Kim}}}, \bibinfo {author} {\bibfnamefont {Y.}~\bibnamefont {{Kim}}}, \bibinfo {author} {\bibfnamefont {C.}~\bibnamefont {{Park}}}, \bibinfo {author} {\bibfnamefont {C.}~\bibnamefont {{Pichon}}}, \bibinfo {author} {\bibfnamefont {J.}~\bibnamefont {{Shin}}}, \bibinfo {author} {\bibfnamefont {O.~N.}\ \bibnamefont {{Snaith}}}, \bibinfo {author} {\bibfnamefont {M.~C.}\
  \bibnamefont {{Artale}}}, \bibinfo {author} {\bibfnamefont {E.}~\bibnamefont {{Gawiser}}}, \bibinfo {author} {\bibfnamefont {L.}~\bibnamefont {{Guaita}}}, \bibinfo {author} {\bibfnamefont {W.-S.}\ \bibnamefont {{Jeong}}}, \bibinfo {author} {\bibfnamefont {K.-S.}\ \bibnamefont {{Lee}}}, \bibinfo {author} {\bibfnamefont {N.}~\bibnamefont {{Padilla}}}, \bibinfo {author} {\bibfnamefont {V.}~\bibnamefont {{Ramakrishnan}}}, \bibinfo {author} {\bibfnamefont {P.}~\bibnamefont {{Troncoso}}}, \ and\ \bibinfo {author} {\bibfnamefont {Y.}~\bibnamefont {{Yang}}},\ }\href {\doibase 10.3847/1538-4357/ad67d2} {\bibfield  {journal} {\bibinfo  {journal} {\apj}\ }\textbf {\bibinfo {volume} {972}},\ \bibinfo {eid} {196} (\bibinfo {year} {2024})},\ \Eprint {http://arxiv.org/abs/2407.18602} {arXiv:2407.18602 [astro-ph.GA]} \BibitemShut {NoStop}%
\bibitem [{\citenamefont {{Lee}}\ \emph {et~al.}(2021)\citenamefont {{Lee}}, \citenamefont {{Shin}}, \citenamefont {{Snaith}}, \citenamefont {{Kim}}, \citenamefont {{Few}}, \citenamefont {{Devriendt}}, \citenamefont {{Dubois}}, \citenamefont {{Cox}}, \citenamefont {{Hong}}, \citenamefont {{Kwon}}, \citenamefont {{Park}}, \citenamefont {{Pichon}}, \citenamefont {{Kim}}, \citenamefont {{Gibson}},\ and\ \citenamefont {{Park}}}]{Lee:2021_horizon_5}%
  \BibitemOpen
  \bibfield  {author} {\bibinfo {author} {\bibfnamefont {J.}~\bibnamefont {{Lee}}}, \bibinfo {author} {\bibfnamefont {J.}~\bibnamefont {{Shin}}}, \bibinfo {author} {\bibfnamefont {O.~N.}\ \bibnamefont {{Snaith}}}, \bibinfo {author} {\bibfnamefont {Y.}~\bibnamefont {{Kim}}}, \bibinfo {author} {\bibfnamefont {C.~G.}\ \bibnamefont {{Few}}}, \bibinfo {author} {\bibfnamefont {J.}~\bibnamefont {{Devriendt}}}, \bibinfo {author} {\bibfnamefont {Y.}~\bibnamefont {{Dubois}}}, \bibinfo {author} {\bibfnamefont {L.~M.}\ \bibnamefont {{Cox}}}, \bibinfo {author} {\bibfnamefont {S.~E.}\ \bibnamefont {{Hong}}}, \bibinfo {author} {\bibfnamefont {O.-K.}\ \bibnamefont {{Kwon}}}, \bibinfo {author} {\bibfnamefont {C.}~\bibnamefont {{Park}}}, \bibinfo {author} {\bibfnamefont {C.}~\bibnamefont {{Pichon}}}, \bibinfo {author} {\bibfnamefont {J.}~\bibnamefont {{Kim}}}, \bibinfo {author} {\bibfnamefont {B.~K.}\ \bibnamefont {{Gibson}}}, \ and\ \bibinfo {author} {\bibfnamefont {C.}~\bibnamefont {{Park}}},\ }\href {\doibase
  10.3847/1538-4357/abd08b} {\bibfield  {journal} {\bibinfo  {journal} {\apj}\ }\textbf {\bibinfo {volume} {908}},\ \bibinfo {eid} {11} (\bibinfo {year} {2021})},\ \Eprint {http://arxiv.org/abs/2006.01039} {arXiv:2006.01039 [astro-ph.GA]} \BibitemShut {NoStop}%
\bibitem [{\citenamefont {{Ravi}}\ \emph {et~al.}(2024)\citenamefont {{Ravi}}, \citenamefont {{Hadzhiyska}}, \citenamefont {{White}}, \citenamefont {{Hernquist}},\ and\ \citenamefont {{Bose}}}]{Ravi:2024}%
  \BibitemOpen
  \bibfield  {author} {\bibinfo {author} {\bibfnamefont {J.}~\bibnamefont {{Ravi}}}, \bibinfo {author} {\bibfnamefont {B.}~\bibnamefont {{Hadzhiyska}}}, \bibinfo {author} {\bibfnamefont {M.~J.}\ \bibnamefont {{White}}}, \bibinfo {author} {\bibfnamefont {L.}~\bibnamefont {{Hernquist}}}, \ and\ \bibinfo {author} {\bibfnamefont {S.}~\bibnamefont {{Bose}}},\ }\href {\doibase 10.1103/PhysRevD.110.103509} {\bibfield  {journal} {\bibinfo  {journal} {\prd}\ }\textbf {\bibinfo {volume} {110}},\ \bibinfo {eid} {103509} (\bibinfo {year} {2024})},\ \Eprint {http://arxiv.org/abs/2403.02414} {arXiv:2403.02414 [astro-ph.CO]} \BibitemShut {NoStop}%
\bibitem [{\citenamefont {{White}}\ \emph {et~al.}(2024)\citenamefont {{White}}, \citenamefont {{Raichoor}}, \citenamefont {{Dey}}, \citenamefont {{Garrison}}, \citenamefont {{Gawiser}}, \citenamefont {{Lang}}, \citenamefont {{Lee}}, \citenamefont {{Myers}}, \citenamefont {{Schlegel}}, \citenamefont {{Valdes}}, \citenamefont {{Aguilar}}, \citenamefont {{Ahlen}}, \citenamefont {{Brooks}}, \citenamefont {{Chaussidon}}, \citenamefont {{Claybaugh}}, \citenamefont {{Dawson}}, \citenamefont {{de la Macorra}}, \citenamefont {{Dey}}, \citenamefont {{Doel}}, \citenamefont {{Fanning}}, \citenamefont {{Font-Ribera}}, \citenamefont {{Forero-Romero}}, \citenamefont {{Gontcho A Gontcho}}, \citenamefont {{Gutierrez}}, \citenamefont {{Guy}}, \citenamefont {{Honscheid}}, \citenamefont {{Kirkby}}, \citenamefont {{Kremin}}, \citenamefont {{Landriau}}, \citenamefont {{Le Guillou}}, \citenamefont {{Levi}}, \citenamefont {{Magneville}}, \citenamefont {{Manera}}, \citenamefont {{Martini}}, \citenamefont {{Meisner}}, \citenamefont
  {{Miquel}}, \citenamefont {{Moon}}, \citenamefont {{Newman}}, \citenamefont {{Niz}}, \citenamefont {{Palanque-Delabrouille}}, \citenamefont {{Park}}, \citenamefont {{Percival}}, \citenamefont {{Prada}}, \citenamefont {{Rossi}}, \citenamefont {{Ruhlmann-Kleider}}, \citenamefont {{Sanchez}}, \citenamefont {{Schlafly}}, \citenamefont {{Schubnell}}, \citenamefont {{Seo}}, \citenamefont {{Sprayberry}}, \citenamefont {{Tarl{\'e}}}, \citenamefont {{Weaver}}, \citenamefont {{Yang}}, \citenamefont {{Y{\`e}che}},\ and\ \citenamefont {{Zou}}}]{White:2024_odin}%
  \BibitemOpen
  \bibfield  {author} {\bibinfo {author} {\bibfnamefont {M.}~\bibnamefont {{White}}}, \bibinfo {author} {\bibfnamefont {A.}~\bibnamefont {{Raichoor}}}, \bibinfo {author} {\bibfnamefont {A.}~\bibnamefont {{Dey}}}, \bibinfo {author} {\bibfnamefont {L.~H.}\ \bibnamefont {{Garrison}}}, \bibinfo {author} {\bibfnamefont {E.}~\bibnamefont {{Gawiser}}}, \bibinfo {author} {\bibfnamefont {D.}~\bibnamefont {{Lang}}}, \bibinfo {author} {\bibfnamefont {K.-s.}\ \bibnamefont {{Lee}}}, \bibinfo {author} {\bibfnamefont {A.~D.}\ \bibnamefont {{Myers}}}, \bibinfo {author} {\bibfnamefont {D.}~\bibnamefont {{Schlegel}}}, \bibinfo {author} {\bibfnamefont {F.}~\bibnamefont {{Valdes}}}, \bibinfo {author} {\bibfnamefont {J.}~\bibnamefont {{Aguilar}}}, \bibinfo {author} {\bibfnamefont {S.}~\bibnamefont {{Ahlen}}}, \bibinfo {author} {\bibfnamefont {D.}~\bibnamefont {{Brooks}}}, \bibinfo {author} {\bibfnamefont {E.}~\bibnamefont {{Chaussidon}}}, \bibinfo {author} {\bibfnamefont {T.}~\bibnamefont {{Claybaugh}}}, \bibinfo {author}
  {\bibfnamefont {K.}~\bibnamefont {{Dawson}}}, \bibinfo {author} {\bibfnamefont {A.}~\bibnamefont {{de la Macorra}}}, \bibinfo {author} {\bibfnamefont {B.}~\bibnamefont {{Dey}}}, \bibinfo {author} {\bibfnamefont {P.}~\bibnamefont {{Doel}}}, \bibinfo {author} {\bibfnamefont {K.}~\bibnamefont {{Fanning}}}, \bibinfo {author} {\bibfnamefont {A.}~\bibnamefont {{Font-Ribera}}}, \bibinfo {author} {\bibfnamefont {J.~E.}\ \bibnamefont {{Forero-Romero}}}, \bibinfo {author} {\bibfnamefont {S.}~\bibnamefont {{Gontcho A Gontcho}}}, \bibinfo {author} {\bibfnamefont {G.}~\bibnamefont {{Gutierrez}}}, \bibinfo {author} {\bibfnamefont {J.}~\bibnamefont {{Guy}}}, \bibinfo {author} {\bibfnamefont {K.}~\bibnamefont {{Honscheid}}}, \bibinfo {author} {\bibfnamefont {D.}~\bibnamefont {{Kirkby}}}, \bibinfo {author} {\bibfnamefont {A.}~\bibnamefont {{Kremin}}}, \bibinfo {author} {\bibfnamefont {M.}~\bibnamefont {{Landriau}}}, \bibinfo {author} {\bibfnamefont {L.}~\bibnamefont {{Le Guillou}}}, \bibinfo {author} {\bibfnamefont {M.~E.}\
  \bibnamefont {{Levi}}}, \bibinfo {author} {\bibfnamefont {C.}~\bibnamefont {{Magneville}}}, \bibinfo {author} {\bibfnamefont {M.}~\bibnamefont {{Manera}}}, \bibinfo {author} {\bibfnamefont {P.}~\bibnamefont {{Martini}}}, \bibinfo {author} {\bibfnamefont {A.}~\bibnamefont {{Meisner}}}, \bibinfo {author} {\bibfnamefont {R.}~\bibnamefont {{Miquel}}}, \bibinfo {author} {\bibfnamefont {B.}~\bibnamefont {{Moon}}}, \bibinfo {author} {\bibfnamefont {J.~A.}\ \bibnamefont {{Newman}}}, \bibinfo {author} {\bibfnamefont {G.}~\bibnamefont {{Niz}}}, \bibinfo {author} {\bibfnamefont {N.}~\bibnamefont {{Palanque-Delabrouille}}}, \bibinfo {author} {\bibfnamefont {C.}~\bibnamefont {{Park}}}, \bibinfo {author} {\bibfnamefont {W.~J.}\ \bibnamefont {{Percival}}}, \bibinfo {author} {\bibfnamefont {F.}~\bibnamefont {{Prada}}}, \bibinfo {author} {\bibfnamefont {G.}~\bibnamefont {{Rossi}}}, \bibinfo {author} {\bibfnamefont {V.}~\bibnamefont {{Ruhlmann-Kleider}}}, \bibinfo {author} {\bibfnamefont {E.}~\bibnamefont {{Sanchez}}},
  \bibinfo {author} {\bibfnamefont {E.~F.}\ \bibnamefont {{Schlafly}}}, \bibinfo {author} {\bibfnamefont {M.}~\bibnamefont {{Schubnell}}}, \bibinfo {author} {\bibfnamefont {H.}~\bibnamefont {{Seo}}}, \bibinfo {author} {\bibfnamefont {D.}~\bibnamefont {{Sprayberry}}}, \bibinfo {author} {\bibfnamefont {G.}~\bibnamefont {{Tarl{\'e}}}}, \bibinfo {author} {\bibfnamefont {B.~A.}\ \bibnamefont {{Weaver}}}, \bibinfo {author} {\bibfnamefont {Y.}~\bibnamefont {{Yang}}}, \bibinfo {author} {\bibfnamefont {C.}~\bibnamefont {{Y{\`e}che}}}, \ and\ \bibinfo {author} {\bibfnamefont {H.}~\bibnamefont {{Zou}}},\ }\href {\doibase 10.1088/1475-7516/2024/08/020} {\bibfield  {journal} {\bibinfo  {journal} {\jcap}\ }\textbf {\bibinfo {volume} {2024}},\ \bibinfo {eid} {020} (\bibinfo {year} {2024})},\ \Eprint {http://arxiv.org/abs/2406.01803} {arXiv:2406.01803 [astro-ph.CO]} \BibitemShut {NoStop}%
\bibitem [{\citenamefont {Zheng}\ \emph {et~al.}(2007)\citenamefont {Zheng}, \citenamefont {Coil},\ and\ \citenamefont {Zehavi}}]{Zheng:2007zg}%
  \BibitemOpen
  \bibfield  {author} {\bibinfo {author} {\bibfnamefont {Z.}~\bibnamefont {Zheng}}, \bibinfo {author} {\bibfnamefont {A.~L.}\ \bibnamefont {Coil}}, \ and\ \bibinfo {author} {\bibfnamefont {I.}~\bibnamefont {Zehavi}},\ }\href {\doibase 10.1086/521074} {\bibfield  {journal} {\bibinfo  {journal} {Astrophys. J.}\ }\textbf {\bibinfo {volume} {667}},\ \bibinfo {pages} {760} (\bibinfo {year} {2007})},\ \Eprint {http://arxiv.org/abs/astro-ph/0703457} {arXiv:astro-ph/0703457} \BibitemShut {NoStop}%
\bibitem [{\citenamefont {Smith}\ \emph {et~al.}(2022)\citenamefont {Smith}, \citenamefont {Kannan}, \citenamefont {Garaldi}, \citenamefont {Vogelsberger}, \citenamefont {Pakmor}, \citenamefont {Springel},\ and\ \citenamefont {Hernquist}}]{Smith:2021hqg}%
  \BibitemOpen
  \bibfield  {author} {\bibinfo {author} {\bibfnamefont {A.}~\bibnamefont {Smith}}, \bibinfo {author} {\bibfnamefont {R.}~\bibnamefont {Kannan}}, \bibinfo {author} {\bibfnamefont {E.}~\bibnamefont {Garaldi}}, \bibinfo {author} {\bibfnamefont {M.}~\bibnamefont {Vogelsberger}}, \bibinfo {author} {\bibfnamefont {R.}~\bibnamefont {Pakmor}}, \bibinfo {author} {\bibfnamefont {V.}~\bibnamefont {Springel}}, \ and\ \bibinfo {author} {\bibfnamefont {L.}~\bibnamefont {Hernquist}},\ }\href {\doibase 10.1093/mnras/stac713} {\bibfield  {journal} {\bibinfo  {journal} {Mon. Not. Roy. Astron. Soc.}\ }\textbf {\bibinfo {volume} {512}},\ \bibinfo {pages} {3243} (\bibinfo {year} {2022})},\ \Eprint {http://arxiv.org/abs/2110.02966} {arXiv:2110.02966 [astro-ph.CO]} \BibitemShut {NoStop}%
\bibitem [{\citenamefont {{Nelson}}\ \emph {et~al.}(2019)\citenamefont {{Nelson}}, \citenamefont {{Springel}}, \citenamefont {{Pillepich}}, \citenamefont {{Rodriguez-Gomez}}, \citenamefont {{Torrey}}, \citenamefont {{Genel}}, \citenamefont {{Vogelsberger}}, \citenamefont {{Pakmor}}, \citenamefont {{Marinacci}}, \citenamefont {{Weinberger}}, \citenamefont {{Kelley}}, \citenamefont {{Lovell}}, \citenamefont {{Diemer}},\ and\ \citenamefont {{Hernquist}}}]{Nelson_TNG}%
  \BibitemOpen
  \bibfield  {author} {\bibinfo {author} {\bibfnamefont {D.}~\bibnamefont {{Nelson}}}, \bibinfo {author} {\bibfnamefont {V.}~\bibnamefont {{Springel}}}, \bibinfo {author} {\bibfnamefont {A.}~\bibnamefont {{Pillepich}}}, \bibinfo {author} {\bibfnamefont {V.}~\bibnamefont {{Rodriguez-Gomez}}}, \bibinfo {author} {\bibfnamefont {P.}~\bibnamefont {{Torrey}}}, \bibinfo {author} {\bibfnamefont {S.}~\bibnamefont {{Genel}}}, \bibinfo {author} {\bibfnamefont {M.}~\bibnamefont {{Vogelsberger}}}, \bibinfo {author} {\bibfnamefont {R.}~\bibnamefont {{Pakmor}}}, \bibinfo {author} {\bibfnamefont {F.}~\bibnamefont {{Marinacci}}}, \bibinfo {author} {\bibfnamefont {R.}~\bibnamefont {{Weinberger}}}, \bibinfo {author} {\bibfnamefont {L.}~\bibnamefont {{Kelley}}}, \bibinfo {author} {\bibfnamefont {M.}~\bibnamefont {{Lovell}}}, \bibinfo {author} {\bibfnamefont {B.}~\bibnamefont {{Diemer}}}, \ and\ \bibinfo {author} {\bibfnamefont {L.}~\bibnamefont {{Hernquist}}},\ }\href {\doibase 10.1186/s40668-019-0028-x} {\bibfield  {journal}
  {\bibinfo  {journal} {Computational Astrophysics and Cosmology}\ }\textbf {\bibinfo {volume} {6}},\ \bibinfo {eid} {2} (\bibinfo {year} {2019})},\ \Eprint {http://arxiv.org/abs/1812.05609} {arXiv:1812.05609 [astro-ph.GA]} \BibitemShut {NoStop}%
\bibitem [{\citenamefont {{Kerutt}}\ \emph {et~al.}(2022)\citenamefont {{Kerutt}}, \citenamefont {{Wisotzki}}, \citenamefont {{Verhamme}}, \citenamefont {{Schmidt}}, \citenamefont {{Leclercq}}, \citenamefont {{Herenz}}, \citenamefont {{Urrutia}}, \citenamefont {{Garel}}, \citenamefont {{Hashimoto}}, \citenamefont {{Maseda}}, \citenamefont {{Matthee}}, \citenamefont {{Kusakabe}}, \citenamefont {{Schaye}}, \citenamefont {{Richard}}, \citenamefont {{Guiderdoni}}, \citenamefont {{Mauerhofer}}, \citenamefont {{Nanayakkara}},\ and\ \citenamefont {{Vitte}}}]{Kerutt2022:EW_metallicity_LAE}%
  \BibitemOpen
  \bibfield  {author} {\bibinfo {author} {\bibfnamefont {J.}~\bibnamefont {{Kerutt}}}, \bibinfo {author} {\bibfnamefont {L.}~\bibnamefont {{Wisotzki}}}, \bibinfo {author} {\bibfnamefont {A.}~\bibnamefont {{Verhamme}}}, \bibinfo {author} {\bibfnamefont {K.~B.}\ \bibnamefont {{Schmidt}}}, \bibinfo {author} {\bibfnamefont {F.}~\bibnamefont {{Leclercq}}}, \bibinfo {author} {\bibfnamefont {E.~C.}\ \bibnamefont {{Herenz}}}, \bibinfo {author} {\bibfnamefont {T.}~\bibnamefont {{Urrutia}}}, \bibinfo {author} {\bibfnamefont {T.}~\bibnamefont {{Garel}}}, \bibinfo {author} {\bibfnamefont {T.}~\bibnamefont {{Hashimoto}}}, \bibinfo {author} {\bibfnamefont {M.}~\bibnamefont {{Maseda}}}, \bibinfo {author} {\bibfnamefont {J.}~\bibnamefont {{Matthee}}}, \bibinfo {author} {\bibfnamefont {H.}~\bibnamefont {{Kusakabe}}}, \bibinfo {author} {\bibfnamefont {J.}~\bibnamefont {{Schaye}}}, \bibinfo {author} {\bibfnamefont {J.}~\bibnamefont {{Richard}}}, \bibinfo {author} {\bibfnamefont {B.}~\bibnamefont {{Guiderdoni}}}, \bibinfo
  {author} {\bibfnamefont {V.}~\bibnamefont {{Mauerhofer}}}, \bibinfo {author} {\bibfnamefont {T.}~\bibnamefont {{Nanayakkara}}}, \ and\ \bibinfo {author} {\bibfnamefont {E.}~\bibnamefont {{Vitte}}},\ }\href {\doibase 10.1051/0004-6361/202141900} {\bibfield  {journal} {\bibinfo  {journal} {\aap}\ }\textbf {\bibinfo {volume} {659}},\ \bibinfo {eid} {A183} (\bibinfo {year} {2022})},\ \Eprint {http://arxiv.org/abs/2202.06642} {arXiv:2202.06642 [astro-ph.GA]} \BibitemShut {NoStop}%
\bibitem [{\citenamefont {{Dijkstra}}\ and\ \citenamefont {{Westra}}(2010)}]{Dijkstra:2010}%
  \BibitemOpen
  \bibfield  {author} {\bibinfo {author} {\bibfnamefont {M.}~\bibnamefont {{Dijkstra}}}\ and\ \bibinfo {author} {\bibfnamefont {E.}~\bibnamefont {{Westra}}},\ }\href {\doibase 10.1111/j.1365-2966.2009.15859.x} {\bibfield  {journal} {\bibinfo  {journal} {\mnras}\ }\textbf {\bibinfo {volume} {401}},\ \bibinfo {pages} {2343} (\bibinfo {year} {2010})},\ \Eprint {http://arxiv.org/abs/0911.1357} {arXiv:0911.1357 [astro-ph.CO]} \BibitemShut {NoStop}%
\bibitem [{\citenamefont {{Kennicutt}}(1998)}]{Kennnicut:1998}%
  \BibitemOpen
  \bibfield  {author} {\bibinfo {author} {\bibfnamefont {R.~C.}\ \bibnamefont {{Kennicutt}}, \bibfnamefont {Jr.}},\ }\href {\doibase 10.1146/annurev.astro.36.1.189} {\bibfield  {journal} {\bibinfo  {journal} {\araa}\ }\textbf {\bibinfo {volume} {36}},\ \bibinfo {pages} {189} (\bibinfo {year} {1998})},\ \Eprint {http://arxiv.org/abs/astro-ph/9807187} {arXiv:astro-ph/9807187 [astro-ph]} \BibitemShut {NoStop}%
\bibitem [{\citenamefont {{Salpeter}}(1955)}]{Salpeter:1955}%
  \BibitemOpen
  \bibfield  {author} {\bibinfo {author} {\bibfnamefont {E.~E.}\ \bibnamefont {{Salpeter}}},\ }\href {\doibase 10.1086/145971} {\bibfield  {journal} {\bibinfo  {journal} {\apj}\ }\textbf {\bibinfo {volume} {121}},\ \bibinfo {pages} {161} (\bibinfo {year} {1955})}\BibitemShut {NoStop}%
\bibitem [{\citenamefont {{Schaerer}}(2003)}]{Schaerer:2003}%
  \BibitemOpen
  \bibfield  {author} {\bibinfo {author} {\bibfnamefont {D.}~\bibnamefont {{Schaerer}}},\ }\href {\doibase 10.1051/0004-6361:20021525} {\bibfield  {journal} {\bibinfo  {journal} {\aap}\ }\textbf {\bibinfo {volume} {397}},\ \bibinfo {pages} {527} (\bibinfo {year} {2003})},\ \Eprint {http://arxiv.org/abs/astro-ph/0210462} {arXiv:astro-ph/0210462 [astro-ph]} \BibitemShut {NoStop}%
\bibitem [{\citenamefont {{Dijkstra}}(2017)}]{Dijkstra:2017_notes}%
  \BibitemOpen
  \bibfield  {author} {\bibinfo {author} {\bibfnamefont {M.}~\bibnamefont {{Dijkstra}}},\ }\href {\doibase 10.48550/arXiv.1704.03416} {\bibfield  {journal} {\bibinfo  {journal} {arXiv e-prints}\ ,\ \bibinfo {eid} {arXiv:1704.03416}} (\bibinfo {year} {2017})},\ \Eprint {http://arxiv.org/abs/1704.03416} {arXiv:1704.03416 [astro-ph.GA]} \BibitemShut {NoStop}%
\bibitem [{\citenamefont {{Finkelstein}}\ \emph {et~al.}(2011)\citenamefont {{Finkelstein}}, \citenamefont {{Hill}}, \citenamefont {{Gebhardt}}, \citenamefont {{Adams}}, \citenamefont {{Blanc}}, \citenamefont {{Papovich}}, \citenamefont {{Ciardullo}}, \citenamefont {{Drory}}, \citenamefont {{Gawiser}}, \citenamefont {{Gronwall}}, \citenamefont {{Schneider}},\ and\ \citenamefont {{Tran}}}]{Finkelstein2011:low_ZZsun_LyaE}%
  \BibitemOpen
  \bibfield  {author} {\bibinfo {author} {\bibfnamefont {S.~L.}\ \bibnamefont {{Finkelstein}}}, \bibinfo {author} {\bibfnamefont {G.~J.}\ \bibnamefont {{Hill}}}, \bibinfo {author} {\bibfnamefont {K.}~\bibnamefont {{Gebhardt}}}, \bibinfo {author} {\bibfnamefont {J.}~\bibnamefont {{Adams}}}, \bibinfo {author} {\bibfnamefont {G.~A.}\ \bibnamefont {{Blanc}}}, \bibinfo {author} {\bibfnamefont {C.}~\bibnamefont {{Papovich}}}, \bibinfo {author} {\bibfnamefont {R.}~\bibnamefont {{Ciardullo}}}, \bibinfo {author} {\bibfnamefont {N.}~\bibnamefont {{Drory}}}, \bibinfo {author} {\bibfnamefont {E.}~\bibnamefont {{Gawiser}}}, \bibinfo {author} {\bibfnamefont {C.}~\bibnamefont {{Gronwall}}}, \bibinfo {author} {\bibfnamefont {D.~P.}\ \bibnamefont {{Schneider}}}, \ and\ \bibinfo {author} {\bibfnamefont {K.-V.}\ \bibnamefont {{Tran}}},\ }\href {\doibase 10.1088/0004-637X/729/2/140} {\bibfield  {journal} {\bibinfo  {journal} {\apj}\ }\textbf {\bibinfo {volume} {729}},\ \bibinfo {eid} {140} (\bibinfo {year} {2011})},\ \Eprint
  {http://arxiv.org/abs/1011.0431} {arXiv:1011.0431 [astro-ph.CO]} \BibitemShut {NoStop}%
\bibitem [{\citenamefont {{Hildebrandt}}\ \emph {et~al.}(2009)\citenamefont {{Hildebrandt}}, \citenamefont {{Pielorz}}, \citenamefont {{Erben}}, \citenamefont {{van Waerbeke}}, \citenamefont {{Simon}},\ and\ \citenamefont {{Capak}}}]{Hildebrandt:2009_CARS_LBG}%
  \BibitemOpen
  \bibfield  {author} {\bibinfo {author} {\bibfnamefont {H.}~\bibnamefont {{Hildebrandt}}}, \bibinfo {author} {\bibfnamefont {J.}~\bibnamefont {{Pielorz}}}, \bibinfo {author} {\bibfnamefont {T.}~\bibnamefont {{Erben}}}, \bibinfo {author} {\bibfnamefont {L.}~\bibnamefont {{van Waerbeke}}}, \bibinfo {author} {\bibfnamefont {P.}~\bibnamefont {{Simon}}}, \ and\ \bibinfo {author} {\bibfnamefont {P.}~\bibnamefont {{Capak}}},\ }\href {\doibase 10.48550/arXiv.0903.3951} {\bibfield  {journal} {\bibinfo  {journal} {arXiv e-prints}\ ,\ \bibinfo {eid} {arXiv:0903.3951}} (\bibinfo {year} {2009})},\ \Eprint {http://arxiv.org/abs/0903.3951} {arXiv:0903.3951 [astro-ph.CO]} \BibitemShut {NoStop}%
\bibitem [{\citenamefont {{Hill}}\ \emph {et~al.}(2008)\citenamefont {{Hill}}, \citenamefont {{Gebhardt}}, \citenamefont {{Komatsu}}, \citenamefont {{Drory}}, \citenamefont {{MacQueen}}, \citenamefont {{Adams}}, \citenamefont {{Blanc}}, \citenamefont {{Koehler}}, \citenamefont {{Rafal}}, \citenamefont {{Roth}}, \citenamefont {{Kelz}}, \citenamefont {{Gronwall}}, \citenamefont {{Ciardullo}},\ and\ \citenamefont {{Schneider}}}]{2008:Hill_hetdex}%
  \BibitemOpen
  \bibfield  {author} {\bibinfo {author} {\bibfnamefont {G.~J.}\ \bibnamefont {{Hill}}}, \bibinfo {author} {\bibfnamefont {K.}~\bibnamefont {{Gebhardt}}}, \bibinfo {author} {\bibfnamefont {E.}~\bibnamefont {{Komatsu}}}, \bibinfo {author} {\bibfnamefont {N.}~\bibnamefont {{Drory}}}, \bibinfo {author} {\bibfnamefont {P.~J.}\ \bibnamefont {{MacQueen}}}, \bibinfo {author} {\bibfnamefont {J.}~\bibnamefont {{Adams}}}, \bibinfo {author} {\bibfnamefont {G.~A.}\ \bibnamefont {{Blanc}}}, \bibinfo {author} {\bibfnamefont {R.}~\bibnamefont {{Koehler}}}, \bibinfo {author} {\bibfnamefont {M.}~\bibnamefont {{Rafal}}}, \bibinfo {author} {\bibfnamefont {M.~M.}\ \bibnamefont {{Roth}}}, \bibinfo {author} {\bibfnamefont {A.}~\bibnamefont {{Kelz}}}, \bibinfo {author} {\bibfnamefont {C.}~\bibnamefont {{Gronwall}}}, \bibinfo {author} {\bibfnamefont {R.}~\bibnamefont {{Ciardullo}}}, \ and\ \bibinfo {author} {\bibfnamefont {D.~P.}\ \bibnamefont {{Schneider}}},\ }in\ \href {\doibase 10.48550/arXiv.0806.0183} {\emph {\bibinfo
  {booktitle} {Panoramic Views of Galaxy Formation and Evolution}}},\ \bibinfo {series} {Astronomical Society of the Pacific Conference Series}, Vol.\ \bibinfo {volume} {399},\ \bibinfo {editor} {edited by\ \bibinfo {editor} {\bibfnamefont {T.}~\bibnamefont {{Kodama}}}, \bibinfo {editor} {\bibfnamefont {T.}~\bibnamefont {{Yamada}}}, \ and\ \bibinfo {editor} {\bibfnamefont {K.}~\bibnamefont {{Aoki}}}}\ (\bibinfo {year} {2008})\ p.\ \bibinfo {pages} {115},\ \Eprint {http://arxiv.org/abs/0806.0183} {arXiv:0806.0183 [astro-ph]} \BibitemShut {NoStop}%
\bibitem [{\citenamefont {{Kannan}}\ \emph {et~al.}(2025)\citenamefont {{Kannan}}, \citenamefont {{Puchwein}}, \citenamefont {{Smith}}, \citenamefont {{Borrow}}, \citenamefont {{Garaldi}}, \citenamefont {{Keating}}, \citenamefont {{Vogelsberger}}, \citenamefont {{Zier}}, \citenamefont {{McClymont}}, \citenamefont {{Shen}}, \citenamefont {{Popovic}}, \citenamefont {{Tacchella}}, \citenamefont {{Hernquist}},\ and\ \citenamefont {{Springel}}}]{2025:Kannan_THESANZOOM}%
  \BibitemOpen
  \bibfield  {author} {\bibinfo {author} {\bibfnamefont {R.}~\bibnamefont {{Kannan}}}, \bibinfo {author} {\bibfnamefont {E.}~\bibnamefont {{Puchwein}}}, \bibinfo {author} {\bibfnamefont {A.}~\bibnamefont {{Smith}}}, \bibinfo {author} {\bibfnamefont {J.}~\bibnamefont {{Borrow}}}, \bibinfo {author} {\bibfnamefont {E.}~\bibnamefont {{Garaldi}}}, \bibinfo {author} {\bibfnamefont {L.}~\bibnamefont {{Keating}}}, \bibinfo {author} {\bibfnamefont {M.}~\bibnamefont {{Vogelsberger}}}, \bibinfo {author} {\bibfnamefont {O.}~\bibnamefont {{Zier}}}, \bibinfo {author} {\bibfnamefont {W.}~\bibnamefont {{McClymont}}}, \bibinfo {author} {\bibfnamefont {X.}~\bibnamefont {{Shen}}}, \bibinfo {author} {\bibfnamefont {F.}~\bibnamefont {{Popovic}}}, \bibinfo {author} {\bibfnamefont {S.}~\bibnamefont {{Tacchella}}}, \bibinfo {author} {\bibfnamefont {L.}~\bibnamefont {{Hernquist}}}, \ and\ \bibinfo {author} {\bibfnamefont {V.}~\bibnamefont {{Springel}}},\ }\href@noop {} {\bibfield  {journal} {\bibinfo  {journal} {arXiv e-prints}\ ,\
  \bibinfo {eid} {arXiv:2502.20437}} (\bibinfo {year} {2025})},\ \Eprint {http://arxiv.org/abs/2502.20437} {arXiv:2502.20437 [astro-ph.GA]} \BibitemShut {NoStop}%
\bibitem [{\citenamefont {{Oyarz{\'u}n}}\ \emph {et~al.}(2016)\citenamefont {{Oyarz{\'u}n}}, \citenamefont {{Blanc}}, \citenamefont {{Gonz{\'a}lez}}, \citenamefont {{Mateo}}, \citenamefont {{Bailey}}, \citenamefont {{Finkelstein}}, \citenamefont {{Lira}}, \citenamefont {{Crane}},\ and\ \citenamefont {{Olszewski}}}]{Oyarzun:2016stellarmasslae}%
  \BibitemOpen
  \bibfield  {author} {\bibinfo {author} {\bibfnamefont {G.~A.}\ \bibnamefont {{Oyarz{\'u}n}}}, \bibinfo {author} {\bibfnamefont {G.~A.}\ \bibnamefont {{Blanc}}}, \bibinfo {author} {\bibfnamefont {V.}~\bibnamefont {{Gonz{\'a}lez}}}, \bibinfo {author} {\bibfnamefont {M.}~\bibnamefont {{Mateo}}}, \bibinfo {author} {\bibfnamefont {J.~I.}\ \bibnamefont {{Bailey}}, \bibfnamefont {III}}, \bibinfo {author} {\bibfnamefont {S.~L.}\ \bibnamefont {{Finkelstein}}}, \bibinfo {author} {\bibfnamefont {P.}~\bibnamefont {{Lira}}}, \bibinfo {author} {\bibfnamefont {J.~D.}\ \bibnamefont {{Crane}}}, \ and\ \bibinfo {author} {\bibfnamefont {E.~W.}\ \bibnamefont {{Olszewski}}},\ }\href {\doibase 10.3847/2041-8205/821/1/L14} {\bibfield  {journal} {\bibinfo  {journal} {\apjl}\ }\textbf {\bibinfo {volume} {821}},\ \bibinfo {eid} {L14} (\bibinfo {year} {2016})},\ \Eprint {http://arxiv.org/abs/1604.03113} {arXiv:1604.03113 [astro-ph.GA]} \BibitemShut {NoStop}%
\bibitem [{\citenamefont {Yuan}\ \emph {et~al.}(2022{\natexlab{a}})\citenamefont {Yuan}, \citenamefont {Hadzhiyska}, \citenamefont {Bose},\ and\ \citenamefont {Eisenstein}}]{Yuan:2022rsc}%
  \BibitemOpen
  \bibfield  {author} {\bibinfo {author} {\bibfnamefont {S.}~\bibnamefont {Yuan}}, \bibinfo {author} {\bibfnamefont {B.}~\bibnamefont {Hadzhiyska}}, \bibinfo {author} {\bibfnamefont {S.}~\bibnamefont {Bose}}, \ and\ \bibinfo {author} {\bibfnamefont {D.~J.}\ \bibnamefont {Eisenstein}},\ }\href {\doibase 10.1093/mnras/stac830} {\bibfield  {journal} {\bibinfo  {journal} {Mon. Not. Roy. Astron. Soc.}\ }\textbf {\bibinfo {volume} {512}},\ \bibinfo {pages} {5793} (\bibinfo {year} {2022}{\natexlab{a}})},\ \Eprint {http://arxiv.org/abs/2202.12911} {arXiv:2202.12911 [astro-ph.CO]} \BibitemShut {NoStop}%
\bibitem [{\citenamefont {{Hadzhiyska}}\ \emph {et~al.}(2021)\citenamefont {{Hadzhiyska}}, \citenamefont {{Tacchella}}, \citenamefont {{Bose}},\ and\ \citenamefont {{Eisenstein}}}]{2021:Hadzhiyska_tng_elg}%
  \BibitemOpen
  \bibfield  {author} {\bibinfo {author} {\bibfnamefont {B.}~\bibnamefont {{Hadzhiyska}}}, \bibinfo {author} {\bibfnamefont {S.}~\bibnamefont {{Tacchella}}}, \bibinfo {author} {\bibfnamefont {S.}~\bibnamefont {{Bose}}}, \ and\ \bibinfo {author} {\bibfnamefont {D.~J.}\ \bibnamefont {{Eisenstein}}},\ }\href {\doibase 10.1093/mnras/stab243} {\bibfield  {journal} {\bibinfo  {journal} {\mnras}\ }\textbf {\bibinfo {volume} {502}},\ \bibinfo {pages} {3599} (\bibinfo {year} {2021})},\ \Eprint {http://arxiv.org/abs/2011.05331} {arXiv:2011.05331 [astro-ph.GA]} \BibitemShut {NoStop}%
\bibitem [{\citenamefont {Hadzhiyska}\ \emph {et~al.}(2021)\citenamefont {Hadzhiyska}, \citenamefont {Eisenstein}, \citenamefont {Bose}, \citenamefont {Garrison},\ and\ \citenamefont {Maksimova}}]{Hadzhiyska:2021zbd}%
  \BibitemOpen
  \bibfield  {author} {\bibinfo {author} {\bibfnamefont {B.}~\bibnamefont {Hadzhiyska}}, \bibinfo {author} {\bibfnamefont {D.}~\bibnamefont {Eisenstein}}, \bibinfo {author} {\bibfnamefont {S.}~\bibnamefont {Bose}}, \bibinfo {author} {\bibfnamefont {L.~H.}\ \bibnamefont {Garrison}}, \ and\ \bibinfo {author} {\bibfnamefont {N.}~\bibnamefont {Maksimova}},\ }\href {\doibase 10.1093/mnras/stab2980} {\bibfield  {journal} {\bibinfo  {journal} {Mon. Not. Roy. Astron. Soc.}\ }\textbf {\bibinfo {volume} {509}},\ \bibinfo {pages} {501} (\bibinfo {year} {2021})},\ \Eprint {http://arxiv.org/abs/2110.11408} {arXiv:2110.11408 [astro-ph.CO]} \BibitemShut {NoStop}%
\bibitem [{\citenamefont {{Davis}}\ \emph {et~al.}(1985)\citenamefont {{Davis}}, \citenamefont {{Efstathiou}}, \citenamefont {{Frenk}},\ and\ \citenamefont {{White}}}]{1985ApJ...292..371D}%
  \BibitemOpen
  \bibfield  {author} {\bibinfo {author} {\bibfnamefont {M.}~\bibnamefont {{Davis}}}, \bibinfo {author} {\bibfnamefont {G.}~\bibnamefont {{Efstathiou}}}, \bibinfo {author} {\bibfnamefont {C.~S.}\ \bibnamefont {{Frenk}}}, \ and\ \bibinfo {author} {\bibfnamefont {S.~D.~M.}\ \bibnamefont {{White}}},\ }\href {\doibase 10.1086/163168} {\bibfield  {journal} {\bibinfo  {journal} {APJ}\ }\textbf {\bibinfo {volume} {292}},\ \bibinfo {pages} {371} (\bibinfo {year} {1985})}\BibitemShut {NoStop}%
\bibitem [{\citenamefont {{Springel}}\ \emph {et~al.}(2001)\citenamefont {{Springel}}, \citenamefont {{White}}, \citenamefont {{Tormen}},\ and\ \citenamefont {{Kauffmann}}}]{springel_subfind}%
  \BibitemOpen
  \bibfield  {author} {\bibinfo {author} {\bibfnamefont {V.}~\bibnamefont {{Springel}}}, \bibinfo {author} {\bibfnamefont {S.~D.~M.}\ \bibnamefont {{White}}}, \bibinfo {author} {\bibfnamefont {G.}~\bibnamefont {{Tormen}}}, \ and\ \bibinfo {author} {\bibfnamefont {G.}~\bibnamefont {{Kauffmann}}},\ }\href {\doibase 10.1046/j.1365-8711.2001.04912.x} {\bibfield  {journal} {\bibinfo  {journal} {\mnras}\ }\textbf {\bibinfo {volume} {328}},\ \bibinfo {pages} {726} (\bibinfo {year} {2001})},\ \Eprint {http://arxiv.org/abs/astro-ph/0012055} {arXiv:astro-ph/0012055 [astro-ph]} \BibitemShut {NoStop}%
\bibitem [{\citenamefont {Schmittfull}\ \emph {et~al.}(2019)\citenamefont {Schmittfull}, \citenamefont {Simonović}, \citenamefont {Assassi},\ and\ \citenamefont {Zaldarriaga}}]{Schmittfull:2018yuk}%
  \BibitemOpen
  \bibfield  {author} {\bibinfo {author} {\bibfnamefont {M.}~\bibnamefont {Schmittfull}}, \bibinfo {author} {\bibfnamefont {M.}~\bibnamefont {Simonović}}, \bibinfo {author} {\bibfnamefont {V.}~\bibnamefont {Assassi}}, \ and\ \bibinfo {author} {\bibfnamefont {M.}~\bibnamefont {Zaldarriaga}},\ }\href {\doibase 10.1103/PhysRevD.100.043514} {\bibfield  {journal} {\bibinfo  {journal} {Phys.\ Rev.\ D}\ }\textbf {\bibinfo {volume} {100}},\ \bibinfo {pages} {043514} (\bibinfo {year} {2019})},\ \Eprint {http://arxiv.org/abs/1811.10640} {arXiv:1811.10640 [astro-ph.CO]} \BibitemShut {NoStop}%
\bibitem [{\citenamefont {Schmittfull}\ \emph {et~al.}(2021)\citenamefont {Schmittfull}, \citenamefont {Simonovi\'c}, \citenamefont {Ivanov}, \citenamefont {Philcox},\ and\ \citenamefont {Zaldarriaga}}]{Schmittfull:2020trd}%
  \BibitemOpen
  \bibfield  {author} {\bibinfo {author} {\bibfnamefont {M.}~\bibnamefont {Schmittfull}}, \bibinfo {author} {\bibfnamefont {M.}~\bibnamefont {Simonovi\'c}}, \bibinfo {author} {\bibfnamefont {M.~M.}\ \bibnamefont {Ivanov}}, \bibinfo {author} {\bibfnamefont {O.~H.~E.}\ \bibnamefont {Philcox}}, \ and\ \bibinfo {author} {\bibfnamefont {M.}~\bibnamefont {Zaldarriaga}},\ }\href {\doibase 10.1088/1475-7516/2021/05/059} {\bibfield  {journal} {\bibinfo  {journal} {JCAP}\ }\textbf {\bibinfo {volume} {05}},\ \bibinfo {pages} {059} (\bibinfo {year} {2021})},\ \Eprint {http://arxiv.org/abs/2012.03334} {arXiv:2012.03334 [astro-ph.CO]} \BibitemShut {NoStop}%
\bibitem [{\citenamefont {Assassi}\ \emph {et~al.}(2014)\citenamefont {Assassi}, \citenamefont {Baumann}, \citenamefont {Green},\ and\ \citenamefont {Zaldarriaga}}]{Assassi:2014fva}%
  \BibitemOpen
  \bibfield  {author} {\bibinfo {author} {\bibfnamefont {V.}~\bibnamefont {Assassi}}, \bibinfo {author} {\bibfnamefont {D.}~\bibnamefont {Baumann}}, \bibinfo {author} {\bibfnamefont {D.}~\bibnamefont {Green}}, \ and\ \bibinfo {author} {\bibfnamefont {M.}~\bibnamefont {Zaldarriaga}},\ }\href {\doibase 10.1088/1475-7516/2014/08/056} {\bibfield  {journal} {\bibinfo  {journal} {JCAP}\ }\textbf {\bibinfo {volume} {1408}},\ \bibinfo {pages} {056} (\bibinfo {year} {2014})},\ \Eprint {http://arxiv.org/abs/1402.5916} {arXiv:1402.5916 [astro-ph.CO]} \BibitemShut {NoStop}%
\bibitem [{\citenamefont {Ivanov}\ \emph {et~al.}(2020)\citenamefont {Ivanov}, \citenamefont {Simonovi\'c},\ and\ \citenamefont {Zaldarriaga}}]{Ivanov:2019pdj}%
  \BibitemOpen
  \bibfield  {author} {\bibinfo {author} {\bibfnamefont {M.~M.}\ \bibnamefont {Ivanov}}, \bibinfo {author} {\bibfnamefont {M.}~\bibnamefont {Simonovi\'c}}, \ and\ \bibinfo {author} {\bibfnamefont {M.}~\bibnamefont {Zaldarriaga}},\ }\href {\doibase 10.1088/1475-7516/2020/05/042} {\bibfield  {journal} {\bibinfo  {journal} {JCAP}\ }\textbf {\bibinfo {volume} {05}},\ \bibinfo {pages} {042} (\bibinfo {year} {2020})},\ \Eprint {http://arxiv.org/abs/1909.05277} {arXiv:1909.05277 [astro-ph.CO]} \BibitemShut {NoStop}%
\bibitem [{\citenamefont {Bernardeau}\ \emph {et~al.}(2002)\citenamefont {Bernardeau}, \citenamefont {Colombi}, \citenamefont {Gaztanaga},\ and\ \citenamefont {Scoccimarro}}]{Bernardeau:2001qr}%
  \BibitemOpen
  \bibfield  {author} {\bibinfo {author} {\bibfnamefont {F.}~\bibnamefont {Bernardeau}}, \bibinfo {author} {\bibfnamefont {S.}~\bibnamefont {Colombi}}, \bibinfo {author} {\bibfnamefont {E.}~\bibnamefont {Gaztanaga}}, \ and\ \bibinfo {author} {\bibfnamefont {R.}~\bibnamefont {Scoccimarro}},\ }\href {\doibase 10.1016/S0370-1573(02)00135-7} {\bibfield  {journal} {\bibinfo  {journal} {Phys. Rept.}\ }\textbf {\bibinfo {volume} {367}},\ \bibinfo {pages} {1} (\bibinfo {year} {2002})},\ \Eprint {http://arxiv.org/abs/astro-ph/0112551} {arXiv:astro-ph/0112551 [astro-ph]} \BibitemShut {NoStop}%
\bibitem [{\citenamefont {Chudaykin}\ \emph {et~al.}(2020)\citenamefont {Chudaykin}, \citenamefont {Ivanov}, \citenamefont {Philcox},\ and\ \citenamefont {Simonovi\'c}}]{Chudaykin:2020aoj}%
  \BibitemOpen
  \bibfield  {author} {\bibinfo {author} {\bibfnamefont {A.}~\bibnamefont {Chudaykin}}, \bibinfo {author} {\bibfnamefont {M.~M.}\ \bibnamefont {Ivanov}}, \bibinfo {author} {\bibfnamefont {O.~H.~E.}\ \bibnamefont {Philcox}}, \ and\ \bibinfo {author} {\bibfnamefont {M.}~\bibnamefont {Simonovi\'c}},\ }\href {\doibase 10.1103/PhysRevD.102.063533} {\bibfield  {journal} {\bibinfo  {journal} {Phys. Rev. D}\ }\textbf {\bibinfo {volume} {102}},\ \bibinfo {pages} {063533} (\bibinfo {year} {2020})},\ \Eprint {http://arxiv.org/abs/2004.10607} {arXiv:2004.10607 [astro-ph.CO]} \BibitemShut {NoStop}%
\bibitem [{\citenamefont {Senatore}\ and\ \citenamefont {Zaldarriaga}(2014)}]{Senatore:2014vja}%
  \BibitemOpen
  \bibfield  {author} {\bibinfo {author} {\bibfnamefont {L.}~\bibnamefont {Senatore}}\ and\ \bibinfo {author} {\bibfnamefont {M.}~\bibnamefont {Zaldarriaga}},\ }\href@noop {} {\  (\bibinfo {year} {2014})},\ \Eprint {http://arxiv.org/abs/1409.1225} {arXiv:1409.1225 [astro-ph.CO]} \BibitemShut {NoStop}%
\bibitem [{\citenamefont {Ivanov}\ and\ \citenamefont {Sibiryakov}(2018)}]{Ivanov:2018gjr}%
  \BibitemOpen
  \bibfield  {author} {\bibinfo {author} {\bibfnamefont {M.~M.}\ \bibnamefont {Ivanov}}\ and\ \bibinfo {author} {\bibfnamefont {S.}~\bibnamefont {Sibiryakov}},\ }\href {\doibase 10.1088/1475-7516/2018/07/053} {\bibfield  {journal} {\bibinfo  {journal} {JCAP}\ }\textbf {\bibinfo {volume} {1807}},\ \bibinfo {pages} {053} (\bibinfo {year} {2018})},\ \Eprint {http://arxiv.org/abs/1804.05080} {arXiv:1804.05080 [astro-ph.CO]} \BibitemShut {NoStop}%
\bibitem [{\citenamefont {Jackson}(1972)}]{Jackson:2008yv}%
  \BibitemOpen
  \bibfield  {author} {\bibinfo {author} {\bibfnamefont {J.~C.}\ \bibnamefont {Jackson}},\ }\href {\doibase 10.1093/mnras/156.1.1P} {\bibfield  {journal} {\bibinfo  {journal} {Mon. Not. Roy. Astron. Soc.}\ }\textbf {\bibinfo {volume} {156}},\ \bibinfo {pages} {1P} (\bibinfo {year} {1972})},\ \Eprint {http://arxiv.org/abs/0810.3908} {arXiv:0810.3908 [astro-ph]} \BibitemShut {NoStop}%
\bibitem [{\citenamefont {Ivanov}(2025)}]{Ivanov:2025qie}%
  \BibitemOpen
  \bibfield  {author} {\bibinfo {author} {\bibfnamefont {M.~M.}\ \bibnamefont {Ivanov}},\ }\href@noop {} {\  (\bibinfo {year} {2025})},\ \Eprint {http://arxiv.org/abs/2503.07270} {arXiv:2503.07270 [astro-ph.CO]} \BibitemShut {NoStop}%
\bibitem [{\citenamefont {Chudaykin}\ \emph {et~al.}(2021)\citenamefont {Chudaykin}, \citenamefont {Ivanov},\ and\ \citenamefont {Simonovi\'c}}]{Chudaykin:2020hbf}%
  \BibitemOpen
  \bibfield  {author} {\bibinfo {author} {\bibfnamefont {A.}~\bibnamefont {Chudaykin}}, \bibinfo {author} {\bibfnamefont {M.~M.}\ \bibnamefont {Ivanov}}, \ and\ \bibinfo {author} {\bibfnamefont {M.}~\bibnamefont {Simonovi\'c}},\ }\href {\doibase 10.1103/PhysRevD.103.043525} {\bibfield  {journal} {\bibinfo  {journal} {Phys. Rev. D}\ }\textbf {\bibinfo {volume} {103}},\ \bibinfo {pages} {043525} (\bibinfo {year} {2021})},\ \Eprint {http://arxiv.org/abs/2009.10724} {arXiv:2009.10724 [astro-ph.CO]} \BibitemShut {NoStop}%
\bibitem [{\citenamefont {Taule}\ and\ \citenamefont {Garny}(2023)}]{Taule:2023izt}%
  \BibitemOpen
  \bibfield  {author} {\bibinfo {author} {\bibfnamefont {P.}~\bibnamefont {Taule}}\ and\ \bibinfo {author} {\bibfnamefont {M.}~\bibnamefont {Garny}},\ }\href {\doibase 10.1088/1475-7516/2023/11/078} {\bibfield  {journal} {\bibinfo  {journal} {JCAP}\ }\textbf {\bibinfo {volume} {11}},\ \bibinfo {pages} {078} (\bibinfo {year} {2023})},\ \Eprint {http://arxiv.org/abs/2308.07379} {arXiv:2308.07379 [astro-ph.CO]} \BibitemShut {NoStop}%
\bibitem [{\citenamefont {Ivanov}\ \emph {et~al.}(2022)\citenamefont {Ivanov}, \citenamefont {Philcox}, \citenamefont {Simonovi\'c}, \citenamefont {Zaldarriaga}, \citenamefont {Nischimichi},\ and\ \citenamefont {Takada}}]{Ivanov:2021fbu}%
  \BibitemOpen
  \bibfield  {author} {\bibinfo {author} {\bibfnamefont {M.~M.}\ \bibnamefont {Ivanov}}, \bibinfo {author} {\bibfnamefont {O.~H.~E.}\ \bibnamefont {Philcox}}, \bibinfo {author} {\bibfnamefont {M.}~\bibnamefont {Simonovi\'c}}, \bibinfo {author} {\bibfnamefont {M.}~\bibnamefont {Zaldarriaga}}, \bibinfo {author} {\bibfnamefont {T.}~\bibnamefont {Nischimichi}}, \ and\ \bibinfo {author} {\bibfnamefont {M.}~\bibnamefont {Takada}},\ }\href {\doibase 10.1103/PhysRevD.105.043531} {\bibfield  {journal} {\bibinfo  {journal} {Phys. Rev. D}\ }\textbf {\bibinfo {volume} {105}},\ \bibinfo {pages} {043531} (\bibinfo {year} {2022})},\ \Eprint {http://arxiv.org/abs/2110.00006} {arXiv:2110.00006 [astro-ph.CO]} \BibitemShut {NoStop}%
\bibitem [{\citenamefont {Ivanov}(2021)}]{Ivanov:2021zmi}%
  \BibitemOpen
  \bibfield  {author} {\bibinfo {author} {\bibfnamefont {M.~M.}\ \bibnamefont {Ivanov}},\ }\href@noop {} {\  (\bibinfo {year} {2021})},\ \Eprint {http://arxiv.org/abs/2106.12580} {arXiv:2106.12580 [astro-ph.CO]} \BibitemShut {NoStop}%
\bibitem [{\citenamefont {Senatore}\ and\ \citenamefont {Zaldarriaga}(2015)}]{Senatore:2014via}%
  \BibitemOpen
  \bibfield  {author} {\bibinfo {author} {\bibfnamefont {L.}~\bibnamefont {Senatore}}\ and\ \bibinfo {author} {\bibfnamefont {M.}~\bibnamefont {Zaldarriaga}},\ }\href {\doibase 10.1088/1475-7516/2015/02/013} {\bibfield  {journal} {\bibinfo  {journal} {JCAP}\ }\textbf {\bibinfo {volume} {1502}},\ \bibinfo {pages} {013} (\bibinfo {year} {2015})},\ \Eprint {http://arxiv.org/abs/1404.5954} {arXiv:1404.5954 [astro-ph.CO]} \BibitemShut {NoStop}%
\bibitem [{\citenamefont {Baldauf}\ \emph {et~al.}(2015)\citenamefont {Baldauf}, \citenamefont {Mirbabayi}, \citenamefont {Simonović},\ and\ \citenamefont {Zaldarriaga}}]{Baldauf:2015xfa}%
  \BibitemOpen
  \bibfield  {author} {\bibinfo {author} {\bibfnamefont {T.}~\bibnamefont {Baldauf}}, \bibinfo {author} {\bibfnamefont {M.}~\bibnamefont {Mirbabayi}}, \bibinfo {author} {\bibfnamefont {M.}~\bibnamefont {Simonović}}, \ and\ \bibinfo {author} {\bibfnamefont {M.}~\bibnamefont {Zaldarriaga}},\ }\href {\doibase 10.1103/PhysRevD.92.043514} {\bibfield  {journal} {\bibinfo  {journal} {Phys. Rev.}\ }\textbf {\bibinfo {volume} {D92}},\ \bibinfo {pages} {043514} (\bibinfo {year} {2015})},\ \Eprint {http://arxiv.org/abs/1504.04366} {arXiv:1504.04366 [astro-ph.CO]} \BibitemShut {NoStop}%
\bibitem [{\citenamefont {Vlah}\ \emph {et~al.}(2016)\citenamefont {Vlah}, \citenamefont {Seljak}, \citenamefont {Chu},\ and\ \citenamefont {Feng}}]{Vlah:2015zda}%
  \BibitemOpen
  \bibfield  {author} {\bibinfo {author} {\bibfnamefont {Z.}~\bibnamefont {Vlah}}, \bibinfo {author} {\bibfnamefont {U.}~\bibnamefont {Seljak}}, \bibinfo {author} {\bibfnamefont {M.~Y.}\ \bibnamefont {Chu}}, \ and\ \bibinfo {author} {\bibfnamefont {Y.}~\bibnamefont {Feng}},\ }\href {\doibase 10.1088/1475-7516/2016/03/057} {\bibfield  {journal} {\bibinfo  {journal} {JCAP}\ }\textbf {\bibinfo {volume} {1603}},\ \bibinfo {pages} {057} (\bibinfo {year} {2016})},\ \Eprint {http://arxiv.org/abs/1509.02120} {arXiv:1509.02120 [astro-ph.CO]} \BibitemShut {NoStop}%
\bibitem [{\citenamefont {Blas}\ \emph {et~al.}(2016{\natexlab{a}})\citenamefont {Blas}, \citenamefont {Garny}, \citenamefont {Ivanov},\ and\ \citenamefont {Sibiryakov}}]{Blas:2015qsi}%
  \BibitemOpen
  \bibfield  {author} {\bibinfo {author} {\bibfnamefont {D.}~\bibnamefont {Blas}}, \bibinfo {author} {\bibfnamefont {M.}~\bibnamefont {Garny}}, \bibinfo {author} {\bibfnamefont {M.~M.}\ \bibnamefont {Ivanov}}, \ and\ \bibinfo {author} {\bibfnamefont {S.}~\bibnamefont {Sibiryakov}},\ }\href {\doibase 10.1088/1475-7516/2016/07/052} {\bibfield  {journal} {\bibinfo  {journal} {JCAP}\ }\textbf {\bibinfo {volume} {1607}},\ \bibinfo {pages} {052} (\bibinfo {year} {2016}{\natexlab{a}})},\ \Eprint {http://arxiv.org/abs/1512.05807} {arXiv:1512.05807 [astro-ph.CO]} \BibitemShut {NoStop}%
\bibitem [{\citenamefont {Blas}\ \emph {et~al.}(2016{\natexlab{b}})\citenamefont {Blas}, \citenamefont {Garny}, \citenamefont {Ivanov},\ and\ \citenamefont {Sibiryakov}}]{Blas:2016sfa}%
  \BibitemOpen
  \bibfield  {author} {\bibinfo {author} {\bibfnamefont {D.}~\bibnamefont {Blas}}, \bibinfo {author} {\bibfnamefont {M.}~\bibnamefont {Garny}}, \bibinfo {author} {\bibfnamefont {M.~M.}\ \bibnamefont {Ivanov}}, \ and\ \bibinfo {author} {\bibfnamefont {S.}~\bibnamefont {Sibiryakov}},\ }\href {\doibase 10.1088/1475-7516/2016/07/028} {\bibfield  {journal} {\bibinfo  {journal} {JCAP}\ }\textbf {\bibinfo {volume} {1607}},\ \bibinfo {pages} {028} (\bibinfo {year} {2016}{\natexlab{b}})},\ \Eprint {http://arxiv.org/abs/1605.02149} {arXiv:1605.02149 [astro-ph.CO]} \BibitemShut {NoStop}%
\bibitem [{\citenamefont {Vasudevan}\ \emph {et~al.}(2019)\citenamefont {Vasudevan}, \citenamefont {Ivanov}, \citenamefont {Sibiryakov},\ and\ \citenamefont {Lesgourgues}}]{Vasudevan:2019ewf}%
  \BibitemOpen
  \bibfield  {author} {\bibinfo {author} {\bibfnamefont {A.}~\bibnamefont {Vasudevan}}, \bibinfo {author} {\bibfnamefont {M.~M.}\ \bibnamefont {Ivanov}}, \bibinfo {author} {\bibfnamefont {S.}~\bibnamefont {Sibiryakov}}, \ and\ \bibinfo {author} {\bibfnamefont {J.}~\bibnamefont {Lesgourgues}},\ }\href {\doibase 10.1088/1475-7516/2019/09/037} {\bibfield  {journal} {\bibinfo  {journal} {JCAP}\ }\textbf {\bibinfo {volume} {09}},\ \bibinfo {pages} {037} (\bibinfo {year} {2019})},\ \Eprint {http://arxiv.org/abs/1906.08697} {arXiv:1906.08697 [astro-ph.CO]} \BibitemShut {NoStop}%
\bibitem [{\citenamefont {Philcox}\ and\ \citenamefont {Ivanov}(2022)}]{Philcox:2021kcw}%
  \BibitemOpen
  \bibfield  {author} {\bibinfo {author} {\bibfnamefont {O.~H.~E.}\ \bibnamefont {Philcox}}\ and\ \bibinfo {author} {\bibfnamefont {M.~M.}\ \bibnamefont {Ivanov}},\ }\href {\doibase 10.1103/PhysRevD.105.043517} {\bibfield  {journal} {\bibinfo  {journal} {Phys. Rev. D}\ }\textbf {\bibinfo {volume} {105}},\ \bibinfo {pages} {043517} (\bibinfo {year} {2022})},\ \Eprint {http://arxiv.org/abs/2112.04515} {arXiv:2112.04515 [astro-ph.CO]} \BibitemShut {NoStop}%
\bibitem [{\citenamefont {Maus}\ \emph {et~al.}(2024)\citenamefont {Maus} \emph {et~al.}}]{Maus:2024dzi}%
  \BibitemOpen
  \bibfield  {author} {\bibinfo {author} {\bibfnamefont {M.}~\bibnamefont {Maus}} \emph {et~al.},\ }\href@noop {} {\  (\bibinfo {year} {2024})},\ \Eprint {http://arxiv.org/abs/2404.07312} {arXiv:2404.07312 [astro-ph.CO]} \BibitemShut {NoStop}%
\bibitem [{\citenamefont {Obuljen}\ \emph {et~al.}(2023)\citenamefont {Obuljen}, \citenamefont {Simonovi\'c}, \citenamefont {Schneider},\ and\ \citenamefont {Feldmann}}]{Obuljen:2022cjo}%
  \BibitemOpen
  \bibfield  {author} {\bibinfo {author} {\bibfnamefont {A.}~\bibnamefont {Obuljen}}, \bibinfo {author} {\bibfnamefont {M.}~\bibnamefont {Simonovi\'c}}, \bibinfo {author} {\bibfnamefont {A.}~\bibnamefont {Schneider}}, \ and\ \bibinfo {author} {\bibfnamefont {R.}~\bibnamefont {Feldmann}},\ }\href {\doibase 10.1103/PhysRevD.108.083528} {\bibfield  {journal} {\bibinfo  {journal} {Phys. Rev. D}\ }\textbf {\bibinfo {volume} {108}},\ \bibinfo {pages} {083528} (\bibinfo {year} {2023})},\ \Eprint {http://arxiv.org/abs/2207.12398} {arXiv:2207.12398 [astro-ph.CO]} \BibitemShut {NoStop}%
\bibitem [{\citenamefont {Eggemeier}\ \emph {et~al.}(2018)\citenamefont {Eggemeier}, \citenamefont {Scoccimarro},\ and\ \citenamefont {Smith}}]{Eggemeier:2018qae}%
  \BibitemOpen
  \bibfield  {author} {\bibinfo {author} {\bibfnamefont {A.}~\bibnamefont {Eggemeier}}, \bibinfo {author} {\bibfnamefont {R.}~\bibnamefont {Scoccimarro}}, \ and\ \bibinfo {author} {\bibfnamefont {R.~E.}\ \bibnamefont {Smith}},\ }\href@noop {} {\  (\bibinfo {year} {2018})},\ \Eprint {http://arxiv.org/abs/1812.03208} {arXiv:1812.03208 [astro-ph.CO]} \BibitemShut {NoStop}%
\bibitem [{\citenamefont {Philcox}\ \emph {et~al.}(2022)\citenamefont {Philcox}, \citenamefont {Ivanov}, \citenamefont {Cabass}, \citenamefont {Simonovi\'c}, \citenamefont {Zaldarriaga},\ and\ \citenamefont {Nishimichi}}]{Philcox:2022frc}%
  \BibitemOpen
  \bibfield  {author} {\bibinfo {author} {\bibfnamefont {O.~H.~E.}\ \bibnamefont {Philcox}}, \bibinfo {author} {\bibfnamefont {M.~M.}\ \bibnamefont {Ivanov}}, \bibinfo {author} {\bibfnamefont {G.}~\bibnamefont {Cabass}}, \bibinfo {author} {\bibfnamefont {M.}~\bibnamefont {Simonovi\'c}}, \bibinfo {author} {\bibfnamefont {M.}~\bibnamefont {Zaldarriaga}}, \ and\ \bibinfo {author} {\bibfnamefont {T.}~\bibnamefont {Nishimichi}},\ }\href {\doibase 10.1103/PhysRevD.106.043530} {\bibfield  {journal} {\bibinfo  {journal} {Phys. Rev. D}\ }\textbf {\bibinfo {volume} {106}},\ \bibinfo {pages} {043530} (\bibinfo {year} {2022})},\ \Eprint {http://arxiv.org/abs/2206.02800} {arXiv:2206.02800 [astro-ph.CO]} \BibitemShut {NoStop}%
\bibitem [{\citenamefont {Obuljen}\ \emph {et~al.}(2020)\citenamefont {Obuljen}, \citenamefont {Percival},\ and\ \citenamefont {Dalal}}]{Obuljen:2020ypy}%
  \BibitemOpen
  \bibfield  {author} {\bibinfo {author} {\bibfnamefont {A.}~\bibnamefont {Obuljen}}, \bibinfo {author} {\bibfnamefont {W.~J.}\ \bibnamefont {Percival}}, \ and\ \bibinfo {author} {\bibfnamefont {N.}~\bibnamefont {Dalal}},\ }\href {\doibase 10.1088/1475-7516/2020/10/058} {\bibfield  {journal} {\bibinfo  {journal} {JCAP}\ }\textbf {\bibinfo {volume} {10}},\ \bibinfo {pages} {058} (\bibinfo {year} {2020})},\ \Eprint {http://arxiv.org/abs/2004.07240} {arXiv:2004.07240 [astro-ph.CO]} \BibitemShut {NoStop}%
\bibitem [{\citenamefont {Ivanov}\ \emph {et~al.}(2024{\natexlab{a}})\citenamefont {Ivanov}, \citenamefont {Cuesta-Lazaro}, \citenamefont {Mishra-Sharma}, \citenamefont {Obuljen},\ and\ \citenamefont {Toomey}}]{Ivanov:2024hgq}%
  \BibitemOpen
  \bibfield  {author} {\bibinfo {author} {\bibfnamefont {M.~M.}\ \bibnamefont {Ivanov}}, \bibinfo {author} {\bibfnamefont {C.}~\bibnamefont {Cuesta-Lazaro}}, \bibinfo {author} {\bibfnamefont {S.}~\bibnamefont {Mishra-Sharma}}, \bibinfo {author} {\bibfnamefont {A.}~\bibnamefont {Obuljen}}, \ and\ \bibinfo {author} {\bibfnamefont {M.~W.}\ \bibnamefont {Toomey}},\ }\href {\doibase 10.1103/PhysRevD.110.063538} {\bibfield  {journal} {\bibinfo  {journal} {Phys. Rev. D}\ }\textbf {\bibinfo {volume} {110}},\ \bibinfo {pages} {063538} (\bibinfo {year} {2024}{\natexlab{a}})},\ \Eprint {http://arxiv.org/abs/2402.13310} {arXiv:2402.13310 [astro-ph.CO]} \BibitemShut {NoStop}%
\bibitem [{\citenamefont {Ivanov}\ \emph {et~al.}(2024{\natexlab{b}})\citenamefont {Ivanov}, \citenamefont {Obuljen}, \citenamefont {Cuesta-Lazaro},\ and\ \citenamefont {Toomey}}]{Ivanov:2024xgb}%
  \BibitemOpen
  \bibfield  {author} {\bibinfo {author} {\bibfnamefont {M.~M.}\ \bibnamefont {Ivanov}}, \bibinfo {author} {\bibfnamefont {A.}~\bibnamefont {Obuljen}}, \bibinfo {author} {\bibfnamefont {C.}~\bibnamefont {Cuesta-Lazaro}}, \ and\ \bibinfo {author} {\bibfnamefont {M.~W.}\ \bibnamefont {Toomey}},\ }\href@noop {} {\  (\bibinfo {year} {2024}{\natexlab{b}})},\ \Eprint {http://arxiv.org/abs/2409.10609} {arXiv:2409.10609 [astro-ph.CO]} \BibitemShut {NoStop}%
\bibitem [{\citenamefont {Baldauf}\ \emph {et~al.}(2013)\citenamefont {Baldauf}, \citenamefont {Seljak}, \citenamefont {Smith}, \citenamefont {Hamaus},\ and\ \citenamefont {Desjacques}}]{Baldauf:2013hka}%
  \BibitemOpen
  \bibfield  {author} {\bibinfo {author} {\bibfnamefont {T.}~\bibnamefont {Baldauf}}, \bibinfo {author} {\bibfnamefont {U.}~\bibnamefont {Seljak}}, \bibinfo {author} {\bibfnamefont {R.~E.}\ \bibnamefont {Smith}}, \bibinfo {author} {\bibfnamefont {N.}~\bibnamefont {Hamaus}}, \ and\ \bibinfo {author} {\bibfnamefont {V.}~\bibnamefont {Desjacques}},\ }\href {\doibase 10.1103/PhysRevD.88.083507} {\bibfield  {journal} {\bibinfo  {journal} {Phys. Rev. D}\ }\textbf {\bibinfo {volume} {88}},\ \bibinfo {pages} {083507} (\bibinfo {year} {2013})},\ \Eprint {http://arxiv.org/abs/1305.2917} {arXiv:1305.2917 [astro-ph.CO]} \BibitemShut {NoStop}%
\bibitem [{\citenamefont {Kaiser}(1987)}]{Kaiser:1987qv}%
  \BibitemOpen
  \bibfield  {author} {\bibinfo {author} {\bibfnamefont {N.}~\bibnamefont {Kaiser}},\ }\href@noop {} {\bibfield  {journal} {\bibinfo  {journal} {Mon. Not. Roy. Astron. Soc.}\ }\textbf {\bibinfo {volume} {227}},\ \bibinfo {pages} {1} (\bibinfo {year} {1987})}\BibitemShut {NoStop}%
\bibitem [{\citenamefont {{Umeda}}\ \emph {et~al.}(2024)\citenamefont {{Umeda}}, \citenamefont {{Ouchi}}, \citenamefont {{Kikuta}}, \citenamefont {{Harikane}}, \citenamefont {{Ono}}, \citenamefont {{Shibuya}}, \citenamefont {{Inoue}}, \citenamefont {{Shimasaku}}, \citenamefont {{Liang}}, \citenamefont {{Matsumoto}}, \citenamefont {{Saito}}, \citenamefont {{Kusakabe}}, \citenamefont {{Kageura}},\ and\ \citenamefont {{Nakane}}}]{Umeda:2024_SILVERRUSH_LF_ACF}%
  \BibitemOpen
  \bibfield  {author} {\bibinfo {author} {\bibfnamefont {H.}~\bibnamefont {{Umeda}}}, \bibinfo {author} {\bibfnamefont {M.}~\bibnamefont {{Ouchi}}}, \bibinfo {author} {\bibfnamefont {S.}~\bibnamefont {{Kikuta}}}, \bibinfo {author} {\bibfnamefont {Y.}~\bibnamefont {{Harikane}}}, \bibinfo {author} {\bibfnamefont {Y.}~\bibnamefont {{Ono}}}, \bibinfo {author} {\bibfnamefont {T.}~\bibnamefont {{Shibuya}}}, \bibinfo {author} {\bibfnamefont {A.~K.}\ \bibnamefont {{Inoue}}}, \bibinfo {author} {\bibfnamefont {K.}~\bibnamefont {{Shimasaku}}}, \bibinfo {author} {\bibfnamefont {Y.}~\bibnamefont {{Liang}}}, \bibinfo {author} {\bibfnamefont {A.}~\bibnamefont {{Matsumoto}}}, \bibinfo {author} {\bibfnamefont {S.}~\bibnamefont {{Saito}}}, \bibinfo {author} {\bibfnamefont {H.}~\bibnamefont {{Kusakabe}}}, \bibinfo {author} {\bibfnamefont {Y.}~\bibnamefont {{Kageura}}}, \ and\ \bibinfo {author} {\bibfnamefont {M.}~\bibnamefont {{Nakane}}},\ }\href {\doibase 10.48550/arXiv.2411.15495} {\bibfield  {journal} {\bibinfo  {journal}
  {arXiv e-prints}\ ,\ \bibinfo {eid} {arXiv:2411.15495}} (\bibinfo {year} {2024})},\ \Eprint {http://arxiv.org/abs/2411.15495} {arXiv:2411.15495 [astro-ph.GA]} \BibitemShut {NoStop}%
\bibitem [{\citenamefont {Paillas}\ \emph {et~al.}(2023)\citenamefont {Paillas} \emph {et~al.}}]{Paillas:2023cpk}%
  \BibitemOpen
  \bibfield  {author} {\bibinfo {author} {\bibfnamefont {E.}~\bibnamefont {Paillas}} \emph {et~al.},\ }\href@noop {} {\  (\bibinfo {year} {2023})},\ \Eprint {http://arxiv.org/abs/2309.16541} {arXiv:2309.16541 [astro-ph.CO]} \BibitemShut {NoStop}%
\bibitem [{\citenamefont {Hearin}\ \emph {et~al.}(2016)\citenamefont {Hearin}, \citenamefont {Zentner}, \citenamefont {van~den Bosch}, \citenamefont {Campbell},\ and\ \citenamefont {Tollerud}}]{Hearin:2015jnf}%
  \BibitemOpen
  \bibfield  {author} {\bibinfo {author} {\bibfnamefont {A.~P.}\ \bibnamefont {Hearin}}, \bibinfo {author} {\bibfnamefont {A.~R.}\ \bibnamefont {Zentner}}, \bibinfo {author} {\bibfnamefont {F.~C.}\ \bibnamefont {van~den Bosch}}, \bibinfo {author} {\bibfnamefont {D.}~\bibnamefont {Campbell}}, \ and\ \bibinfo {author} {\bibfnamefont {E.}~\bibnamefont {Tollerud}},\ }\href {\doibase 10.1093/mnras/stw840} {\bibfield  {journal} {\bibinfo  {journal} {Mon. Not. Roy. Astron. Soc.}\ }\textbf {\bibinfo {volume} {460}},\ \bibinfo {pages} {2552} (\bibinfo {year} {2016})},\ \Eprint {http://arxiv.org/abs/1512.03050} {arXiv:1512.03050 [astro-ph.CO]} \BibitemShut {NoStop}%
\bibitem [{\citenamefont {Yuan}\ \emph {et~al.}(2022{\natexlab{b}})\citenamefont {Yuan}, \citenamefont {Garrison}, \citenamefont {Hadzhiyska}, \citenamefont {Bose},\ and\ \citenamefont {Eisenstein}}]{Yuan:2021izi}%
  \BibitemOpen
  \bibfield  {author} {\bibinfo {author} {\bibfnamefont {S.}~\bibnamefont {Yuan}}, \bibinfo {author} {\bibfnamefont {L.~H.}\ \bibnamefont {Garrison}}, \bibinfo {author} {\bibfnamefont {B.}~\bibnamefont {Hadzhiyska}}, \bibinfo {author} {\bibfnamefont {S.}~\bibnamefont {Bose}}, \ and\ \bibinfo {author} {\bibfnamefont {D.~J.}\ \bibnamefont {Eisenstein}},\ }\href {\doibase 10.1093/mnras/stab3355} {\bibfield  {journal} {\bibinfo  {journal} {Mon. Not. Roy. Astron. Soc.}\ }\textbf {\bibinfo {volume} {510}},\ \bibinfo {pages} {3301} (\bibinfo {year} {2022}{\natexlab{b}})},\ \Eprint {http://arxiv.org/abs/2110.11412} {arXiv:2110.11412 [astro-ph.CO]} \BibitemShut {NoStop}%
\bibitem [{\citenamefont {Yuan}\ \emph {et~al.}(2023)\citenamefont {Yuan} \emph {et~al.}}]{DESI:2023ujh}%
  \BibitemOpen
  \bibfield  {author} {\bibinfo {author} {\bibfnamefont {S.}~\bibnamefont {Yuan}} \emph {et~al.} (\bibinfo {collaboration} {DESI}),\ }\href@noop {} {\  (\bibinfo {year} {2023})},\ \Eprint {http://arxiv.org/abs/2310.09329} {arXiv:2310.09329 [astro-ph.CO]} \BibitemShut {NoStop}%
\bibitem [{\citenamefont {Baldauf}\ \emph {et~al.}(2016{\natexlab{a}})\citenamefont {Baldauf}, \citenamefont {Schaan},\ and\ \citenamefont {Zaldarriaga}}]{Baldauf:2015tla}%
  \BibitemOpen
  \bibfield  {author} {\bibinfo {author} {\bibfnamefont {T.}~\bibnamefont {Baldauf}}, \bibinfo {author} {\bibfnamefont {E.}~\bibnamefont {Schaan}}, \ and\ \bibinfo {author} {\bibfnamefont {M.}~\bibnamefont {Zaldarriaga}},\ }\href {\doibase 10.1088/1475-7516/2016/03/017} {\bibfield  {journal} {\bibinfo  {journal} {JCAP}\ }\textbf {\bibinfo {volume} {03}},\ \bibinfo {pages} {017} (\bibinfo {year} {2016}{\natexlab{a}})},\ \Eprint {http://arxiv.org/abs/1505.07098} {arXiv:1505.07098 [astro-ph.CO]} \BibitemShut {NoStop}%
\bibitem [{\citenamefont {Baldauf}\ \emph {et~al.}(2016{\natexlab{b}})\citenamefont {Baldauf}, \citenamefont {Schaan},\ and\ \citenamefont {Zaldarriaga}}]{Baldauf:2015zga}%
  \BibitemOpen
  \bibfield  {author} {\bibinfo {author} {\bibfnamefont {T.}~\bibnamefont {Baldauf}}, \bibinfo {author} {\bibfnamefont {E.}~\bibnamefont {Schaan}}, \ and\ \bibinfo {author} {\bibfnamefont {M.}~\bibnamefont {Zaldarriaga}},\ }\href {\doibase 10.1088/1475-7516/2016/03/007} {\bibfield  {journal} {\bibinfo  {journal} {JCAP}\ }\textbf {\bibinfo {volume} {03}},\ \bibinfo {pages} {007} (\bibinfo {year} {2016}{\natexlab{b}})},\ \Eprint {http://arxiv.org/abs/1507.02255} {arXiv:1507.02255 [astro-ph.CO]} \BibitemShut {NoStop}%
\bibitem [{\citenamefont {Chudaykin}\ \emph {et~al.}(2024)\citenamefont {Chudaykin}, \citenamefont {Ivanov},\ and\ \citenamefont {Nishimichi}}]{Chudaykin:2024wlw}%
  \BibitemOpen
  \bibfield  {author} {\bibinfo {author} {\bibfnamefont {A.}~\bibnamefont {Chudaykin}}, \bibinfo {author} {\bibfnamefont {M.~M.}\ \bibnamefont {Ivanov}}, \ and\ \bibinfo {author} {\bibfnamefont {T.}~\bibnamefont {Nishimichi}},\ }\href@noop {} {\  (\bibinfo {year} {2024})},\ \Eprint {http://arxiv.org/abs/2410.16358} {arXiv:2410.16358 [astro-ph.CO]} \BibitemShut {NoStop}%
\bibitem [{\citenamefont {Chudaykin}\ and\ \citenamefont {Ivanov}(2019)}]{Chudaykin:2019ock}%
  \BibitemOpen
  \bibfield  {author} {\bibinfo {author} {\bibfnamefont {A.}~\bibnamefont {Chudaykin}}\ and\ \bibinfo {author} {\bibfnamefont {M.~M.}\ \bibnamefont {Ivanov}},\ }\href {\doibase 10.1088/1475-7516/2019/11/034} {\bibfield  {journal} {\bibinfo  {journal} {JCAP}\ }\textbf {\bibinfo {volume} {11}},\ \bibinfo {pages} {034} (\bibinfo {year} {2019})},\ \Eprint {http://arxiv.org/abs/1907.06666} {arXiv:1907.06666 [astro-ph.CO]} \BibitemShut {NoStop}%
\bibitem [{\citenamefont {Sailer}\ \emph {et~al.}(2021)\citenamefont {Sailer}, \citenamefont {Castorina}, \citenamefont {Ferraro},\ and\ \citenamefont {White}}]{Sailer:2021yzm}%
  \BibitemOpen
  \bibfield  {author} {\bibinfo {author} {\bibfnamefont {N.}~\bibnamefont {Sailer}}, \bibinfo {author} {\bibfnamefont {E.}~\bibnamefont {Castorina}}, \bibinfo {author} {\bibfnamefont {S.}~\bibnamefont {Ferraro}}, \ and\ \bibinfo {author} {\bibfnamefont {M.}~\bibnamefont {White}},\ }\href {\doibase 10.1088/1475-7516/2021/12/049} {\bibfield  {journal} {\bibinfo  {journal} {JCAP}\ }\textbf {\bibinfo {volume} {12}},\ \bibinfo {pages} {049} (\bibinfo {year} {2021})},\ \Eprint {http://arxiv.org/abs/2106.09713} {arXiv:2106.09713 [astro-ph.CO]} \BibitemShut {NoStop}%
\bibitem [{\citenamefont {Cabass}\ \emph {et~al.}(2022)\citenamefont {Cabass}, \citenamefont {Ivanov}, \citenamefont {Philcox}, \citenamefont {Simonovic},\ and\ \citenamefont {Zaldarriaga}}]{Cabass:2022epm}%
  \BibitemOpen
  \bibfield  {author} {\bibinfo {author} {\bibfnamefont {G.}~\bibnamefont {Cabass}}, \bibinfo {author} {\bibfnamefont {M.~M.}\ \bibnamefont {Ivanov}}, \bibinfo {author} {\bibfnamefont {O.~H.~E.}\ \bibnamefont {Philcox}}, \bibinfo {author} {\bibfnamefont {M.}~\bibnamefont {Simonovic}}, \ and\ \bibinfo {author} {\bibfnamefont {M.}~\bibnamefont {Zaldarriaga}},\ }\href@noop {} {\  (\bibinfo {year} {2022})},\ \Eprint {http://arxiv.org/abs/2211.14899} {arXiv:2211.14899 [astro-ph.CO]} \BibitemShut {NoStop}%
\bibitem [{\citenamefont {Akitsu}(2024)}]{Akitsu:2024lyt}%
  \BibitemOpen
  \bibfield  {author} {\bibinfo {author} {\bibfnamefont {K.}~\bibnamefont {Akitsu}},\ }\href@noop {} {\  (\bibinfo {year} {2024})},\ \Eprint {http://arxiv.org/abs/2410.08998} {arXiv:2410.08998 [astro-ph.CO]} \BibitemShut {NoStop}%
\bibitem [{\citenamefont {Zhang}\ \emph {et~al.}(2025)\citenamefont {Zhang}, \citenamefont {Bonici}, \citenamefont {D'Amico}, \citenamefont {Paradiso},\ and\ \citenamefont {Percival}}]{Zhang:2024thl}%
  \BibitemOpen
  \bibfield  {author} {\bibinfo {author} {\bibfnamefont {H.}~\bibnamefont {Zhang}}, \bibinfo {author} {\bibfnamefont {M.}~\bibnamefont {Bonici}}, \bibinfo {author} {\bibfnamefont {G.}~\bibnamefont {D'Amico}}, \bibinfo {author} {\bibfnamefont {S.}~\bibnamefont {Paradiso}}, \ and\ \bibinfo {author} {\bibfnamefont {W.~J.}\ \bibnamefont {Percival}},\ }\href {\doibase 10.1088/1475-7516/2025/04/041} {\bibfield  {journal} {\bibinfo  {journal} {JCAP}\ }\textbf {\bibinfo {volume} {04}},\ \bibinfo {pages} {041} (\bibinfo {year} {2025})},\ \Eprint {http://arxiv.org/abs/2409.12937} {arXiv:2409.12937 [astro-ph.CO]} \BibitemShut {NoStop}%
\bibitem [{\citenamefont {{Shiferaw}}\ \emph {et~al.}(2024)\citenamefont {{Shiferaw}}, \citenamefont {{Kokron}},\ and\ \citenamefont {{Wechsler}}}]{Shiferaw:2024_flbias_hydro}%
  \BibitemOpen
  \bibfield  {author} {\bibinfo {author} {\bibfnamefont {M.}~\bibnamefont {{Shiferaw}}}, \bibinfo {author} {\bibfnamefont {N.}~\bibnamefont {{Kokron}}}, \ and\ \bibinfo {author} {\bibfnamefont {R.~H.}\ \bibnamefont {{Wechsler}}},\ }\href {\doibase 10.48550/arXiv.2412.06886} {\bibfield  {journal} {\bibinfo  {journal} {arXiv e-prints}\ ,\ \bibinfo {eid} {arXiv:2412.06886}} (\bibinfo {year} {2024})},\ \Eprint {http://arxiv.org/abs/2412.06886} {arXiv:2412.06886 [astro-ph.CO]} \BibitemShut {NoStop}%
\end{thebibliography}%

\end{document}